\documentclass[A4,12pt]{article}
\usepackage{setspace}
\usepackage{amsmath}
\usepackage[top=1in,bottom=1in,right=1in,left=0.5in]{geometry}
\usepackage{caption}
\usepackage{psfrag}
\usepackage{graphicx}
\usepackage{hyperref}
\usepackage{subfigure}
\usepackage{authblk}
\usepackage{verbatim}
\usepackage{multicol}
\usepackage{bibcheck}
\usepackage[natbib=true,sorting=none,style=nature]{biblatex}
\title{Hydrodynamical properties of baryon rich thermal plasma with flavour quarks}
\author{Rishi Pokhrel\thanks{E-Mail: rishipokhrel.smit@gmail.com and rishi\_20211037@smit.smu.edu.in} }

\author{Tanay K. Dey\thanks{E-mail: tanay.dey@gmail.com and tanay.d@smit.smu.edu.in}}
\affil{Department of Physics, Sikkim Manipal Institute of Technology, Sikkim Manipal University, Majitar, Rangpo, East Sikkim, 737136, India.}

\date{}
\begin{document}
	
	\maketitle

	\begin{abstract}
In this work, we holographically study the hydrodynamical properties of strongly coupled  $\mathcal{N} = 4$ SYM baryon rich thermal plasma with large number of flavour quarks. Specifically, we study the drag force acting on the moving heavy probe quark and corresponding energy loss. We also study the jet quenching parameter, screening length and binding energy of the quark-antiquark pair. Due to the presence of finite baryon density and flavour quarks the drag force, energy loss, jet quenching parameter and binding energy of the quark-antiquark pair are enhanced for the increase in temperature. However, the screening length of the quark-antiquark pair is reduced, leading to the thermal plasma phase being achieved at a lower temperature, which is consistent with the thermal phase diagram of the quark-gluon plasma.
We observed that the perpendicular orientation of quark-antiquark pair with respect to the direction of motion deconfined early compare to the parallel orientation once temperature raises.
	\end{abstract}

\clearpage



\section{Introduction}
The duality between strongly coupled $\mathcal{N}=4$ Super Yang-Mills (SYM) gauge theories residing on the boundary of the weakly coupled gravity theory and vice versa, are known as the gauge/gravity duality \cite{Maldacena1998,Maldacena1999,Witten1998}. Since its inception, researchers have used this duality to understand various classes of strongly coupled gauge theories through corresponding weakly coupled dual gravity descriptions, as the conventional field-theoretic approach to studying strongly coupled gauge theories has proven inadequate for addressing many issues in such systems.
 The Relativistic Heavy Ion Collider (RHIC) and Large Hadron Collider (LHC) suggests the creation of Quark-Gluon Plasma (QGP) at high number density and temperature \cite{Zajc_2008,muller2007,Shuryak_2007,d_Enterria_2007, salgado2006, Shuryak_2007_report,Tannenbaum_2006,Muller_2006,Gyulassy_2005,Adcox_2005,Arsene_2005,Back_2005,Adams_2005}, which is a deconfined state of quarks and gluons. Interaction of an external high energetic parton probe with the QGP medium indicates the strong coupling between them \cite{Baier_1997,Eskola_2005}. Due to the strong coupling nature of the interaction between the probe and the QGP medium, it is difficult to study and estimate various hydrodynamic properties like drag force, jet quenching, energy loss etc., of the probe quark using the field theory approach. However, using gauge/gravity duality various research work has explored different aspects of the gauge theory. In the framework of AdS/CFT duality, end point of a fundamental string attached to the boundary of the AdS black hole represents an external heavy quark. The string end point carry a fundamental $SU(N)$ charge and the string elongates towards the radial direction into the bulk. The length of the string is proportional to the mass of the quark and the trailing of string represents the energy loss by imparting a drag force. The drag force experienced by an external heavy quark moving through the $\mathcal{N}=4$ SYM thermal plasma has been discussed in \cite{Herzog_2006,Gubser_2006,Caceres2006}. Along this line further generalisation is reported in \cite{Casalderrey_Solana_2006,Matsuo_2006,Talavera_2007,Nakano_2007}. The prescription for the calculation of the jet quenching parameter described in \cite{Liu2006} has been used by many to study the jet quenching in various setup \cite{Caceres2006a,Nakano_2007,Bertoldi_2007}. Heavy probe meson moving through $\mathcal{N}=4$ SYM thermal plasma has been considered and the binding energy and screening length of quark-antiquark $(q\bar{q})$ pair has been studied in \cite{Chernicoff_2006,Chernicoff2008,Liu_2007a}. The energy loss of a rotating external quark in $\mathcal{N}=4$ SYM thermal plasma has been studied in \cite{Fadafan_2009,Athanasiou_2010,Herzog_2006,Herzog_2007}.\\

A $5$-dimensional bottom up holographic model was proposed in \cite{DeWolfe:2010he} as a phenomenological gravity dual for the thermal QGP and baryon chemical potential in the strongly coupled regime. In \cite{Chakrabortty2011a}, a gravity dual for the QGP at finite temperature with large number of external heavy quarks (flavour quarks say charm) has been constructed. The external heavy quarks are represented by long strings stretched from boundary to horizon of the gravity dual. The string cloud  back reacted on the dual AdS bulk theory and produces a deformed AdS spacetime. They have studied the drag force experienced by an external probe quark in this deformed spacetime. In \cite{Chakrabortty2016a}, the jet quenching parameter, energy loss, screening length has been holographically studied for the dual back reacted $\mathcal{N}=4$ SYM thermal plasma theory of the above gravity dual. Further in \cite{Zhang_2024}, Gauss-Bonnet AdS black hole with string cloud is considered to study the screening length of quark-antiquark pair in the dual gauge theory. Along the same line, in \cite{Pokhrel2023}, we have considered charged AdS black hole with string cloud as the gravity dual of back reacted $\mathcal{N}=4$ SYM  thermal plasma theory with baryon chemical potential and studied the thermodynamics of it. It has been noticed that there is no existence of thermal AdS space even in zero temperature which leads to the non existence of bound state in the boundary theory. However, by studying the screening length of $q\bar{q}$ pair in the boundary theory against the size of the black hole in the bulk theory it has been shown that at low temperature, bound state of $q\bar{q}$ pair exists and as temperature increases it breaks into an unbound state in the boundary theory.\\

In this work, we have considered baryon rich back reacted thermal QGP in the strongly coupled regime as the gauge theory dual of charged AdS black hole with cloud of strings. Considering such a gravity dual, we intend to study various properties in the corresponding back reacted $\mathcal{N}=4$ SYM  thermal plasma theory with baryon chemical potential. First of all, the drag force experienced by an external moving heavy quark is realised as a string whose one end is attached to the boundary corresponding to the quark and the body of the string hanging into the bulk. The external quark is directed to move on the boundary of the AdS spacetime, such that the body of the string trails behind and the quark moves ahead, generating drag force on the moving quark. We study the behaviour of the drag force as a function of the string cloud density, temperature, chemical potential and velocity. It is observed that the drag force experienced by the quark increases with increase of all the parameters. The drag force enhances due to the thermal agitation and damping effect of the plasma medium and back reaction of the heavy quarks and finite baryon density. 

The jet quenching parameter which is closely related to the energy loss of the heavy $q\bar{q}$ pair in the presence of thermal medium has been studied as a function of the string cloud density, temperature and chemical potential. The $q\bar{q}$ pair is represented by the two ends of the same string hanging from the boundary of the bulk. The jet quenching parameter raises with the raise of string cloud density, temperature and chemical potential which indicates the enhancement of the energy loss due to the suppression of external heavy probe quark moving with high transverse momentum. 

The separation distance between the $q\bar{q}$ pair moving with some speed in the  back reacted $\mathcal{N}=4$ SYM thermal plasma with chemical potential has been studied. It is observed that the $q\bar{q}$ separation distance increases monotonically up to a maximum value for a certain value of constant of motion and then decreases. The maximum value attained by the separation distance between the pair is called as the screening length. Beyond this length the pair gets screened in the plasma medium or they get separated from each other with no binding energy. We have considered both perpendicular and parallel orientation of the $q\bar{q}$ pair with respect to the direction of motion and have studied the screening length. In both the orientations it is observed that the screening length decreases with increase in the rapidity parameter, string cloud density, chemical potential and temperature. Therefore, the chemical potential triggers the phase transition towards the lower temperature which is consistent with the phase diagram of thermal QGP. The reduction of screening length is more prominent for smaller value of rapidity parameter. Comparing both the orientation of the pair, we observe that the value of the screening length for parallel set up is larger than the perpendicular set up and the bound state of the $q\bar{q}$ pair gets screened easily in perpendicular orientation.

Then, we have considered a heavy probe quark rotating with constant angular speed and studied the radial profile of the corresponding string and energy loss with respect to the velocity for different values of the string cloud density and chemical potential. For fixed value of chemical potential and string density, the radius of the probe string is constant from boundary to horizon of the bulk spacetime if the angular velocity less than one. Otherwise the radius increases. The radius becomes smaller as the baryon and flavour quark density increases keeping the angular velocity fixed. The rate of rotational or drag energy loss are increased for the increase of the chemical potential and flavour quark density and velocity.  The rate of drag energy loss also increases with the temperature of the medium. The enhancement rate is more for temperature and chemical potential than the density of flavour quark. Then we have calculated the rate of energy loss at zero temperature called vacuum energy loss in absence of plasma, baryons and flavour quarks. This energy loss is only dependent on the velocity of the external probe quark and is nothing but the Lienard’s electromagnetic radiation. We have further computed the ratio of rotational energy loss with vacuum energy loss and drag energy loss with the vacuum energy loss. For both the cases, it is observed that when the external probe quark moves slowly, the rotational and drag energy loss are dominating over the vacuum energy loss. The dominance is higher for the higher value of chemical potential and flavour quark density and it is lower for the higher angular velocity of the external probe quark. Once the linear speed of the probe quark increases the dominance start to reduce and approaches towards zero value at the speed of light due to the destructive interference of the two radiation with opposite phase. Finally, we have plotted the ratio of energy loss due to rotation and drag and it shows that at the lower speed of the probe quark both the radiation are radiated equally and near to the velocity of light the radiations are interfering constructively.\\

This work is organized as follows: in section (\ref{dragforce}) we first compute the drag force experienced by an external probe quark linearly moving through the finite temperature back reacted plasma with chemical potential. In section (\ref{jetquenching}), the jet quenching parameter has been studied and in section (\ref{screeninglength}) we have studied the screening length for perpendicular and parallel orientations of the $q\bar{q}$ with respect to the direction of motion of the pair. In section (\ref{radialprofile}), we study the radial profile of the rotating quark and in (\ref{energyloss}), we discussed the energy loss of a probe quark due to its linear and circular motion in the quark gluon plasma. Finally, in (\ref{discussion}), we summarise our results.


\section{Dissipative force on an external quark}
\label{dragforce}
In this section we compute the drag force experienced by an external probe quark moving with constant velocity in the  $\mathcal{ N} = 4 $ SYM  finite temperature plasma theory having baryon chemical potential and cloud of massive quarks using gauge/gravity duality. Following this duality, we study the four dimensional gauge theory residing on the boundary of a five dimensional charged AdS black hole spacetime with flat boundary. Following \cite{Pokhrel2023}, the flat metric solution can be expressed as,
\begin{equation}
\label{metricsol}
    ds^2 = f(u) \left[-h(u) dt^2 + dx^2 + dy^2 + dz^2 + \frac{du^2}{h(u)}\right],
\end{equation}
 where,
\begin{align}
    f(u) &= \frac{l^2}{u^2}\\
    h(u) &= 1 - \frac{m}{l^6}u^4 + \frac{q^2}{l^{10}}u^6 - \frac{16\pi G_5 a}{3 l^4} u^3,
\end{align}
where $(x,y,z)$ are the spatial coordinates along which the D3-brane is extended, $ u $ is the bulk radial coordinate with boundary at $ u = 0$, $l$ is the cosmological length scale, $m$ is the constant related to the ADM mass of the black hole and $q$ is the Maxwell charge, $G_5$ is the five dimensional gravitational constant and $a$ is the string cloud density.

\begin{figure}[!h]
	\centering
	\subfigure[]{\includegraphics[scale=0.10]{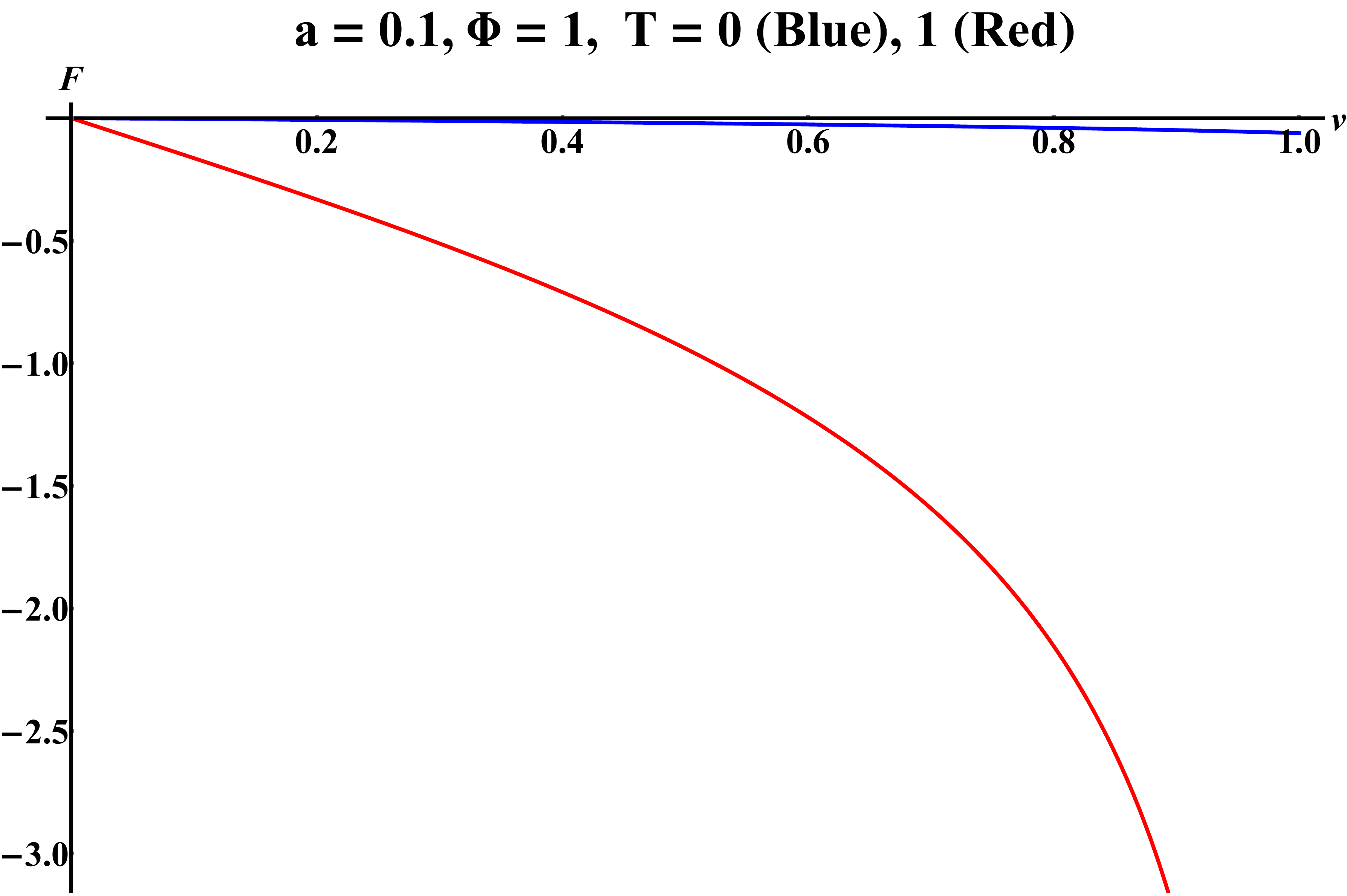}}
	\subfigure[]{\includegraphics[scale=0.09]{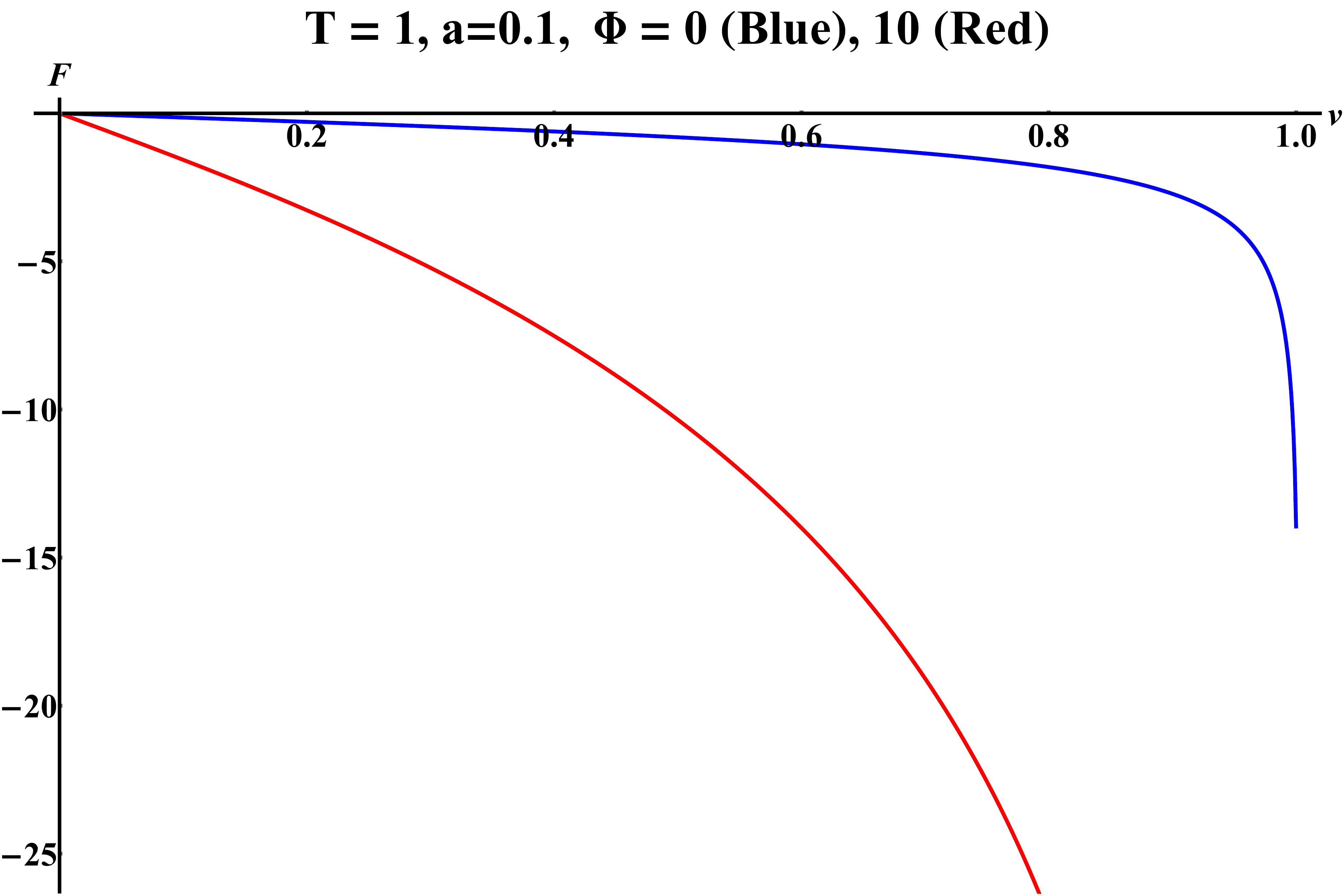}}
    \subfigure[]{\includegraphics[scale=0.09]{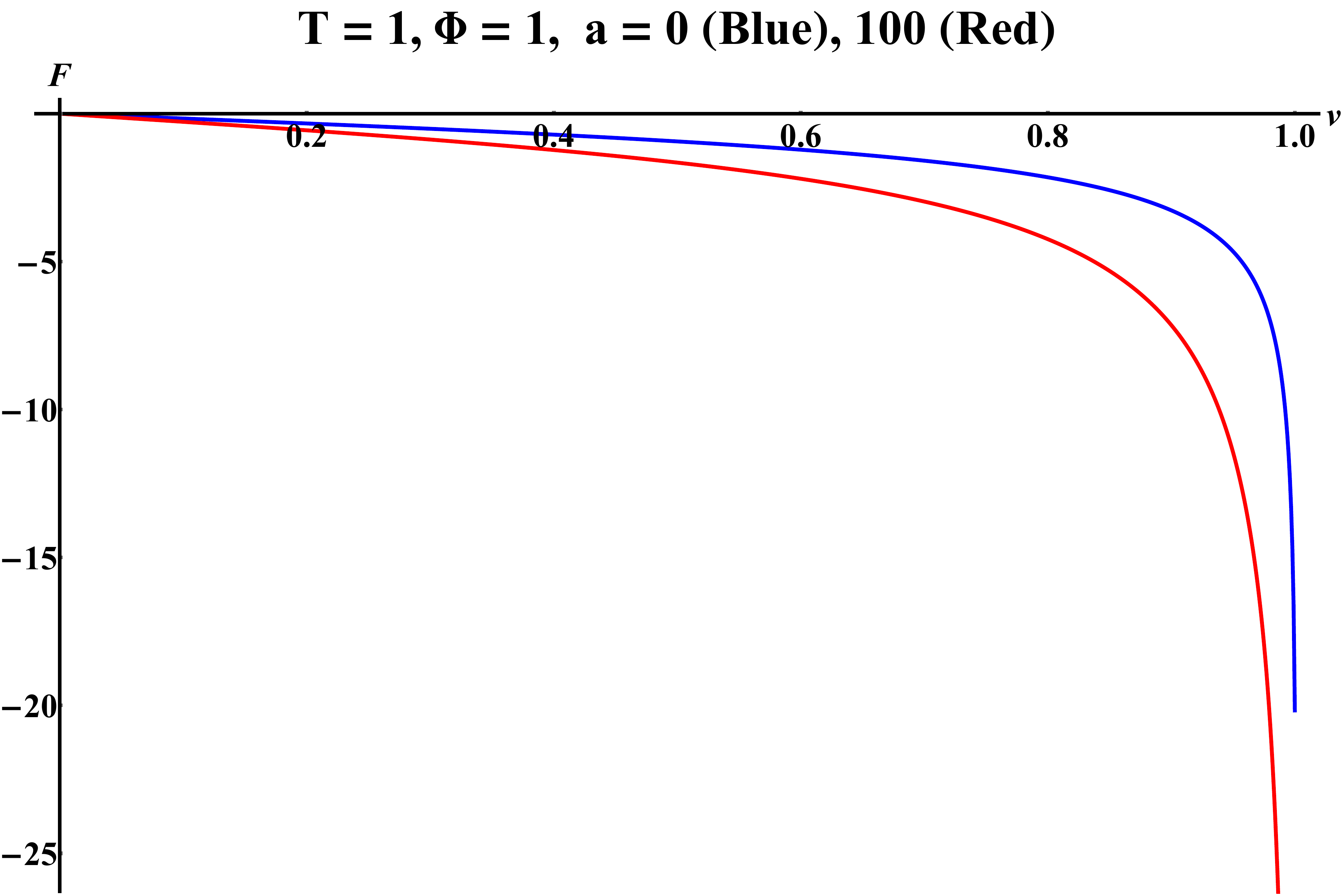}}
	\caption{
	Dissipative Force ($F$) vs $v$ for fixed (a) $a = 0.1$, $\Phi=1$ and different temperature $T$. (b) $T = 1$, $a=0.1$ and different potential $\Phi$. (c) $T = 1$, $\Phi=1$ and different string density $a$.}
\label{F_Vs_v_plots}
\end{figure}

The drag or dissipative force on the probe quark originates due to the loss of momentum of the boundary quark which is flowing along the corresponding string from boundary to the horizon of the bulk geometry. The motion of the hanging string can be described by the Nambu-Goto action as \cite{Gubser_2006, Chakrabortty2011a},
\begin{equation}
	S(\mathcal {C}) = -\frac{1}{2\pi \alpha}\int d^2\sigma \sqrt{-det g_{\gamma \beta }},
	\label{Nambu_Goto_action}
\end{equation}
where $\alpha$ is related with the string tension and $ g_{\gamma \beta} $ is the induced world-sheet metric and
is given as,
\begin{equation}
g_{\gamma \beta } \equiv G_{\mu \nu} \partial_\gamma X^\mu \partial_\beta X^\nu.
\end{equation}
Here $G_{\mu\nu}$  is the background metric given in the equation (\ref{metricsol}) and $x^{\gamma}$ with $\gamma = 0, 1$ are
the world-sheet coordinates $\tau = t $ and $\sigma = u $. In the dual theory, we prefer to work in static gauge along with the following boundary conditions,
\begin{equation}
X^\mu(\tau, \sigma) = (t = \tau, u = \sigma, x = vt + \xi(\sigma), y = 0, z = 0),
\label{gaugefixing} 
\end{equation}
where, $\xi(u)$ depends on the radial coordinate of the bulk theory. This factor arises due to the trailing profile of the string with $\xi(u = 0) = 0$. Here, we consider unidirectional linear motion of the boundary quark along the $x$-direction of the boundary coordinates. The induced action of the string in this set up becomes,
    \begin{equation}
        \begin{split}
            S(\mathcal {C}) = -\frac{1}{2\pi \alpha}\int dt du \left[\frac{f(u)^2}{h(u)}\left[h(u) + h(u)^2 \xi^{\prime 2}-v^2\right]\right]^{\frac{1}{2}}.
            \label{equation_dissipative_force_nambu_goto_action_in_terms_of_xi_and_v}
            \end{split}
        \end{equation}
        Here prime ($\prime$) denotes the derivative with respect to radial coordinate $u$. From (\ref{equation_dissipative_force_nambu_goto_action_in_terms_of_xi_and_v}), it is observed that the action is not an explicit function of $\xi$. Hence the conjugate momenta corresponding to $\xi(u)$ will be a constant and it can be expressed as,
        \begin{equation}
            \Pi_\xi = \frac{\partial L}{\partial \xi'} = -\frac{1}{2\pi \alpha} \frac{f(u)^2 h(u)}{\sqrt{-det g_{\gamma \beta}}} \xi^\prime.
        \end{equation}
        Rearranging the above equation we can write $\xi'$ as,
    \begin{equation}
        \xi^\prime = \frac{2 \pi \alpha \Pi_\xi}{f(u) h(u)}\sqrt{\frac{(h(u)-v^2)}{\left(h(u) - (2\pi \alpha)^2 \frac{\Pi_\xi^2}{f(u)^2}\right)}}.
    \end{equation}
    Here, we consider only positive sign to allow the probe string to maintain the trailing profile. Since $\xi$ should be real, the numerator and denominator of under root have to be zero at the same position of the radial coordinate called $u_v$ which set up the following two constraints, 
    \begin{equation}
        \begin{split}
       h(u_v) - v^2  = 0,\\
       h(u_v)f(u_v)^2 - (2\pi \alpha)^2 \Pi_\xi^2=0.
        \label{equation_dissipative_force_polynomial}
        \end{split} 
    \end{equation}
   From these two constraints $\Pi_\xi$ can be written as, 
    \begin{equation}
        \Pi_\xi = \pm \frac{1}{2\pi \alpha} \frac{v l^2}{u_v^2}.
    \end{equation} 
As discussed in \cite{Gubser_2006}, the dissipative force is equivalent to the rate of loss of momentum of the boundary probe quark during the interaction with the medium. According to the gauge/gravity duality the loss of momentum is mapped with the flow of total string momentum from boundary to the horizon. Using the constraint of conservation of world-sheet current of spacetime energy-momentum around a closed loop on the test string whose one end is fixed and the other one is free, we can write the dissipative force due to the flow of world-sheet current as,
\begin{equation}
 F = -\frac{1}{2\pi \alpha}\frac{v l^2}{u_v^2}.
 \label{equation_dissipative_force_dissipative_force}
 \end{equation}
 In order to write the dissipative force in terms of the boundary gauge theory parameters, we use the standard AdS/CFT relation,
  \begin{equation}
    \frac{l^4}{\alpha^2} = g^2_{YM} N,
\end{equation}
where $g^2_{YM}$ and $N$ are the Yang-Mills gauge coupling and order of the gauge group respectively. The drag force of equation (\ref{equation_dissipative_force_dissipative_force}) now takes the form as,
\begin{equation}
    F =-\frac{\sqrt{g_{YM}^2 N}}{2\pi}  \left(\frac{v}{u_v^2}\right).
	\label{equation_dissipative_force_dissipative_force_in_terms_of_rv_by_rh}
\end{equation} 
 The dissipative force depends on the solution $u_v$  of equation,
\begin{equation}    
h(u_v) - v^2 = 1 - \frac{m}{l^6}u_v^4 + \frac{q^2}{l^{10}}u_v^6 - \frac{16\pi G_5 a}{3 l^4} u_v^3 - v^2 = 0.
\end{equation}  
Since this is a sextic polynomial of $u_v$, it is difficult to find out the analytical solution of this equation. However, the solution can be written as a function of chemical potential, temperature and string density. Therefore, dissipative force can be represented as,
\begin{equation}
    F =-\frac{\sqrt{g_{YM}^2 N}}{2\pi}  \left(\frac{v}{f^2(T,\Phi,a)}\right),
	\label{equation_dissipative_force_dissipative_force_in_terms_of_rv_by_rh}
\end{equation} 
where, chemical potential $\Phi$ is related with the Maxwell charge of the black hole in the following way, 
  \begin{equation}
  \Phi = \frac{\sqrt 3}{2}\frac{q u_h^2}{l^4},
\end{equation}
here, $u_h$ is the solution of equation $h(u_h)=0$ and called black hole horizon. $T$ is the temperature of the black hole and it can be computed from the metric solution (\ref{metricsol}) and it takes the form as,
 \begin{equation}
 T = \frac{6 l^4-a u_h^3-4 l^2 \Phi^2 u_h^2}{6 \pi  l^4 u_h}.
 \end{equation}
Finally, the dissipative force (\ref{equation_dissipative_force_dissipative_force_in_terms_of_rv_by_rh}) can be studied for different values of $T, \Phi , a \text{ and } v$  of the probe quark. In figure (\ref{F_Vs_v_plots}), we plot the dissipative force with respect to the linear velocity $v$ while  keeping two variables fixed among the rest of the three variables and assigning two different values for the remaining one variable. From all the plots, it is observed that dissipative force always increases with the increase of any variables. Dissipative force increases with velocity since damping effect due to the plasma increases and drag force increases with string density and chemical potential since back reaction plays the role here. Thermal agitation of plasma slows down the motion.\\


\section{Jet Quenching Parameter}
\label{jetquenching}
Following \cite{Liu:2006ug}, in this section holographically we compute the jet quenching parameter $(\hat{q})$. Phenomenologically, the jet quenching parameter is connected to the energy loss of probe quark caused by the suppression of heavy quark with high transverse momentum in the QGP medium at finite temperature. In field theoretic point of view, the relation between the jet quenching parameter and the expectation value of light-like Wilson loop in the adjoint representation can be established as \cite{Liu:2006he}
\begin{equation}
	\left< W^A(\mathcal{C})\right> = e^{-\frac{1}{4\sqrt{2}}\hat{q} L^- L^2},
	\label{equation_jet_quenching_Wilson_loop_expect_adjoint}
\end{equation}
where $\mathcal{C}$ is traced out by the $q\bar{q}$ separation length $L$ and length $L^-$ along the light cone of the boundary gauge theory. However the calculation of expectation value of light-like Wilson loop is very difficult caused by the lack of systematic formulation of the technique. But, using the  gauge/gravity duality, we can compute the expectation value of the Wilson loop in the fundamental representation \cite{Liu:2006he,Maldacena1998},
\begin{equation}
	\left< W^F (\mathcal{C})\right> = e^{i S(\mathcal{C})}.
	\label{equation_jet_quenching_Wilson_loop_expect_funda_Nambu_goto}
\end{equation}
Where, $S(\mathcal{C})$ is the Nambu-Goto action of the fundamental string whose two end points are attached to the boundary corresponding to the $q\bar{q}$ pair. The action of equation (\ref{Nambu_Goto_action}) can be represented as the Nambu-Goto action of the aforesaid string. Furthermore, from the group theoretical identity, the relation between the expectation value of fundamental representation and adjoint representation of Wilson loop can be constructed and it takes the simple form as,
 \begin{equation}
	\left< W^F(\mathcal{C})\right>^2 = \left< W^A(\mathcal{C})\right>.
	\label{equation_jet_quenching_Wilson_loop_expect_funda_adjoint_relation}
 \end{equation}
 Now from equations (\ref{equation_jet_quenching_Wilson_loop_expect_adjoint}, \ref{equation_jet_quenching_Wilson_loop_expect_funda_Nambu_goto}) and (\ref{equation_jet_quenching_Wilson_loop_expect_funda_adjoint_relation}), the jet quenching parameter ($\hat{q}$) is defined as,
\begin{equation}
	\hat{q} = -\frac{8 \sqrt{2}i}{L^- L^2}(S-S_0).
\end{equation}
Here $S_0$ is the action for the self energy contribution of the total mass of $q\bar{q}$ pair. To find out the action we express the background metric of equation (\ref{metricsol}) in terms of the light-cone coordinates and it takes the form as,
\begin{equation}
	ds^2 = f(u)\left[-(1+h(u)) dx^+ dx^- + \frac{1}{2}(1-h(u)) \{dx^{+2}+dx^{-2}\}+dy^2 + dz^2 + \frac{du^2}{h(u)}\right].
	\label{equation_jet_quenching_metric_fu_hu}
\end{equation}
Here, $x^\pm$ is the light-cone coordinate defined as,
\begin{equation}
	x^\pm = \frac{t \pm x}{\sqrt{2}}.
\end{equation}
Now, choose the static gauge as follows $\tau = x^- (L^-\ge x^- \ge 0)$, $ \sigma = y (-\frac{L}{2}\le y \le \frac{L}{2} )$, $q\bar{q}$ pair at $y=\pm \frac{L}{2}$,  $x^+ = $ constant and $z = $ constant plane. For these set of gauge choices and by keeping $L^-\gg L$, the string profile becomes an explicit function of $u=u(y)$ and the action of the string obtains the form, 
\begin{equation}
	S = \frac{i L^-}{\sqrt{2} \pi \alpha}\int_{0}^{\frac{L}{2}} dy f\sqrt{(1-h) (1+\frac{u'^2}{h} )},
\label{langrangian}
\end{equation}
where $u'$ is defined in the following way. The action does not explicitly depends on $y$, we can write the Hamiltonian equation as,
\begin{equation}
	\frac{\partial \mathcal{L}}{\partial u'} u' -\mathcal{L}=E.
\end{equation}
Here $\mathcal{L}$ is the Lagrangian density of equation (\ref{langrangian})  and $E$ is the constant of motion. The above equation provide the value of $u'$ as, 
\begin{equation}
	u' = \sqrt{h\left[\frac{f^2 (1-h)}{E^2} -1 \right]}.
\label{extreme}
\end{equation}
Since, all the parameters of equation (\ref{langrangian}) are function of $u$, we convert the integral in terms of $u$ and to set the limit of integral, we apply the boundary conditions on the hanging string along the bulk direction. The boundary conditions at the end points and at the middle of the string respectively are $u(\pm\frac{L}{2})=0$ and  $u' (0)=0$. The second condition arises due to the existence of turning point of the string. Using the second condition in equation (\ref{extreme}), we get the following two conditions as
\begin{equation}
	h = 0 \text{ and } h = 1-\frac{E^2}{f^2}.
\end{equation}
The first condition says that the turning of string occurs at the horizon and the second condition says that the turning point takes place before the horizon and we name it as $u_t$. The depth $u_t$ depends on the value of $E$. It can be shown that the $u'^2$ is always greater or equal to zero  within the range of $u_t$ and horizon.  Whereas, it is negative between the boundary and $u_t$. Therefore, we should neglect the range from boundary to $u_t$. Now using the form of $u'$ of equation (\ref{extreme}) we get the action of equation (\ref{langrangian}) as, 
\begin{equation}
	S = \frac{i L^-}{\sqrt{2} \pi \alpha}\int_{\delta}^{u_h} du f\sqrt{\frac{1-h}{h}}\left[1-\frac{u^4 E^2}{l^4 (1-h)}\right]^{-\frac{1}{2}},
\end{equation}
where $u_t <\delta < u_h $ is the cut-off set on the boundary in the radial direction $u=\delta$. At the end of the calculation we consider $\delta \rightarrow 0$ to get the final result cut-off independent. The self energies of the  $q\bar{q}$ pair must be subtracted  from the above action to get the finite value. The self energy contribution can be calculated by considering world-sheet of the two free straight fundamental string both hanging from boundary to horizon of the spacetime. We also consider the static gauge as $\tau =x^-$ and $\sigma = u$. With these considerations the self energy contribution can be written as,  
\begin{equation}
	S_0 = \frac{i L^-}{ \pi \alpha}\int_{\delta}^{u_h} du \sqrt{G_{uu}G_{--}} = \frac{i L^-}{\sqrt{2} \pi \alpha}\int_{\delta}^{u_h} du f\sqrt{\frac{1-h}{h}}.
\end{equation}
Finally, the regularised action becomes, 
\begin{equation}
	S-S_0 = \frac{i L^-}{\sqrt{2} \pi \alpha}\int_{\delta}^{u_h} du f\sqrt{\frac{1-h}{h}}\left[\left\{1-\frac{u^4 E^2}{l^4 (1-h)}\right\}^{-\frac{1}{2}}-1\right] \equiv \frac{i L^- E^2}{2\sqrt{2} \pi \alpha} I_1,
\label{sszero}
\end{equation}
where, 
\begin{equation}
	I_1 = \int_{\delta}^{u_h} \frac{du}{f\sqrt{h(1-h)}}.
\end{equation}
In the above equation (\ref{sszero}),  we replace the $E$ in terms of $q\bar{q}$ separation distance $L$ to remove the $E$ dependence of the jet quenching parameter. We replace $E$ in the following way,
\begin{equation}
L = \int_{-\frac{L}{2}}^{\frac{L}{2}}dx = 2 \int_{\delta}^{u_h} \frac{du}{u'}= 2\int_{\delta}^{u_h} \frac{du}{\sqrt{h\left[\frac{f^2 (1-h)}{E^2} -1 \right]}}.
\label{qqbardis}
\end{equation}
For small distance of $q\bar{q}$ pair the constant $E$ can be expanded in the power of $L$ by inverting the above equation. $E$ can be estimated in the first order of $L$ as,
\begin{equation}
	E = \frac{L}{2I_1}.
\end{equation}
Using this the regularised action can be written as,
\begin{equation}
	S -S_0 \approx \frac{i L^- L^2}{8 \sqrt{2} \pi \alpha I_1}.
\end{equation}
Corresponding jet quenching parameter becomes
\begin{equation}
	\hat{q} = \frac{1}{\pi \alpha I_1}.
\end{equation}
Using the relation $ \frac{l^4}{\alpha^2} = g^2_{YM} N_c$, the jet quenching parameter is recast as, 
\begin{equation}
	\hat{q} = \frac{\sqrt{g^2_{YM} N_c}}{ \pi I_1'(T,a,\Phi)}.
\end{equation}
Here $I_1' = \int_{\delta}^{u_h(T,a,\Phi)} \frac{u^2 du}{\sqrt{h(1-h)}}$.\\

\begin{figure}[!h]
	\centering
	\subfigure[]{\includegraphics[scale=0.11]{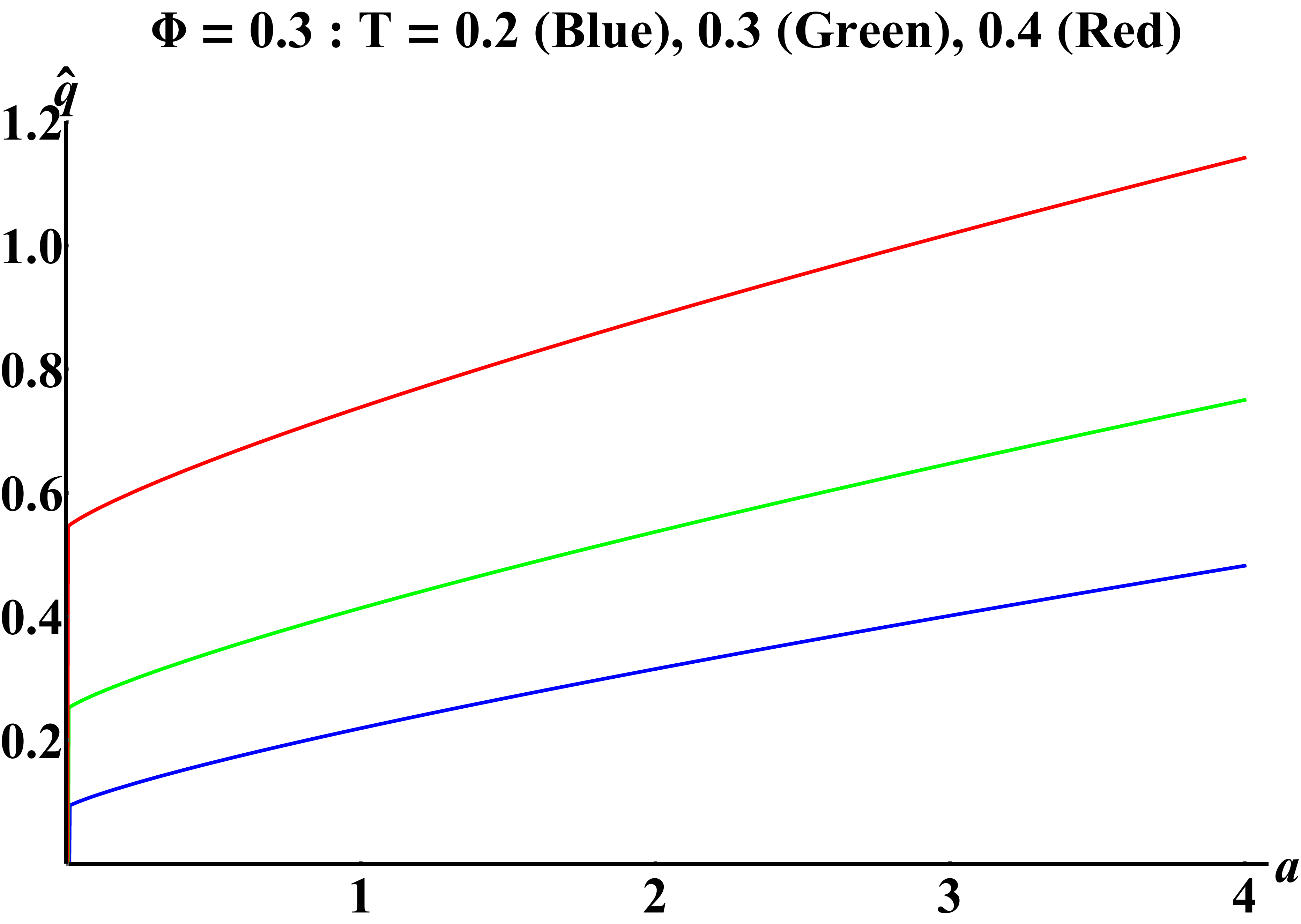}}
	\hspace{.5in}\subfigure[]{\includegraphics[scale=0.11]{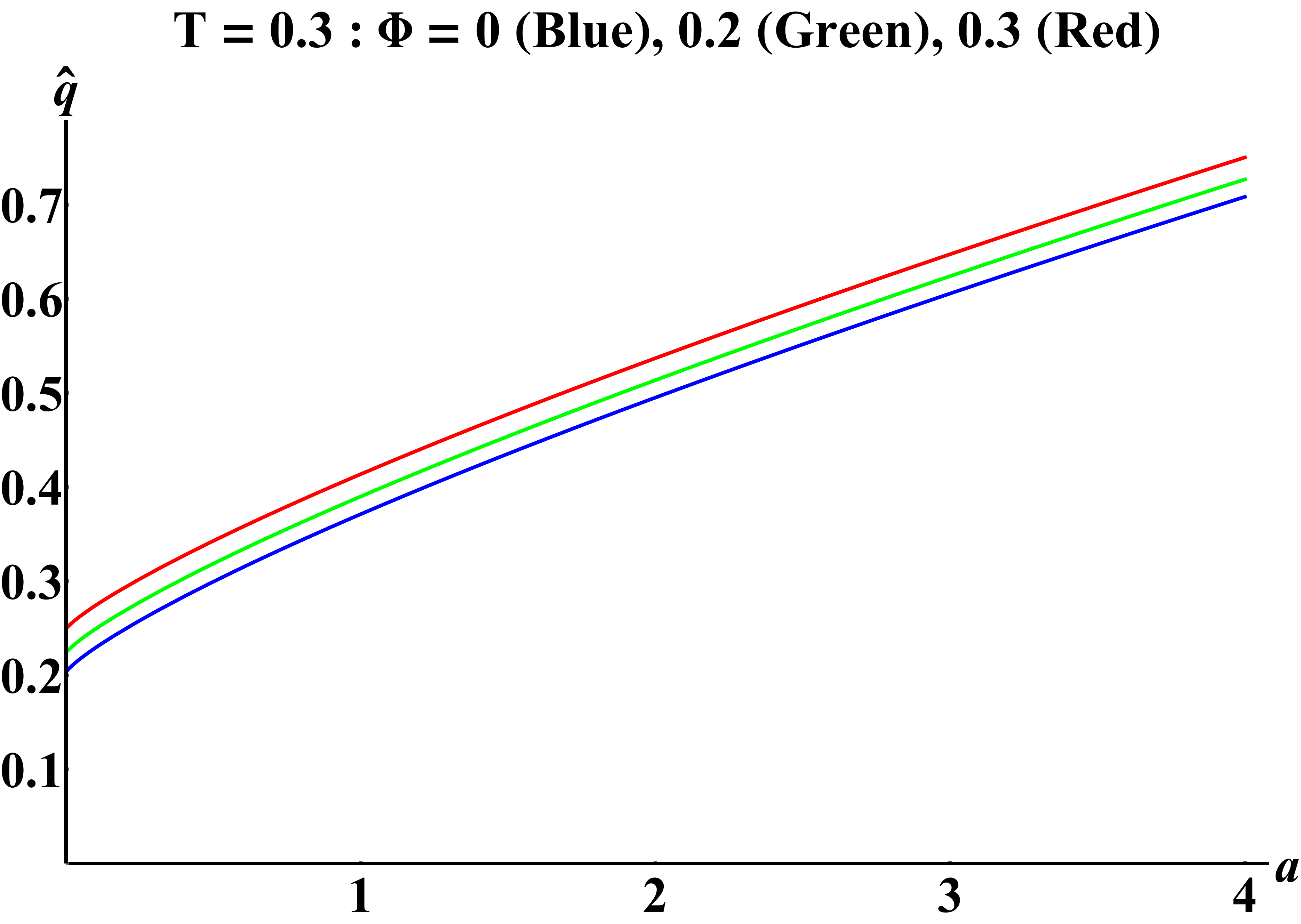}}
	\caption{Jet quenching $\hat{q}$ vs string density $a$ for (a) fixed chemical potential $\Phi$ and different temperature $T$. (b) Fixed temperature and different chemical potential $\Phi$}
	\label{jetQ_vs_a_plot}
\end{figure}

Now we are ready to study the jet quenching parameter with respect to the temperature, string density and chemical potential. Particularly, we study the nature of jet quenching parameter by plotting it with respect to one variable while keeping other parameters fixed in the following figures (\ref{jetQ_vs_phi_plot}, \ref{jetQ_vs_Temp_plot} and \ref{jetQ_vs_a_plot}). The overall observation of the behaviour of the jet quenching parameter is that the jet quenching parameter increases with increase in the back reaction, temperature and chemical potential.
\begin{figure}[!h]
	\centering
	\subfigure[]{\includegraphics[scale=0.11]{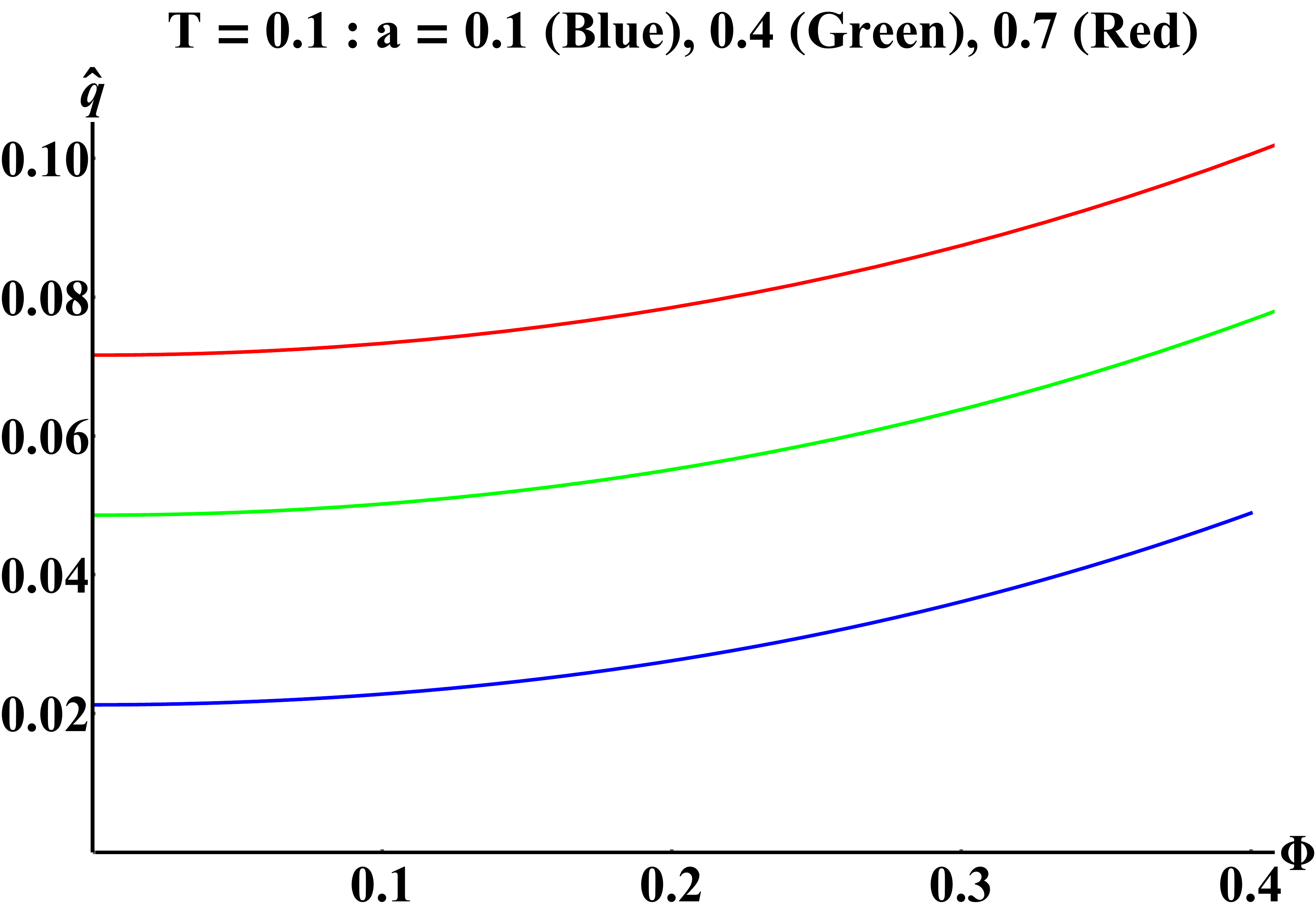}}
	\hspace{.5in}\subfigure[]{\includegraphics[scale=0.11]{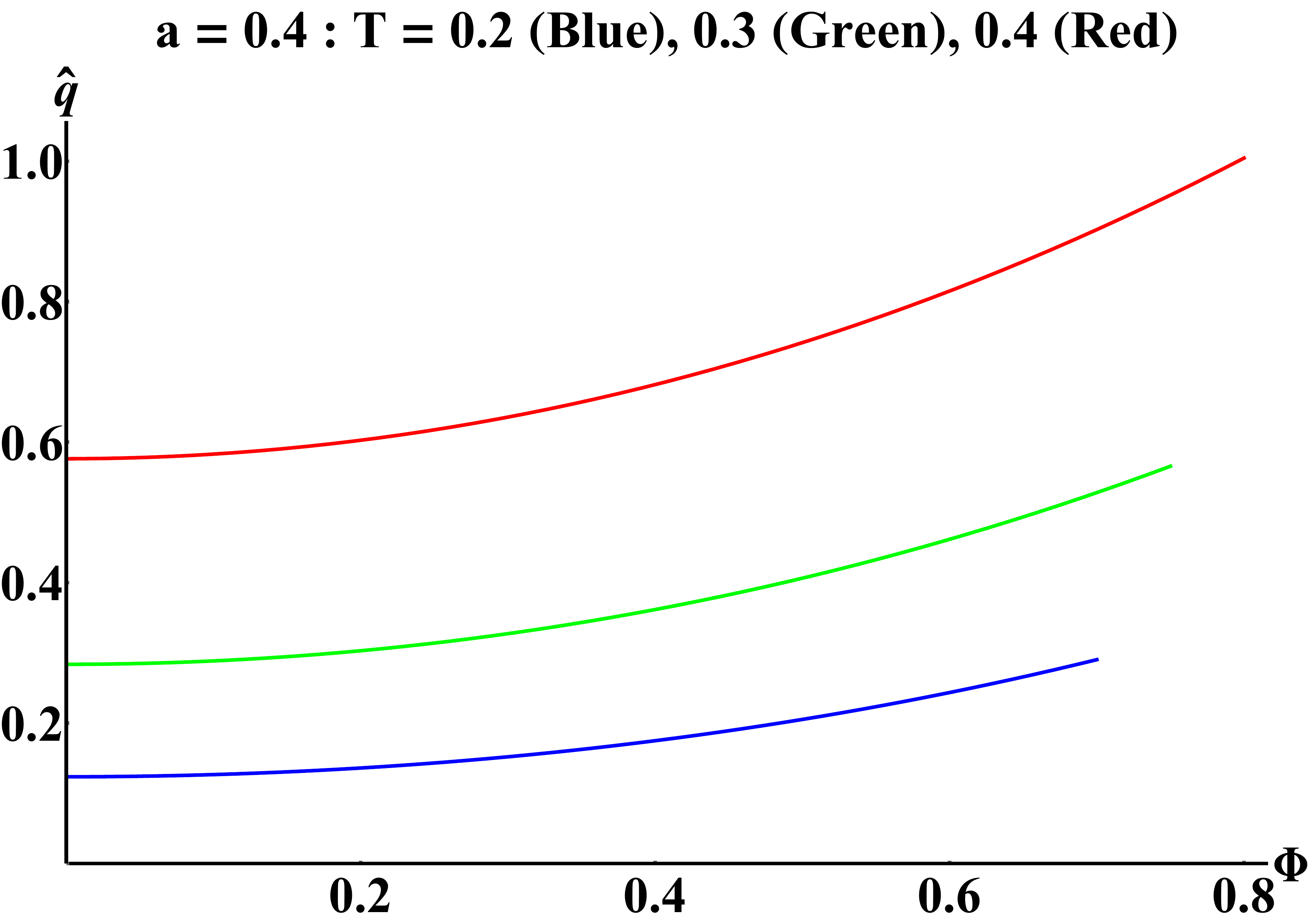}}
	\caption{Jet quenching $\hat{q}$ vs chemical potential $\Phi$ for (a) fixed temperature $T$ and different string density $a$. (b) Fixed string density $a$ and different temperature $T$}
	\label{jetQ_vs_phi_plot}
\end{figure}
\begin{figure}[!h]
	\centering
	\subfigure[]{\includegraphics[scale=0.11]{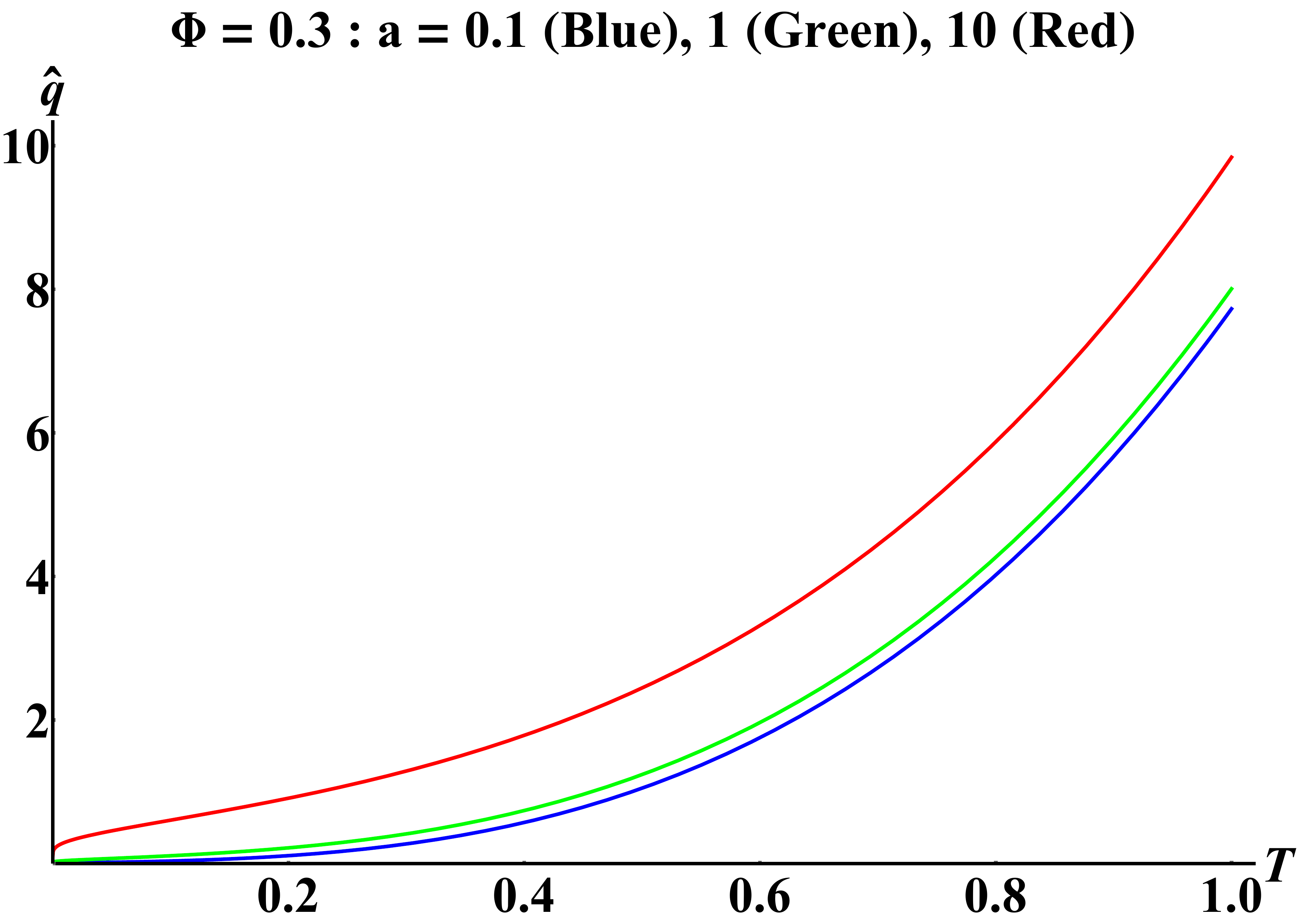}}
	\hspace{.5in}\subfigure[]{\includegraphics[scale=0.11]{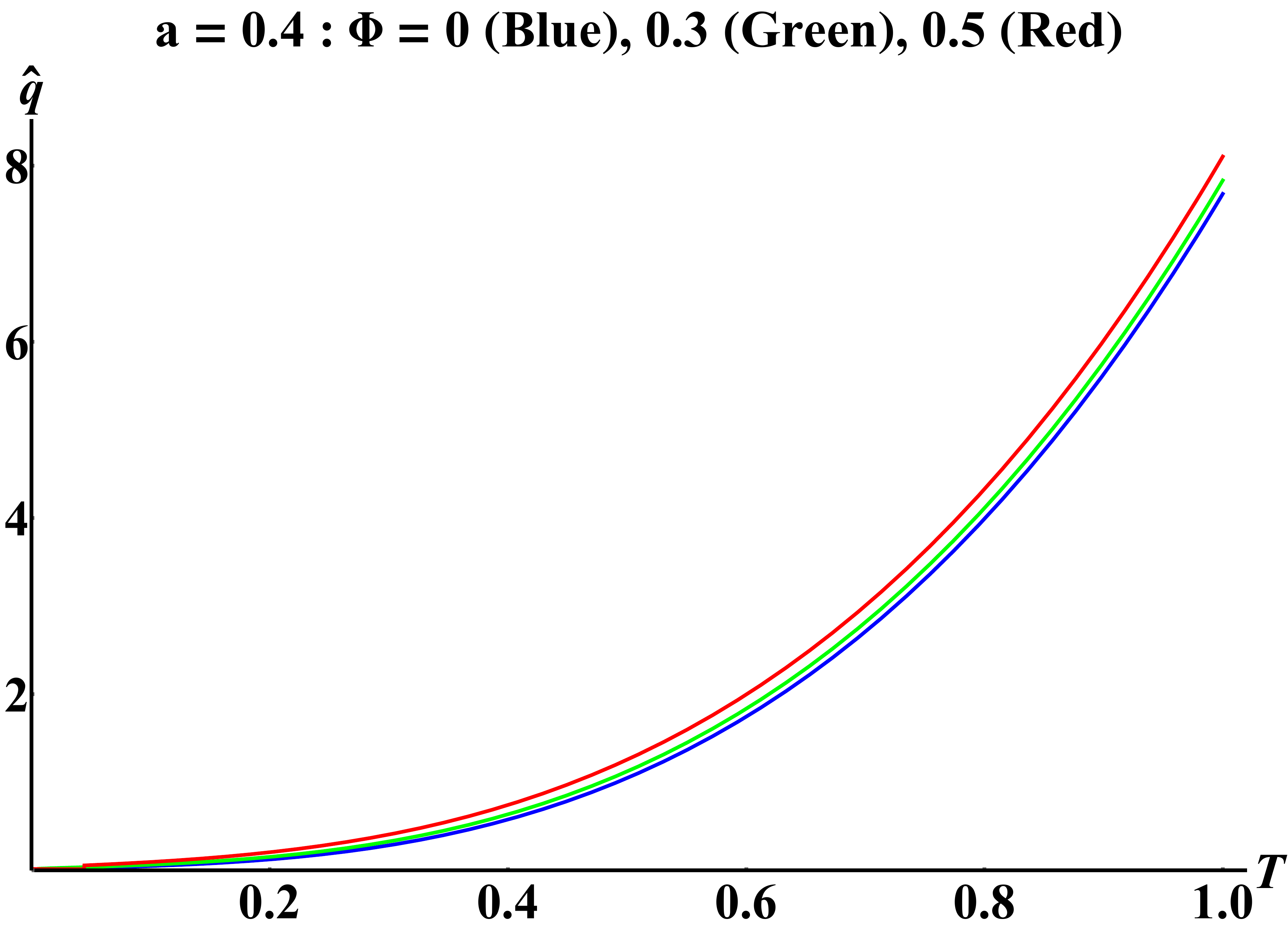}}
	\caption{Jet quenching $\hat{q}$ vs temperature $T$ for (a) fixed chemical potential $\Phi$ and different string density $a$. (b) Fixed string density $a$ and different chemical potential $\Phi$}
	\label{jetQ_vs_Temp_plot}
\end{figure}

\section{Screening length}
\label{screeninglength}
The maximum separation ($L_S$) between the quark-antiquark pair which is moving with a constant speed $v$ in the baryon rich back reacted $\mathcal{N} = 4$ SYM finite temperature plasma has been studied in this section. When the separation distance between the $q\bar{q}$ pair becomes greater than the screening length ($L_S$), the pair gets separated from each other. As a result they do not have binding energy and are screened in the thermal QGP medium. For the study of screening length of the quark-antiquark pair we follow the prescription described in \cite{Liu_2007a,Chakrabortty2016a}. Introducing a boost in the dual gravity background in the following way 
\begin{equation}
	\begin{split}
	dt= cosh(\eta) dt^* - sinh(\eta) dz^*\\
	dz = -sinh(\eta) dt^* + cosh (\eta) dz^*,
	\end{split}
\end{equation}
the metric solution of equation (\ref{metricsol}) recast as,
\begin{equation}
	\begin{split}
	ds^2 = f\left[-\{1-cosh^2(\eta) (1-h)\}dt^{*2} + \{1+ (1-h) sinh^2(\eta)\}dz^{*2}\right. \\ \left.-2(1-h) cosh(\eta) sinh(\eta) dt^* dz^* + dx^2 + dy^2 + \frac{du^2}{h}\right],
	\label{G_Metric_Screening_Length}
	\end{split}
\end{equation}
where $\eta = tanh^{-1} v$ is the rapidity parameter. In the following subsections we will look at two different set up where the orientation of the $q\bar{q}$ pair axis is aligned perpendicular and parallel to the direction of motion of the probe $q\bar{q}$ pair. First we study perpendicular alignment.
\subsection{Perpendicular }
In this subsection, we focus on the set up where the orientation of the $q\bar{q}$ pair axis is along $x$ direction. In the choice of static gauge $\tau = t^*,\, \sigma = x,\, y=z^*=0$ and for the following boundary conditions on the connecting string between $q\bar{q}$ pair,
\begin{equation}
	u(\sigma =\pm \frac{L}{2}) = 0,\, u(\sigma =0) = u_{ext},\, u'(\sigma=0) =0,
\end{equation}
the world-sheet action of the connecting string can be written as, 
\begin{equation}
	S = -\frac{\mathcal{T}}{2 \pi \alpha} \int d\sigma \sqrt{f^2[1 + (h-1) cosh^2(\eta)] + \frac{f^2}{h}[1 + (h-1) cosh^2(\eta)]u'^{2}}.
\label{wsheeta}
\end{equation}
The Lagrangian of the above action does not explicitly depends on $\sigma$. Hence, the Hamilton's like equation can be constructed for the constant of motion $W$ as,
\begin{equation}
	\frac{\partial \mathcal{L}}{\partial u'} u' - \mathcal{L} = W.
\label{hequation}
\end{equation}
Combination of the equation (\ref{wsheeta}) and (\ref{hequation}) gives the solution for $u'^2$ in the following form, 
\begin{equation}
	u'^2 =  \frac{h[f^2 + f^2(h-1) cosh^2(\eta) -{W}^2]}{{W}^2}.
	\label{equation_screening_length_u_dashed_squared}
\end{equation}
Since the connecting string has turning nature, the depth of the string along the radial direction is determined from the solution of $u' = 0 $. There are two conditions for which the turning situation of the string appear. One is $\left.h\right|_{u_{ext1}}=0$ and the other one is $\left.(f^2 + f^2(h-1)cosh^2(\eta) - {W}^2)\right|_{u_{ext2}}=0 \label{ucsol}$. Since the first condition is the solution of the horizon of the black hole geometry, the tip of the connecting string reaches at the horizon and breaks down into two separate straight strings corresponding to a pair of free quark with no binding energy. In the second condition, since, $f^2$, $cosh^2\eta $ and ${W}^2$ all are greater or equal to zero and $h-1 \le 0$, hence, according to the Descartes' sign rules, the above equation have only one real positive solution, written as $u_c$. Therefore the tip of the string can extend in the radial direction from zero to the solution $u_c$ of the second equation.  
As in the previous section, we found the distance $L$ between the $q\bar{q}$ pair in equation (\ref{qqbardis}) from equation (\ref{extreme}),  here also we find it by integrating the equation (\ref{equation_screening_length_u_dashed_squared}) and using the boundary condition $u(\sigma=\pm \frac{L}{2}) = 0$. The distance $L^{\perp}$ has the form,
\begin{equation}
	L^{\perp} = \int_{0}^{u_c} \frac{2 W du}{\sqrt{h}\sqrt{f^2(1+(h-1)cosh^2(\eta))-{W}^2}}.
	\label{screening_length_final_equation}
\end{equation}
Now to study the nature of separation distance of $q\bar{q}$ pair, we plot $L^{\perp}$ with respect to the constant of motion $W$ keeping the three parameters fixed among the four parameters and the remaining one is allowed to take three different values.
\begin{figure}[!h]
	\centering
	\subfigure[]{\includegraphics[scale=0.12]{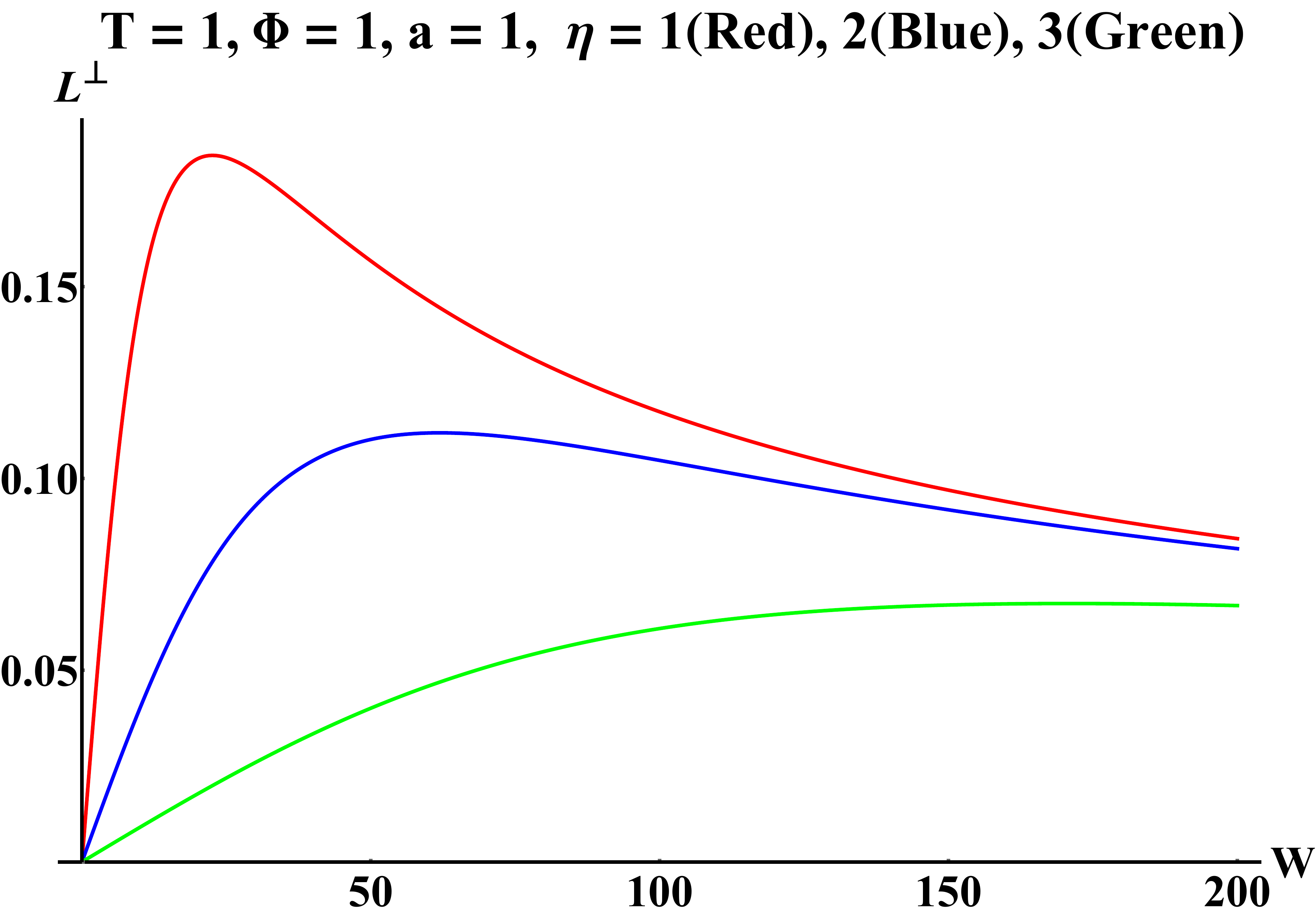}}
	\hspace{.2in}\subfigure[]{\includegraphics[scale=0.12]{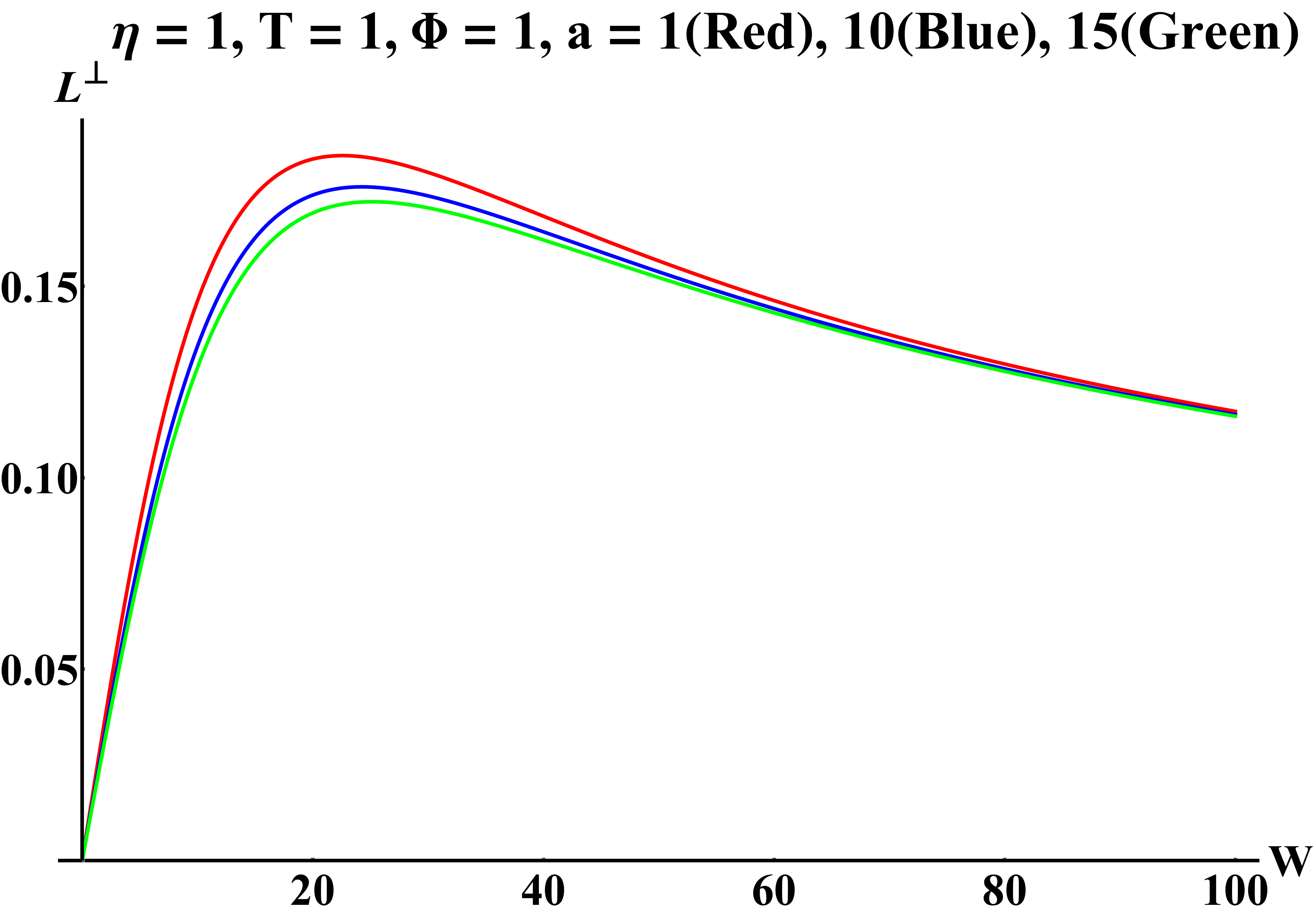}}
	\noindent\subfigure[]{\includegraphics[scale=0.12]{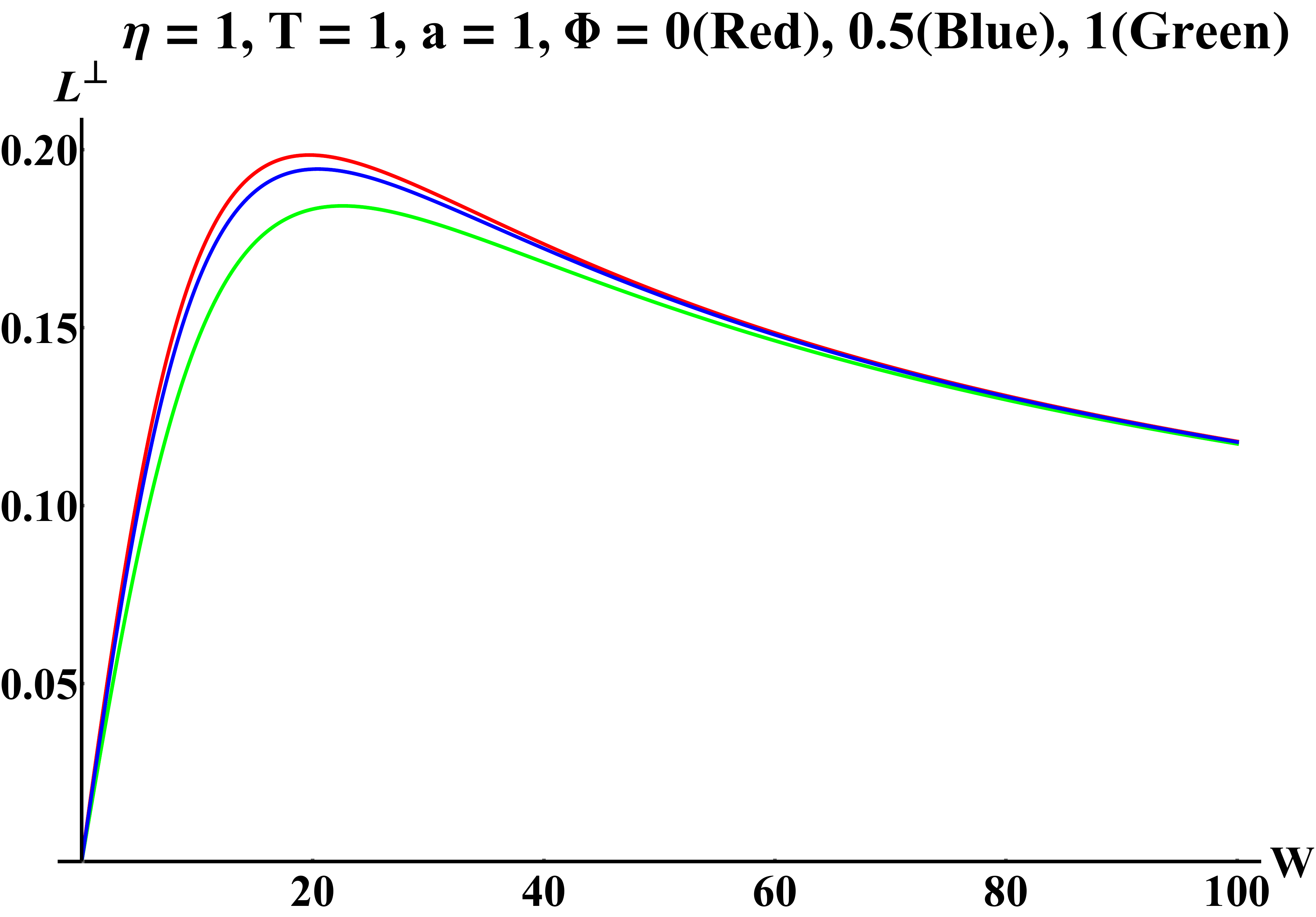}}
	\hspace{.2in}\subfigure[]{\includegraphics[scale=0.12]{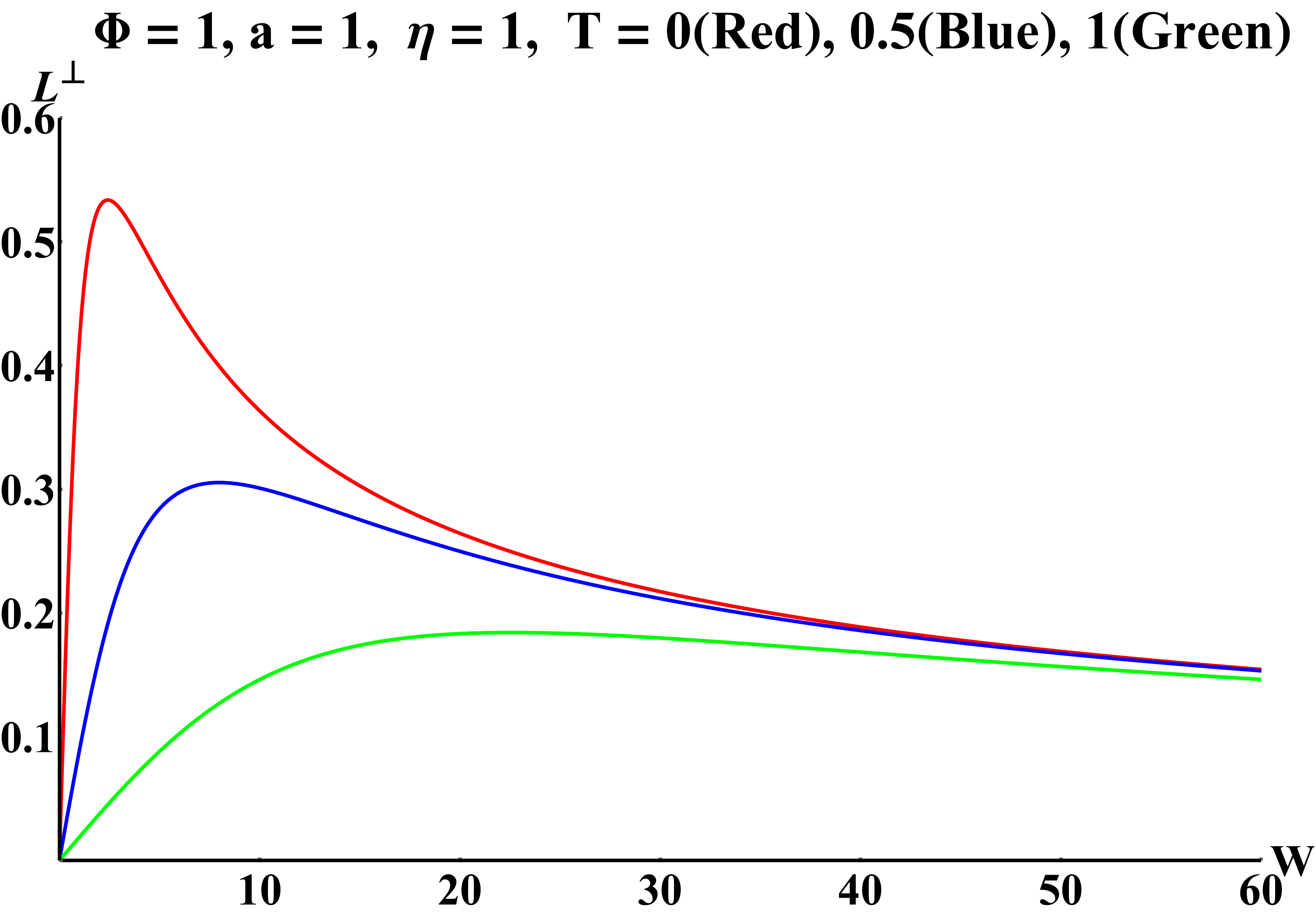}}
	\caption{Distance $L^{\perp}$ vs $W$ for (a) $T = 1,\, \Phi = 1,\, a=1$ for different $\eta = 1$ (red), $2$ (blue) and $3$ (green). (b) $\eta = 1,\,T=1,\, \Phi = 1$ for $a = 1$ (red), $10$ (blue) and $15$ (green). (c) $\eta =1,\,T=1,\, a=1$ different $\Phi = 0$ (red), $0.5$ (blue) and $1$ (green).  (d) $\Phi =1,\, a=1,\, \eta=1$ different $T = 0$ (red), $0.5$ (blue) and $1$ (green).}
	\label{L_vs_W_plot_eta_1}
\end{figure}
From plots (\ref{L_vs_W_plot_eta_1}), it is observed that for constant of motion $W=0$, the separation distance $L^{\perp}$ also becomes zero. Irrespective of the values of rapidity parameter, string density, temperature and chemical potential, $L^{\perp}$ increases monotonically and attains the maximum value for a certain value of constant of motion and starts to decrease for further increase of $W$. The maximum separation length corresponding to the $q\bar{q}$ pair breaks down and produces quark and antiquark with no binding energy beyond this length. The maximum separation length between the $q\bar{q}$ pair is known as the screening length $L_s^{\perp}$. The screening length attains in lower values of constant of motion with the increase of any parameter.  We also observe that the screening length decreases with the increase of any parameter which accelerates the deconfinement process of the bound state of $q\bar{q}$ pair.\\

Further, it is observed that for any value of $L^\perp < L_s$, there are two possible values of the constant of motion $W$. To find the preferred value of the constant of motion $W$, now we move towards the study of potential energy of the $q\bar{q}$ pair for the given values of separation distance $L^\perp $. The binding energy between the $q\bar{q}$ pair is calculated holographically using the following relation \cite{Liu_2007a},
	\begin{equation}
		V = -\frac{S-S_0}{\mathcal{T}},
		\label{equation_screening_length_perpd_potnetial_V}
	\end{equation}
where $S_0$ is the action of the two separate strings hanging from the boundary to the horizon of the bulk spacetime. These strings are representing the two free quarks in the boundary theory.  The action $S_0$ takes the following form,
\begin{equation}
	S_0 = -\frac{2\mathcal{T}}{2\pi \alpha}\int_{0}^{u_h} \sqrt{-det g},
	\label{equation_screening_length_potential_self_energy_action_S0}
\end{equation}
where the induced metric $g_{\gamma\beta} = G_{\mu\nu}\partial_\gamma X^\mu \partial_\beta X^\nu$ and the bulk metric $G_{\mu\nu}$ is provided in (\ref{G_Metric_Screening_Length}). To include the contribution coming from both quark and antiquark we have multiplied the action of single quark by a factor of $2$. The $q\bar{q}$ potential energy is studied using the relation of $L^\perp(W)$ from which we solve for $W(L^\perp)$. Further, we use $W(L^\perp)$ in (\ref{equation_screening_length_perpd_potnetial_V}) through (\ref{equation_screening_length_u_dashed_squared}) and (\ref{wsheeta}) to study the quark-antiquark potential as a function of $L^\perp$.
\begin{figure}[!h]
	\centering
	\subfigure[]{\includegraphics[scale=0.12]{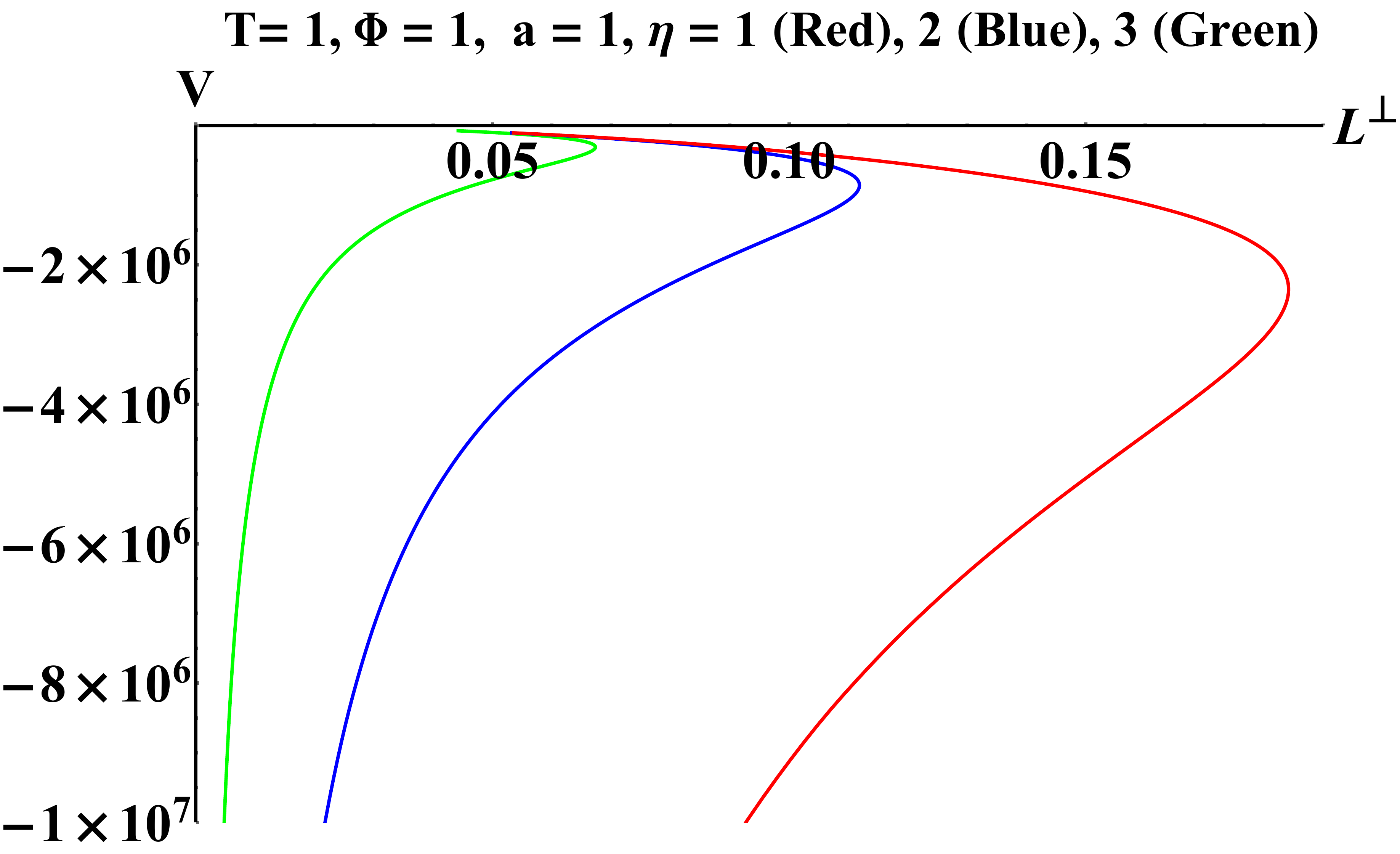}}
	\hspace{.2in}\subfigure[]{\includegraphics[scale=0.12]{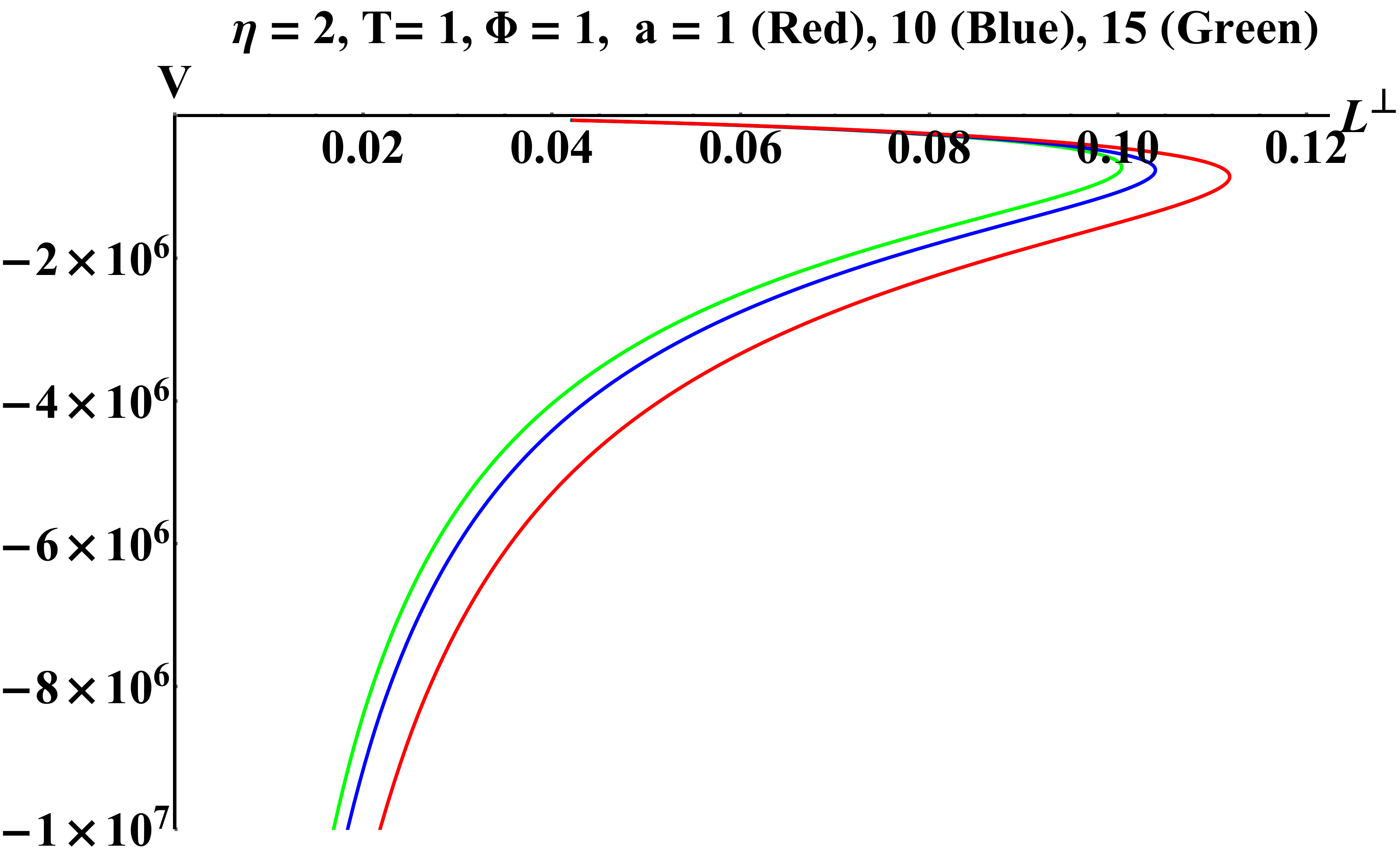}}
	\subfigure[]{\includegraphics[scale=0.12]{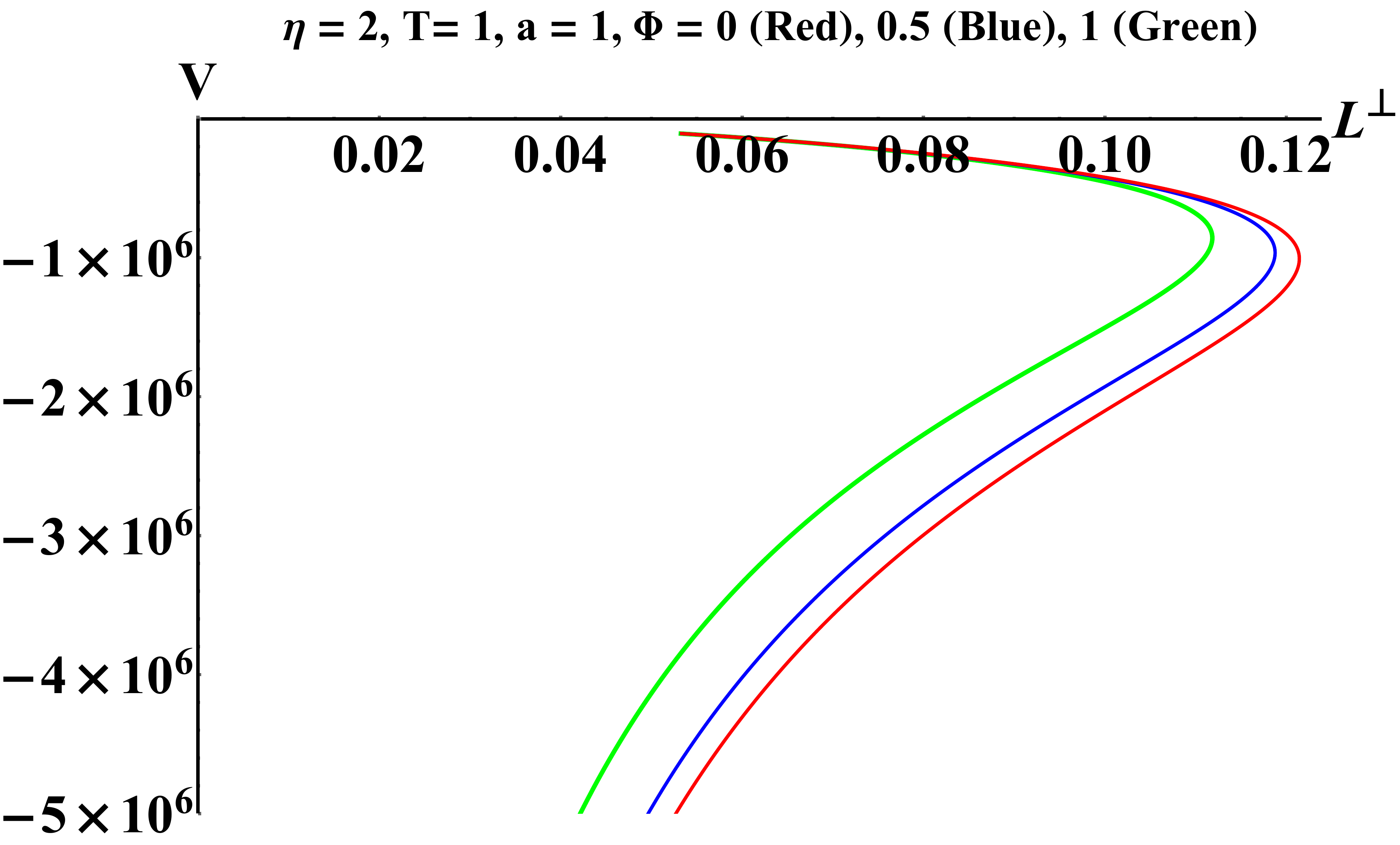}}
	\hspace{.2in}\subfigure[]{\includegraphics[scale=0.12]{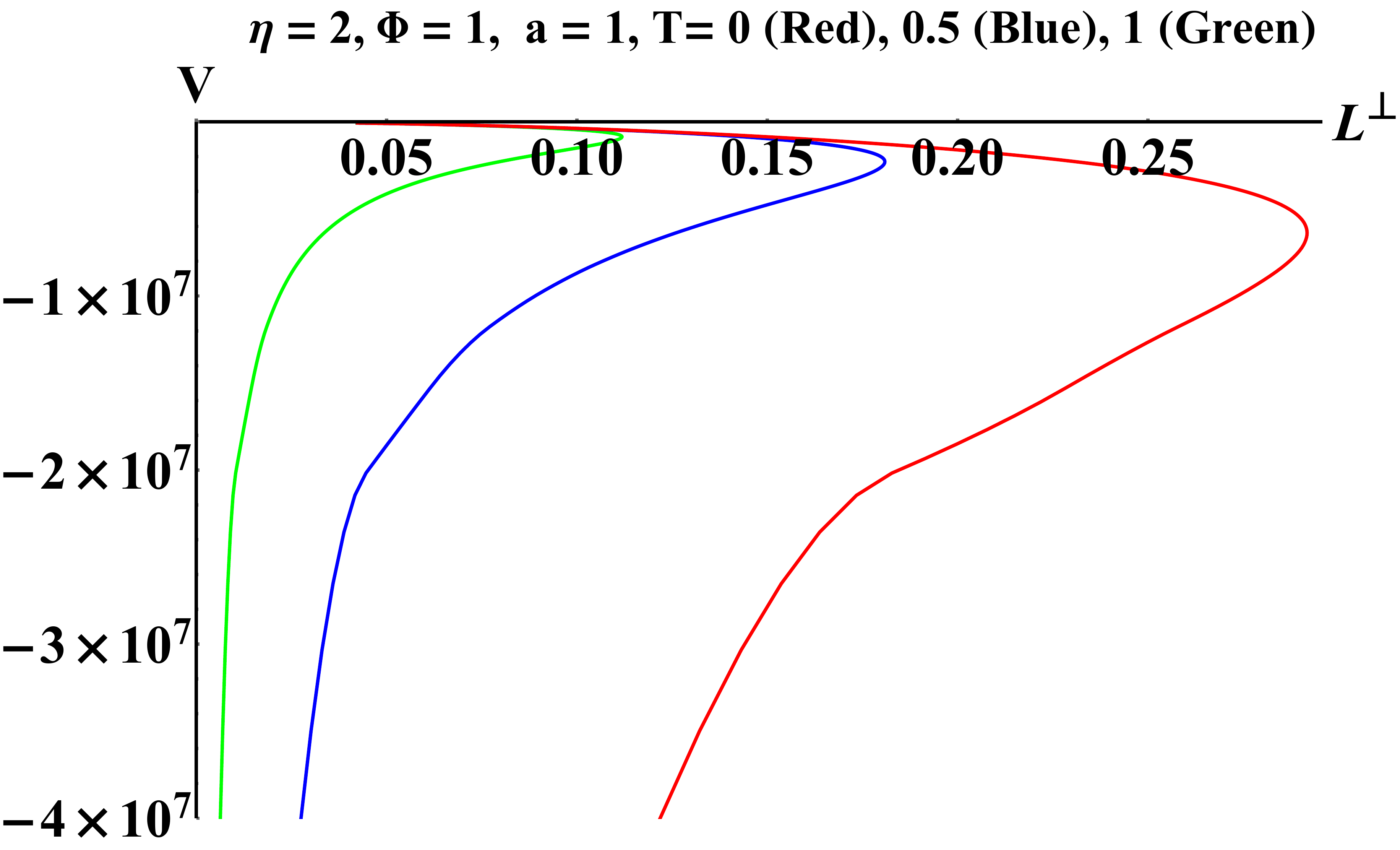}}
	\caption{Perpendicular case: Binding energy $V$ vs the $q\bar{q}$ separation distance $L^\perp$ for for (a) $ T=1,\,\Phi = 1,\, a=1$ for different $\eta = 1$ (Red), $2$ (blue) and $3$ (green) (b) $\eta = 2,\,T=1,\, \Phi = 1$ for $a = 1$ (red), $10$ (blue) and $15$ (green). (c) $\eta=2,\,T=1,\, a=1$ for $\Phi = 0$ (red), $0.5$ (blue) and $1$ (green). (d) $\Phi=1,\, a=1,\, \eta=2$ for $T = 0$ (red), $0.5$ (blue) and $1$ (green).}
	\label{figure_Screening_Length_QAQ_binding_energy_Perpd_case}
\end{figure}
In figure (\ref{figure_Screening_Length_QAQ_binding_energy_Perpd_case}), we have plotted the $q\bar{q}$ potential $V$ as a function of the $q\bar{q}$ separation distance $L^\perp$ for fixed values of three parameters and remaining one taking three different values. It is observed from figure (\ref{figure_Screening_Length_QAQ_binding_energy_Perpd_case}), that irrespective of the values of parameters, the separation distance $L^\perp$ has a maximum value which corresponds to the screening length $L_s$ and for a given value of $L^\perp<L_s$ there are two possible values of the binding energy $V$. The higher value of $W$ corresponds to higher value of the binding energy $V$. Whereas the lower value of the binding energy corresponds to the lower value of $W$. The lower value of the binding energy $V$ is energetically favourable. Potential energy rises as increase of any parameters leads to the unstability of confined phase of $q\bar{q}$ pair which causes the phase transition of the pair to QGP phase.


\subsection{Parallel}
\begin{figure}[!h]
	\centering
	\subfigure[]{\includegraphics[scale=0.11]{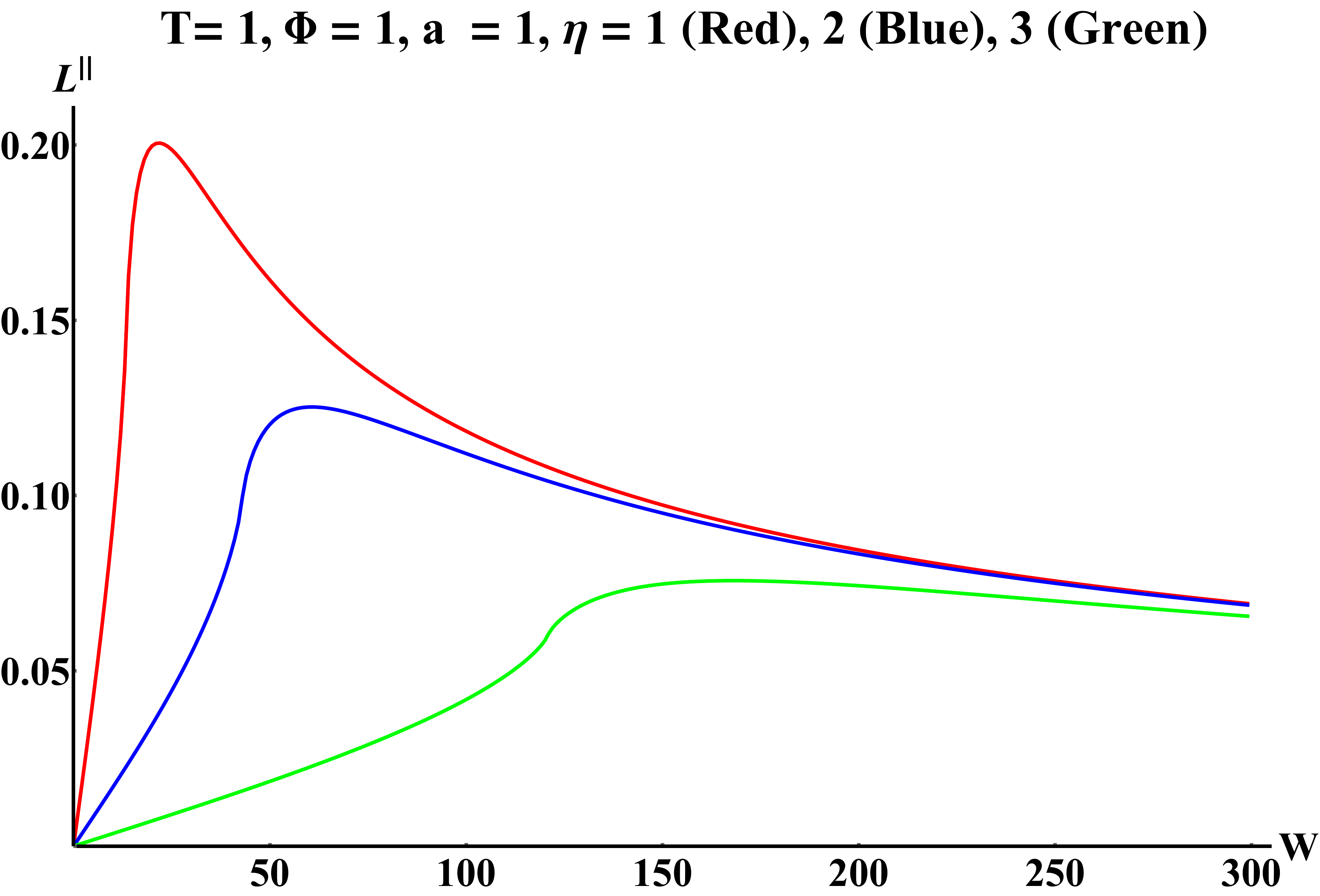}}
	\hspace{.2in}\subfigure[]{\includegraphics[scale=0.11]{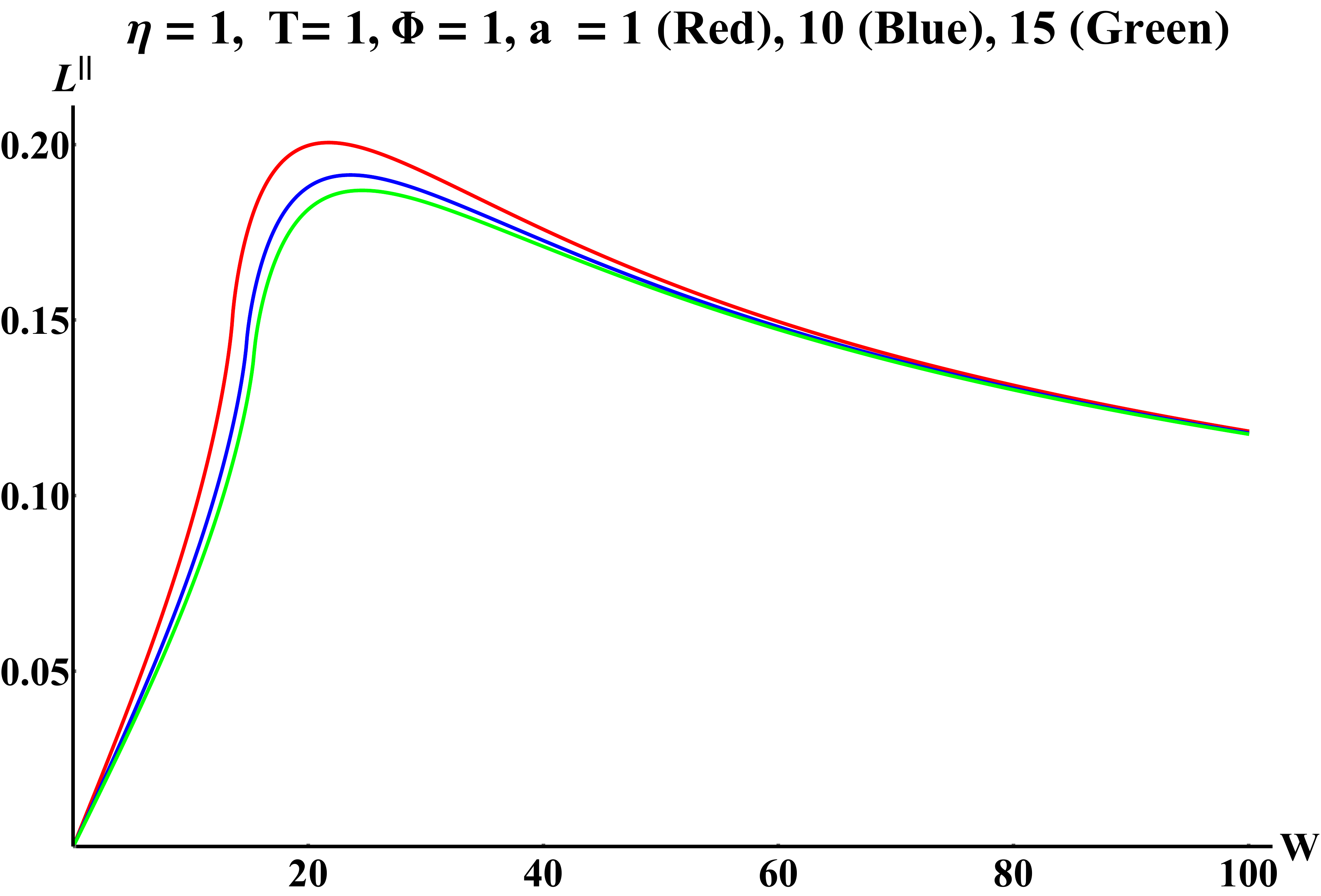}}
	\hspace{.2in}\subfigure[]{\includegraphics[scale=0.11]{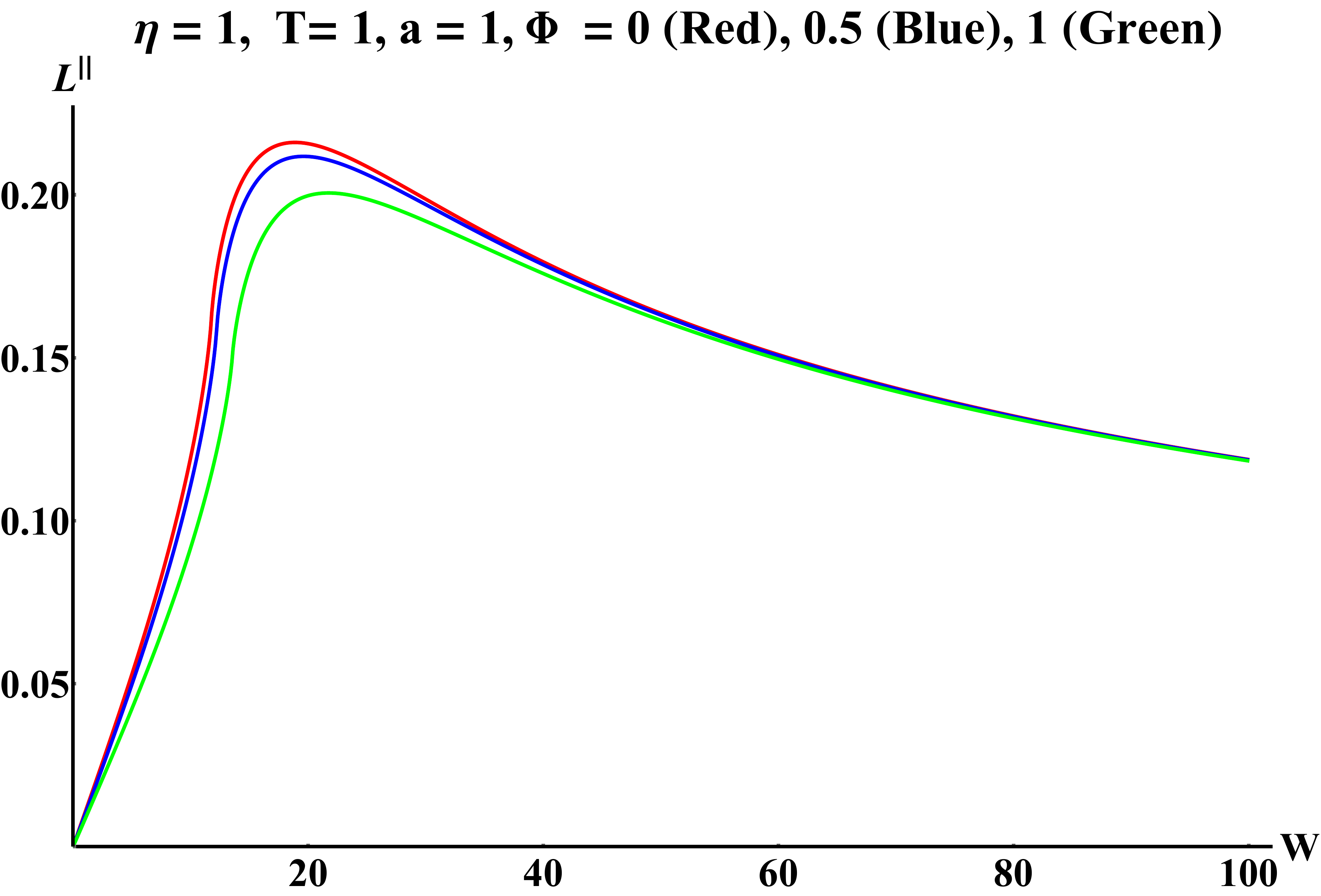}}
	\hspace{.2in}\subfigure[]{\includegraphics[scale=0.11]{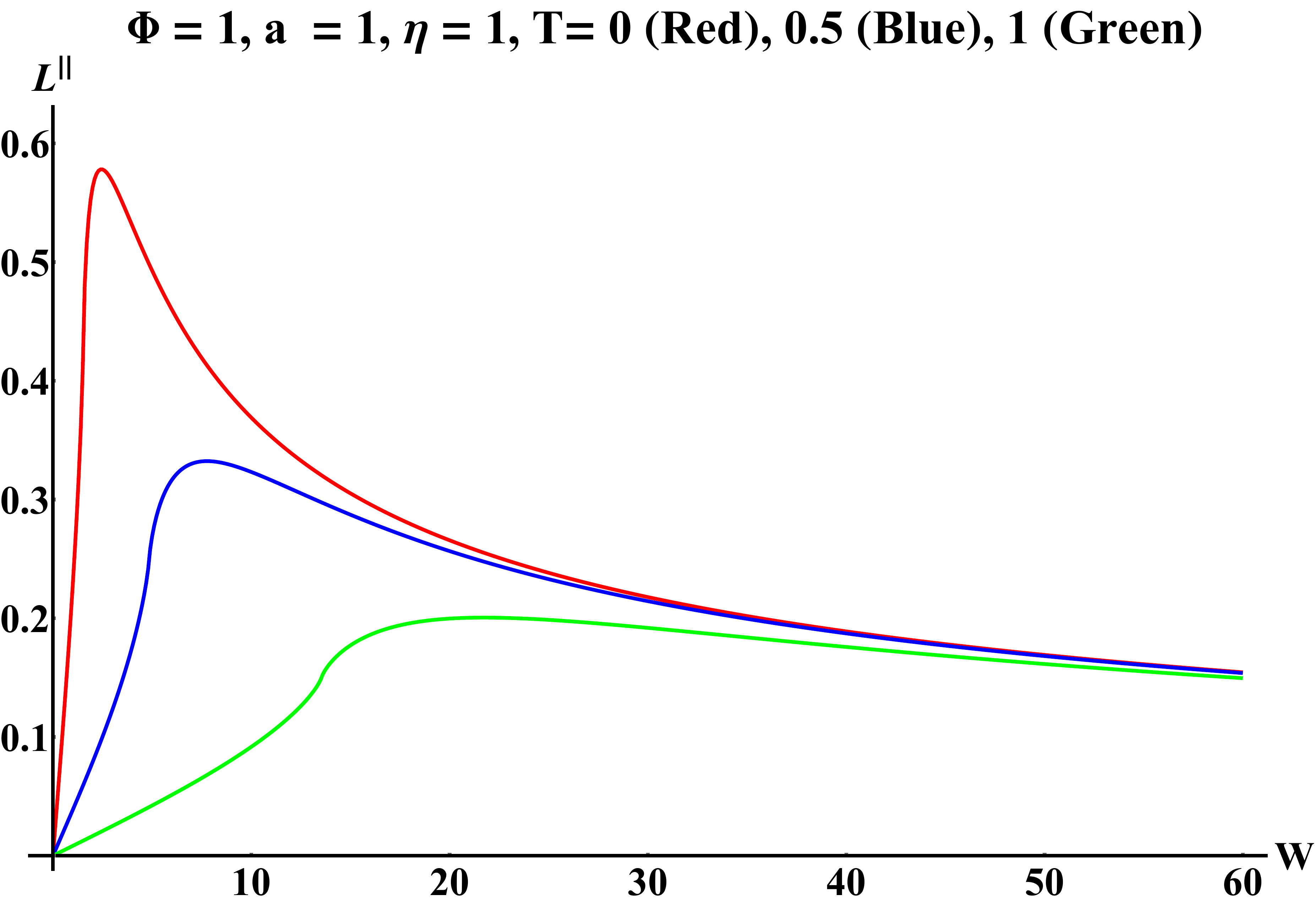}}
	\caption{Distance $L^\parallel$ vs $W$ for (a) $ T=1,\,\Phi = 1,\, a=1$ for different $\eta = 1$ (Red), $2$ (blue) and $3$ (green) (b) $\eta = 1,\,T=1,\, \Phi = 1$ for $a = 1$ (red), $10$ (blue) and $15$ (green). (c) $\eta=1,\,T=1,\, a=1$ for $\Phi = 0$ (red), $0.5$ (blue) and $1$ (green). (d) $\Phi=1,\, a=1,\, \eta=1$ for $T = 0$ (red), $0.5$ (blue) and $1$ (green). }
	\label{L_vs_W_plot_parallel_eta_1}
\end{figure}

In this subsection, we focus on the set up where the orientation of the $q\bar{q}$ pair axis is along $z$ direction.
So the choice of the static gauge is
\begin{equation}
	\tau = t^*,\, \sigma = z,\, x=y=0.
\end{equation}
For the boundary conditions 
\begin{equation}
	u(\sigma =\pm \frac{L}{2}) = 0,\, u(\sigma =0) = u_{ext},\, u'(\sigma=0) =0,
    \label{equation_screening_length_parallel_boundary_conditions}
\end{equation}
the world-sheet action takes the form as 
\begin{equation}
    S = -\frac{\mathcal{T}}{2\pi \alpha}\int d\sigma f\sqrt{h + \left[\frac{1}{h}+\left(1-\frac{1}{h}\right)cosh^2(\eta)\right]u'^2}.
\end{equation}
Like the previous set up, by constructing the Hamilton's equation, we can write
\begin{equation}
    u'^2 = \frac{h(f^2h - {W} ^2)}{[1 + (h-1) cosh^2(\eta)]{W}^2}.
    \label{equation_screening_length_parallel_u_dashed_squared}
\end{equation}
The solution of the equation,
\begin{equation}
    \left. f^2 h - {W} ^2\right|_{u_{ext2}} = 0, 
\end{equation}
provides the depth of the tip of the connecting string $u_c$ along the radial direction. Further, the inter distance between the $q\bar{q}$ pair becomes,
\begin{equation}
    L^\parallel = \int_{0}^{u_c} \frac{2 W  \sqrt{1 + (h-1) cosh^2(\eta)}}{h\sqrt{f^2 h - {W}^2}}.
\end{equation}
As earlier, we now study the nature of $L^\parallel$ by plotting it with respect to constant of motion $W $ for fixed values of three parameters and giving three values to the other one. From the plots (\ref{L_vs_W_plot_parallel_eta_1}), it is clear that the nature of $L^\parallel$ is similar to $L^\perp$. It reduces if rapidity parameter, string density, temperature and the chemical potential raise.

\begin{figure}[!h]
	\centering
	\subfigure[]{\includegraphics[scale=0.12]{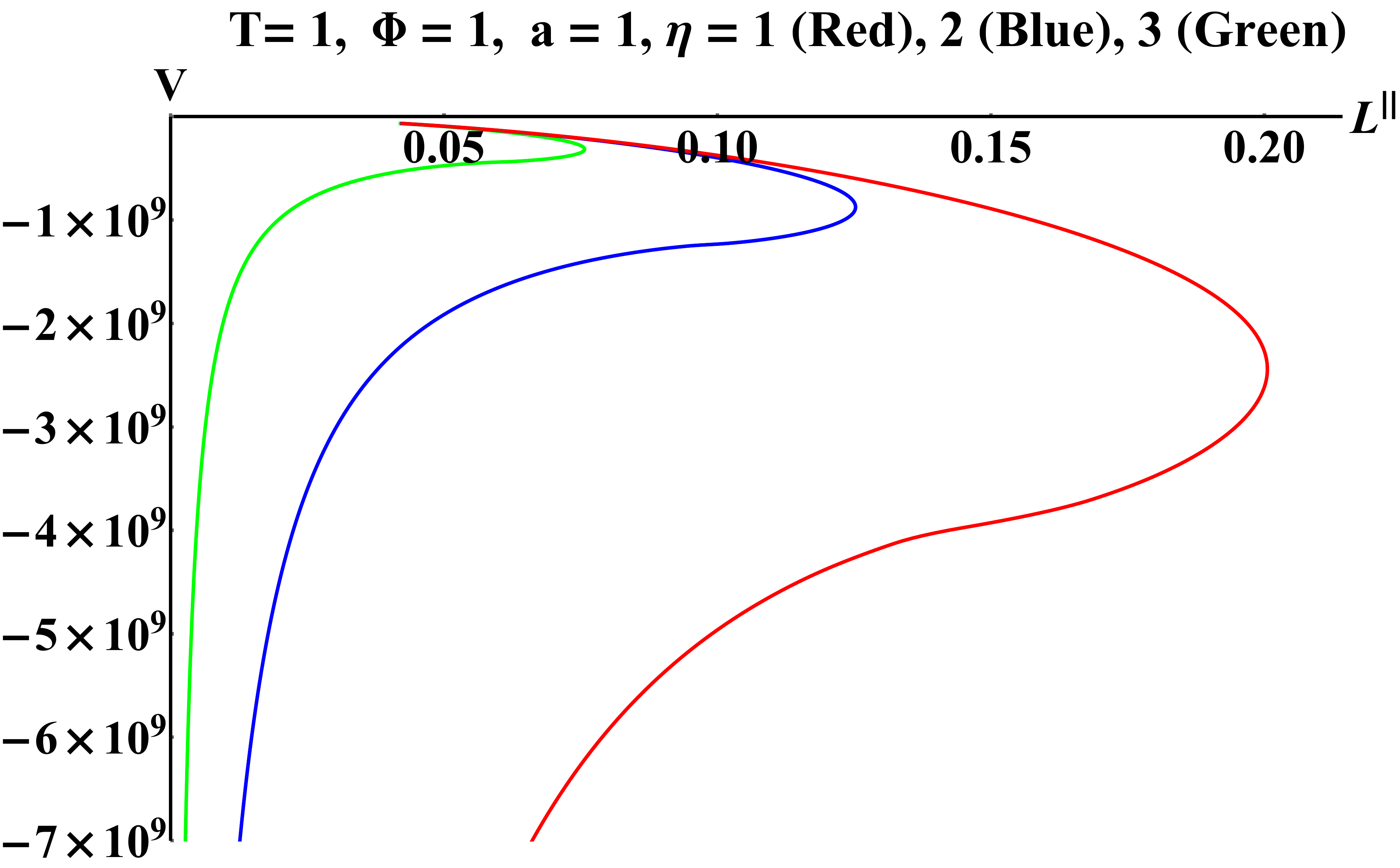}}
	\hspace{.2in}\subfigure[]{\includegraphics[scale=0.12]{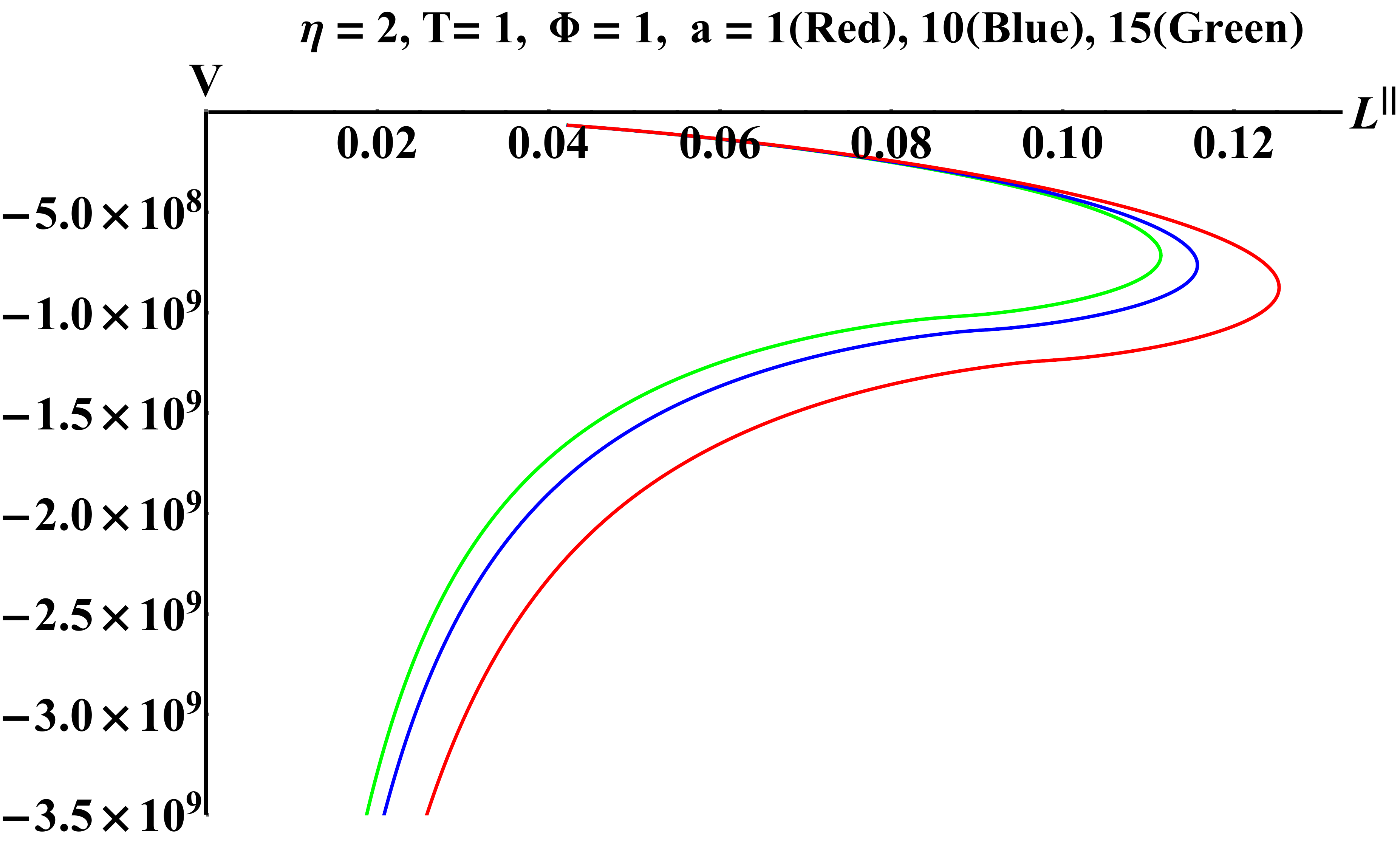}}
	\subfigure[]{\includegraphics[scale=0.12]{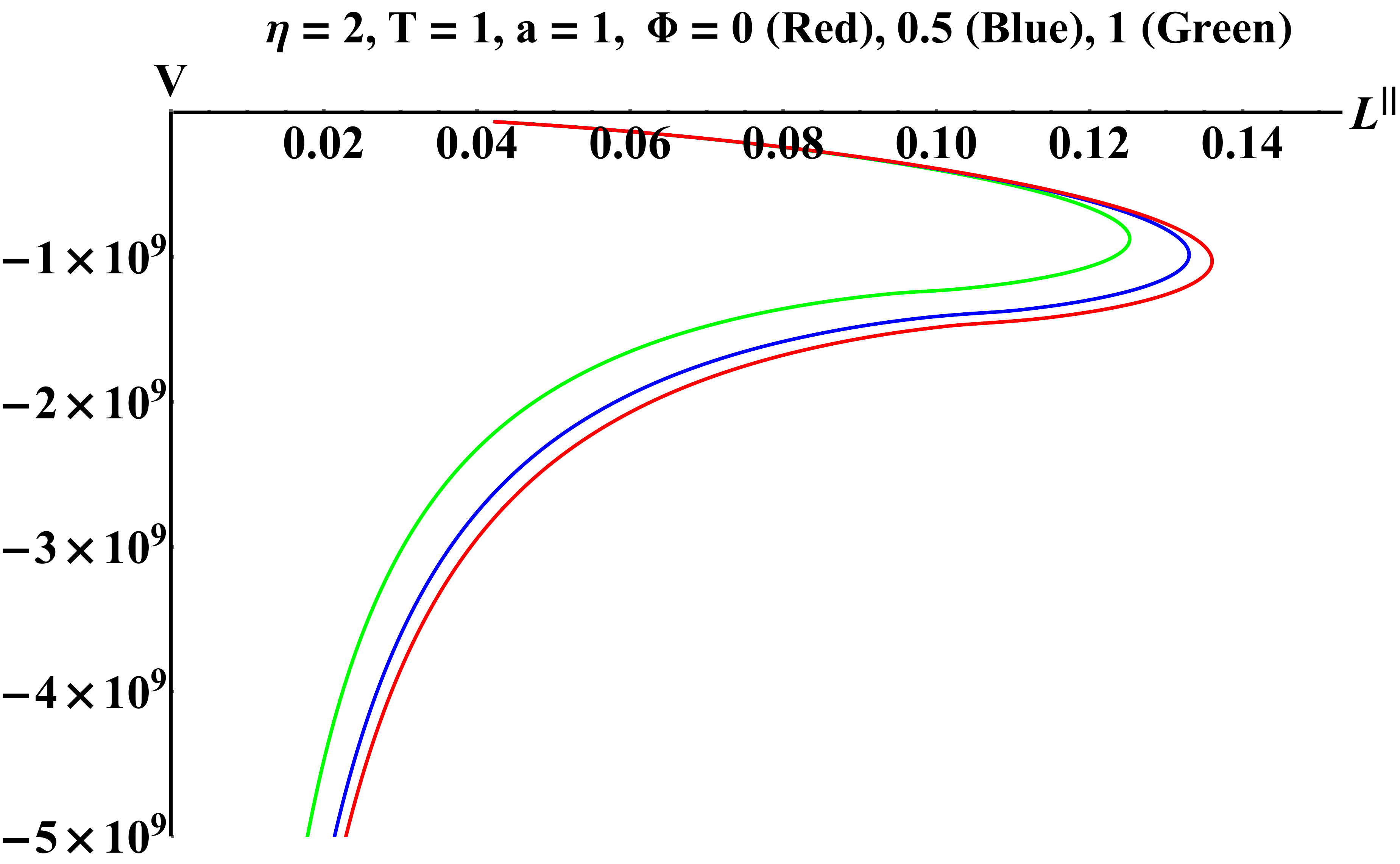}}
	\hspace{.2in}\subfigure[]{\includegraphics[scale=0.12]{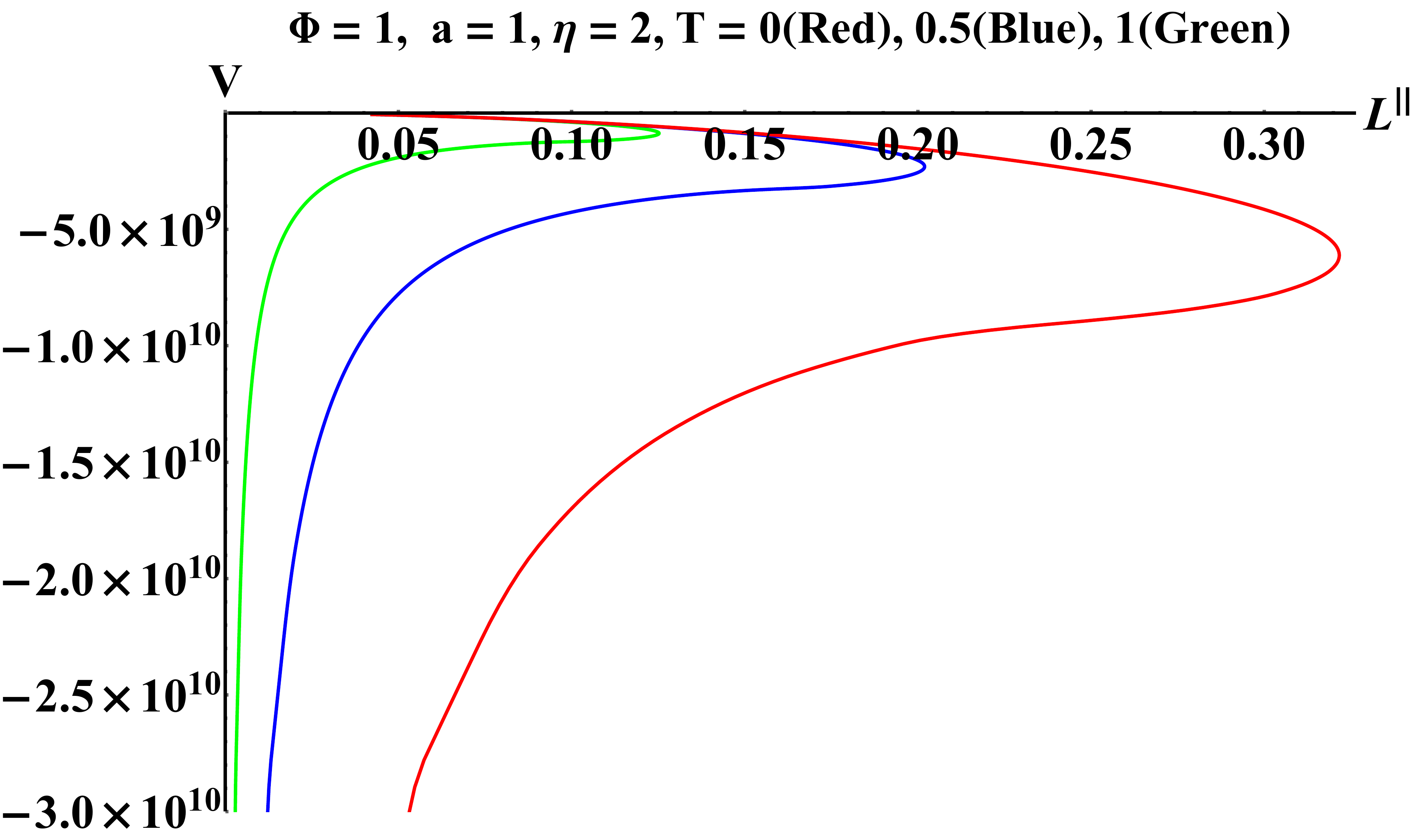}}
	\caption{Parallel case: Binding energy $V$ vs the $q\bar{q}$ separation distance $L^\parallel$ for (a) $ T=1,\,\Phi = 1,\, a=1$ for different $\eta = 1$ (Red), $2$ (blue) and $3$ (green) (b) $\eta = 2,\,T=1,\, \Phi = 1$ for $a = 1$ (red), $10$ (blue) and $15$ (green). (c) $\eta=2,\,T=1,\, a=1$ for $\Phi = 0$ (red), $0.5$ (blue) and $1$ (green). (d) $\Phi=1,\, a=1,\, \eta=2$ for $T = 0$ (red), $0.5$ (blue) and $1$ (green).}
	\label{figure_Screening_Length_QAQ_binding_energy_Parallel_case}
\end{figure}

As in previous subsection, here also we study the binding energy of the quark-antiquark pair. In figure (\ref{figure_Screening_Length_QAQ_binding_energy_Parallel_case}), the binding energy of the $q\bar{q}$ pair has been plotted with respect to the separation distance $L^\parallel$ for fixed values of three parameters while assigning the three values to the other parameters. The qualitative nature of the $q\bar{q}$ potential remains same as in figure (\ref{figure_Screening_Length_QAQ_binding_energy_Perpd_case}).\\

\begin{figure}[!h]
	\centering
	\subfigure[]{\includegraphics[scale=0.15]{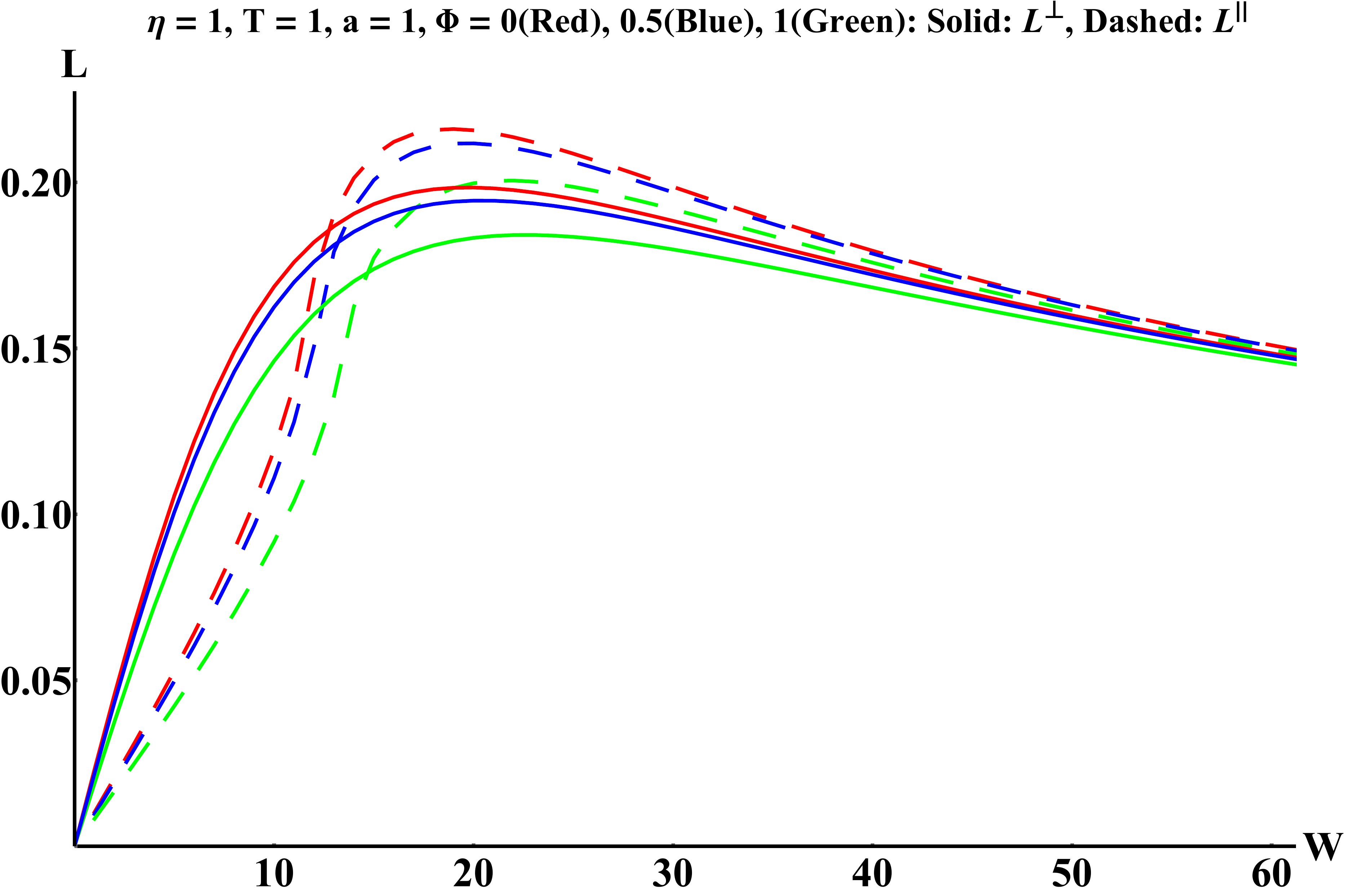}}
	\caption{Screening Length ($L$) vs $W$ for $\eta=1,\,T=1,\,a=1$ and different $\Phi=0$ (Red), $0.5$ (Blue), $1$ (Green). Solid line represents perpendicular case while dashed line represents parallel case.}
	\label{L_vs_W_plot_parallel_and_perp_comparision_different_a}
\end{figure}

Finally, in figure(\ref{L_vs_W_plot_parallel_and_perp_comparision_different_a}), we have compared the nature of the inter-distance between the $q\bar{q}$ pair for perpendicular and parallel alignment, where solid line represents the perpendicular orientation and the dashed line represents the parallel orientation.  It is observed that the value of the screening length for the parallel alignment is larger than the perpendicular one resulting more stable bound state.
	

\section{Radial Profile of a Rotating String}
\label{radialprofile}
\begin{figure}[!t]
	\centering
	\subfigure[]{\includegraphics[scale=0.08]{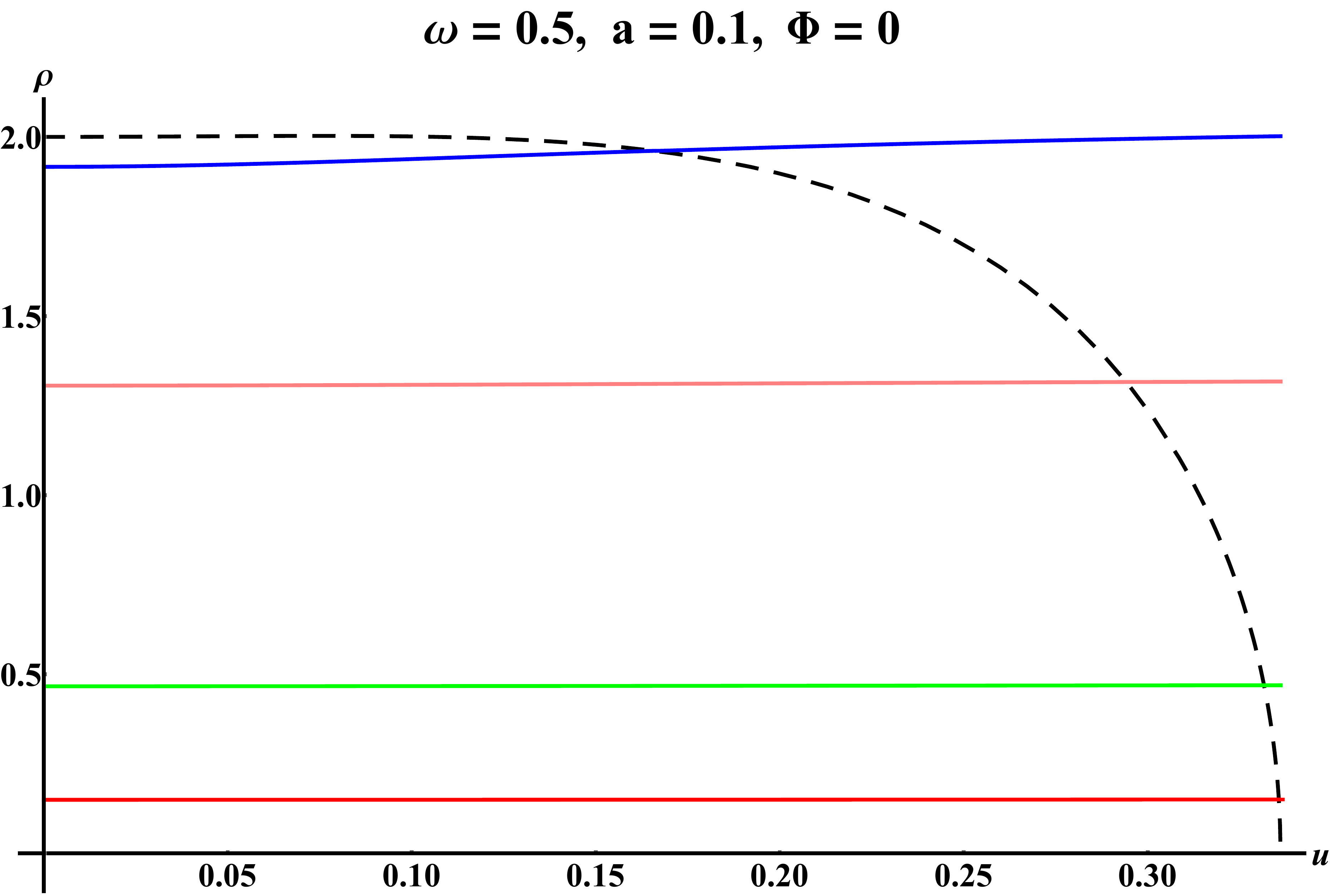}}
	\hspace{.3in}\subfigure[]{\includegraphics[scale=0.08]{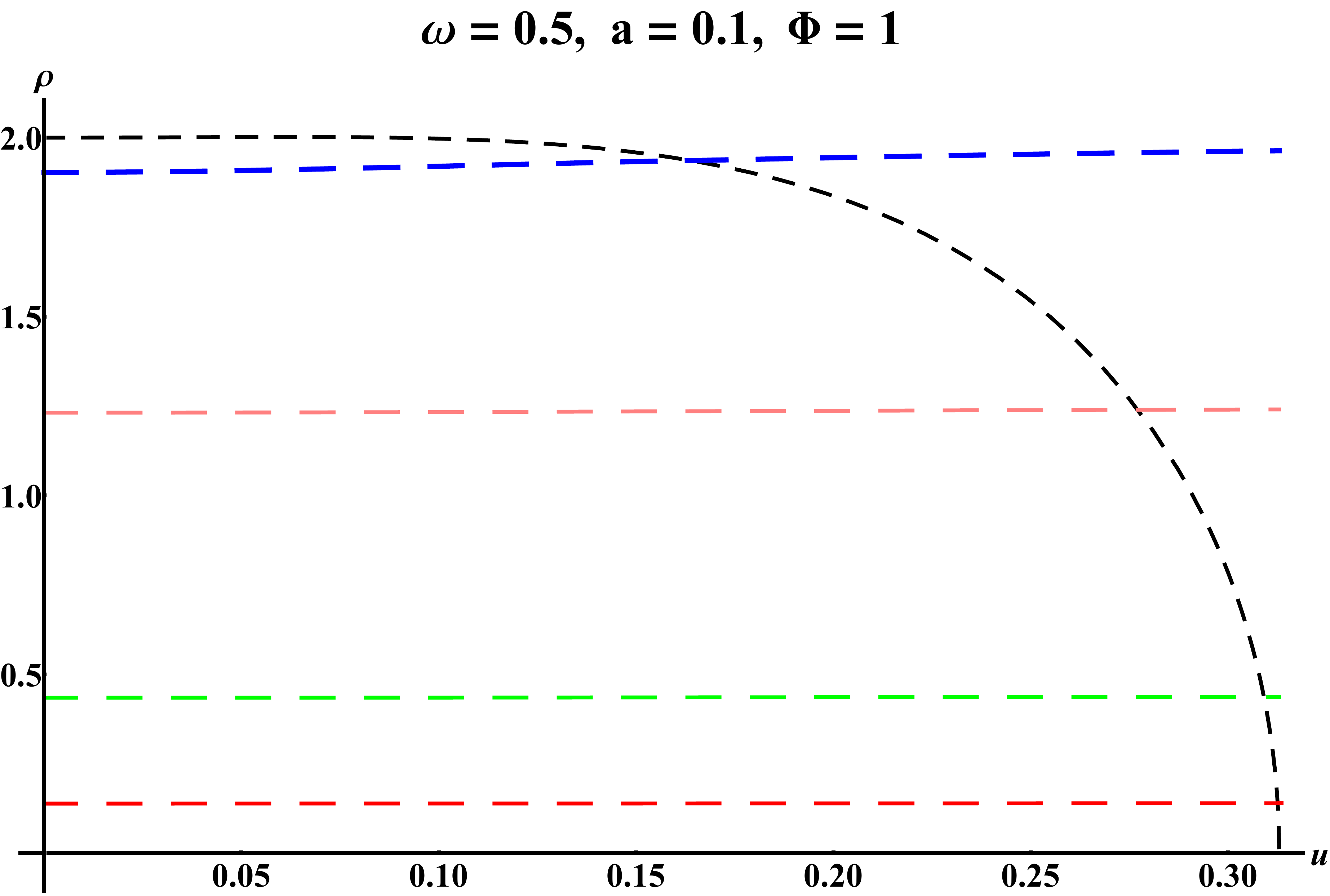}}
	\hspace{.3in}\subfigure[]{\includegraphics[scale=0.08]{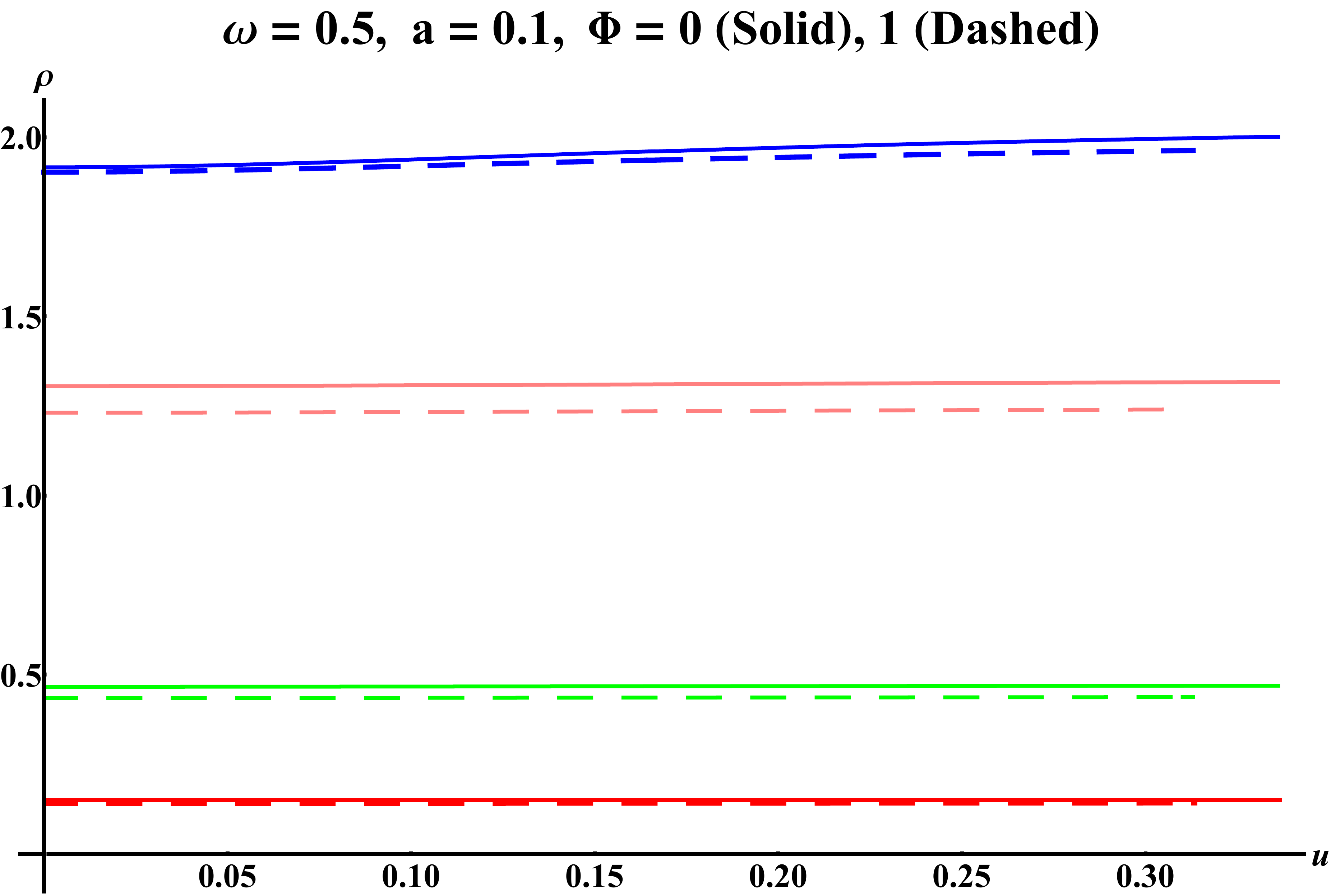}}
	\caption{
	$\rho(u)$ vs $u$ plot for fixed $T = 1$, $\omega = 0.5$, $a=0.1$, $\Phi=0$ (plot a) and $\Phi=1$ (plot b). Coloured lines (bottom to top) corresponds to $\Pi = .1$, $1$, $10$ and $70$ respectively. Plot (c) comparison of $\rho(u)$ vs $u$ for $\Phi=0$ (solid) and $\Phi=1$ (dashed).}
	\label{Plot_energy_loss_of_rotating_quark_a_point1_q_point1_anda_100_q_1_different_PI}
\end{figure}

\begin{figure}[!t]
	\centering
	\subfigure[]{\includegraphics[scale=0.08]{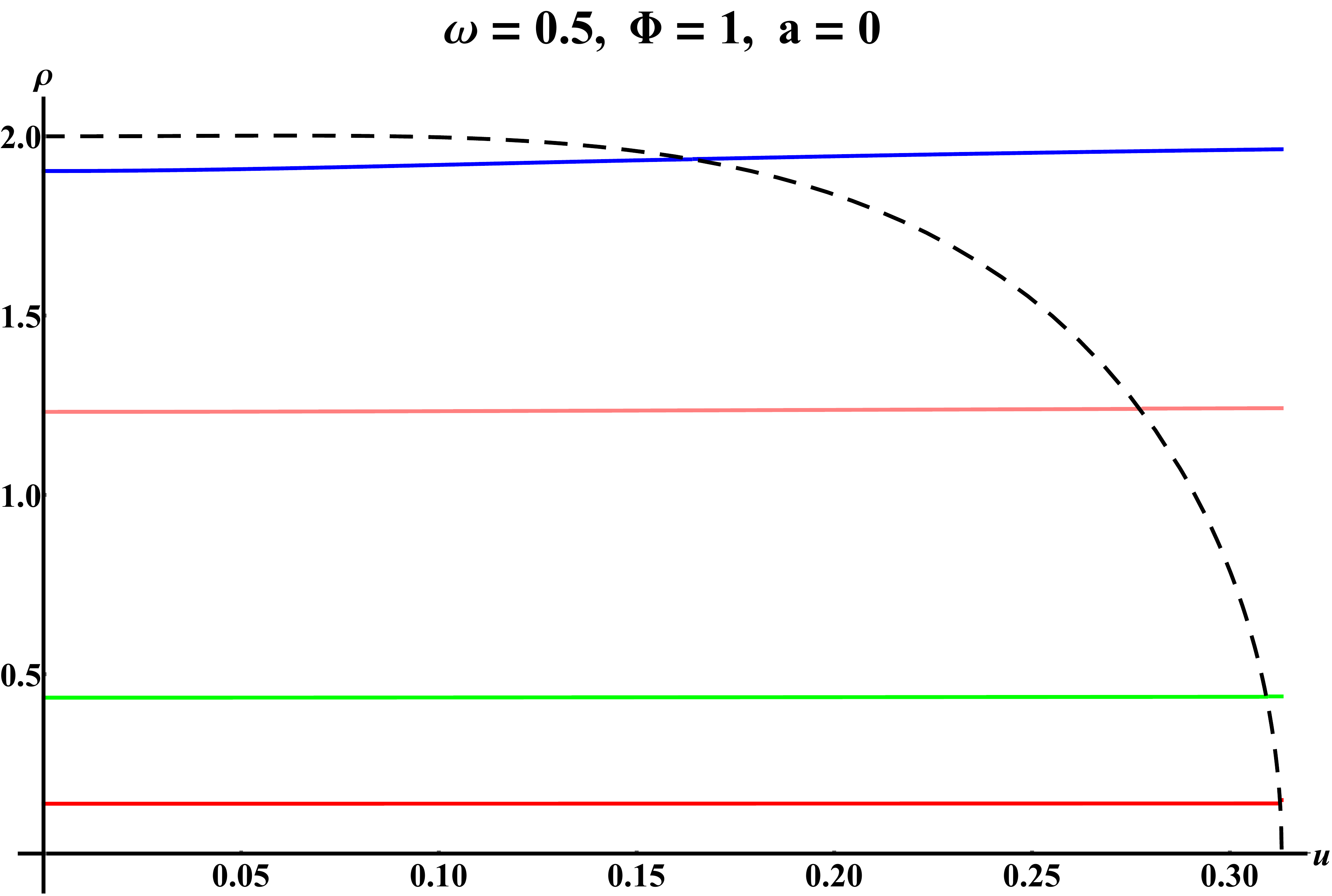}}
	\hspace{.3in}\subfigure[]{\includegraphics[scale=0.08]{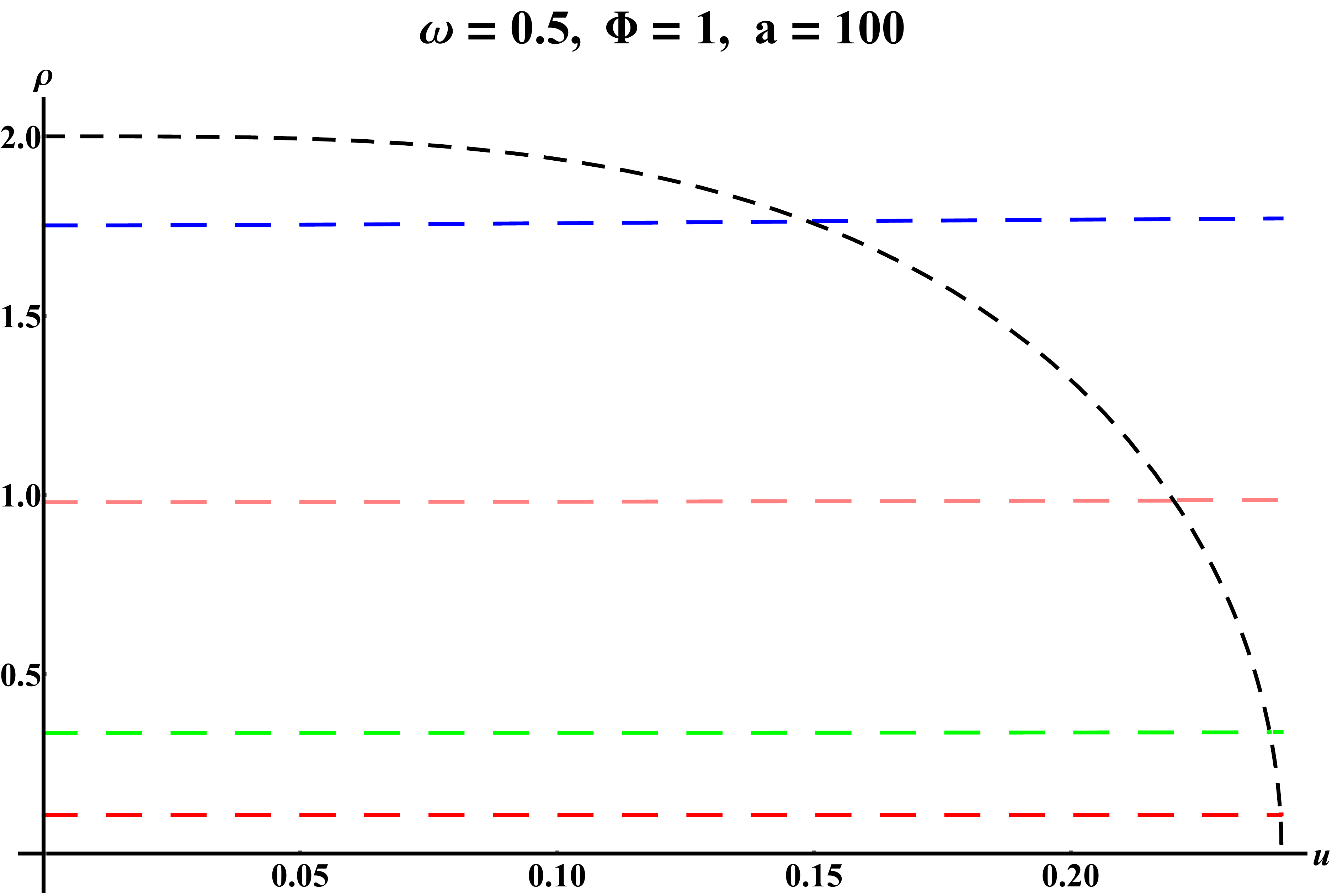}}
	\hspace{.3in}\subfigure[]{\includegraphics[scale=0.08]{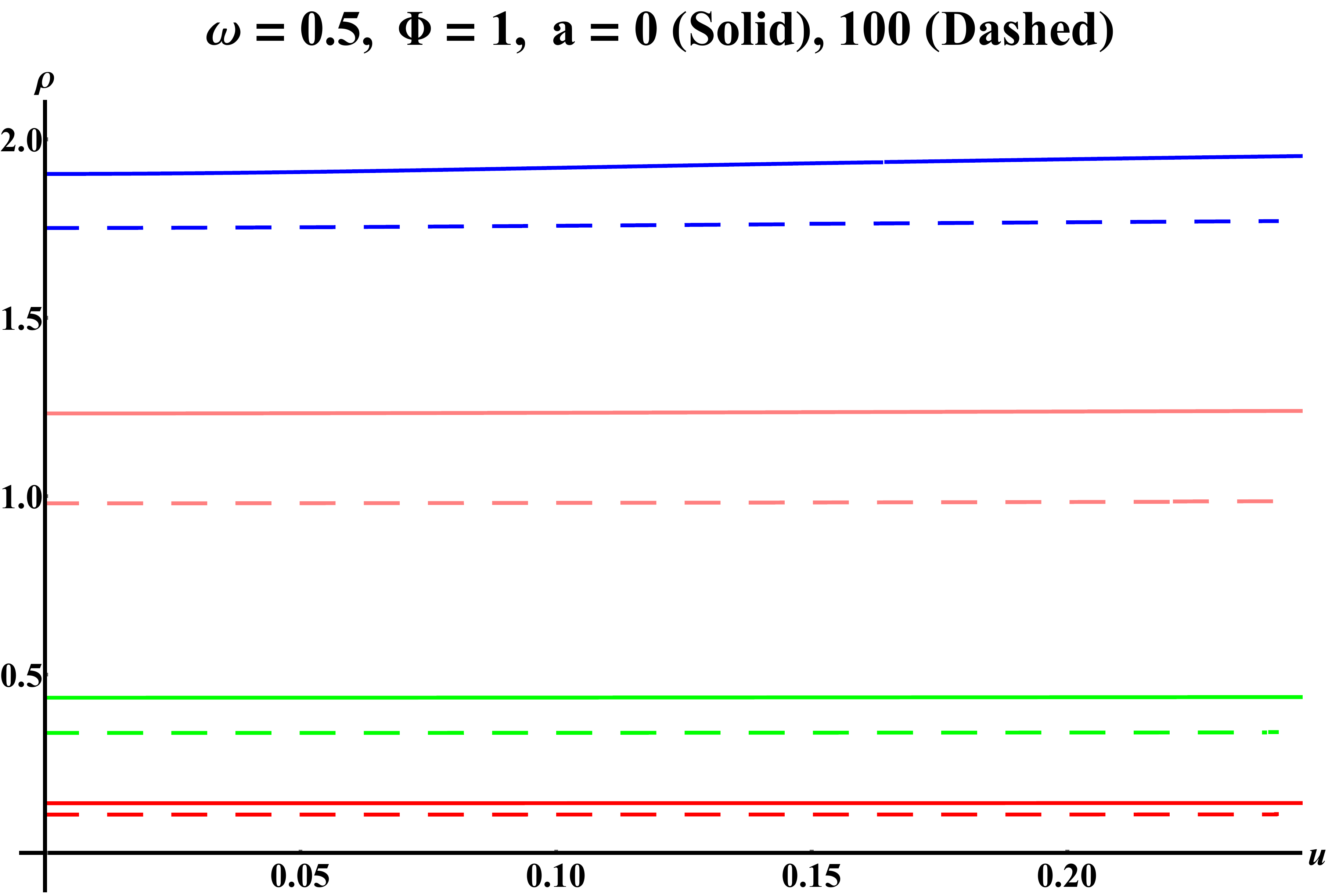}}
	\caption{
	$\rho(u)$ vs $u$ plot for fixed $T = 1$, $\omega = 0.5$, $\Phi=1$, $a=0$ (plot a) and $a=100$ (plot b). Coloured lines (bottom to top) corresponds to $\Pi = .1$, $1$, $10$ and $70$ respectively. Plot (c) comparison of $\rho(u)$ vs $u$ for $a=0$ (solid) and $a=100$ (dashed).}
	\label{Plot_energy_loss_of_rotating_quark_a_point1_q_point1_anda_100_q_1_different_PI_Omega_point5}
\end{figure}

\begin{figure}[!h]
	\centering
	\subfigure[]{\includegraphics[scale=0.08]{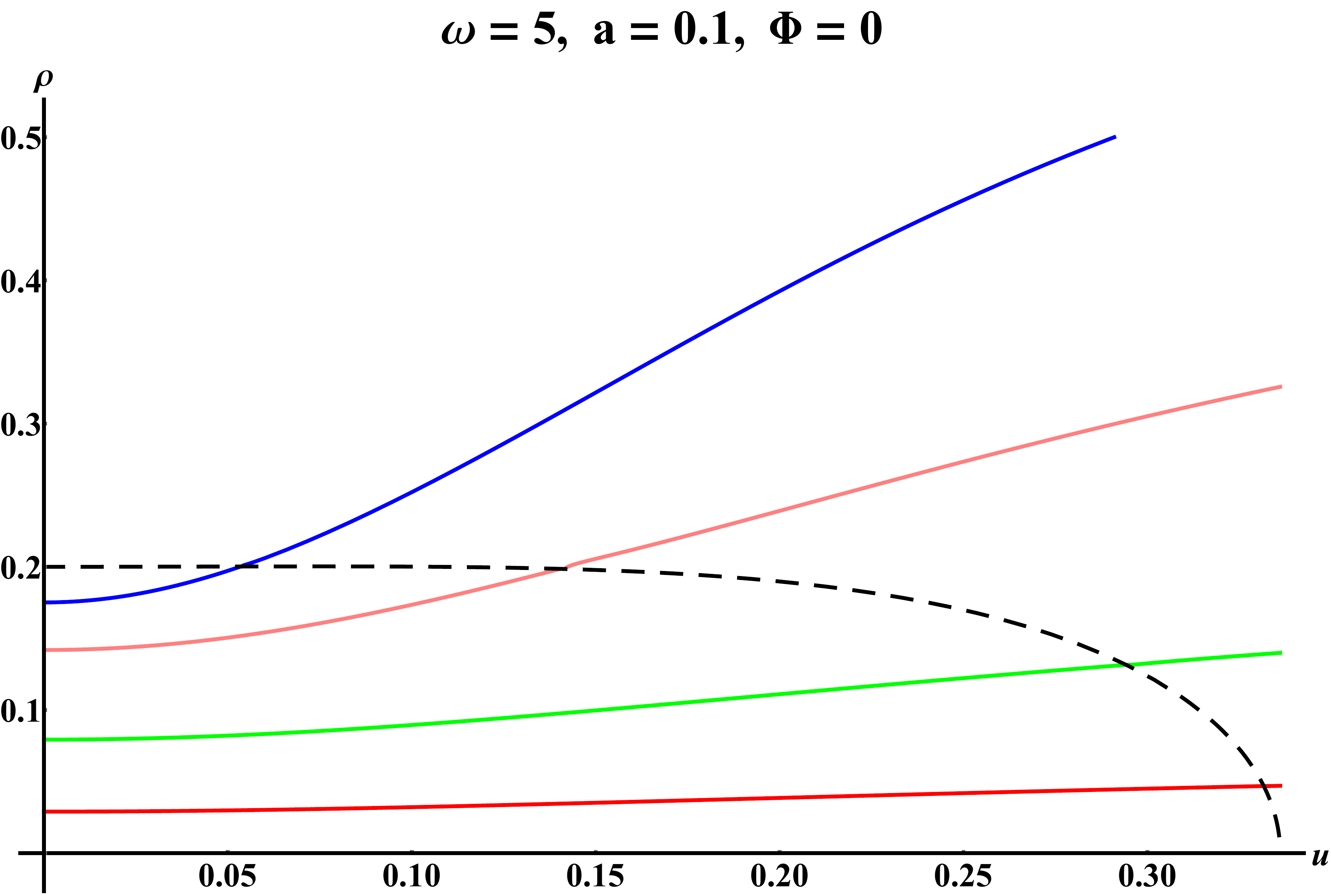}}
	\hspace{.3in}\subfigure[]{\includegraphics[scale=0.08]{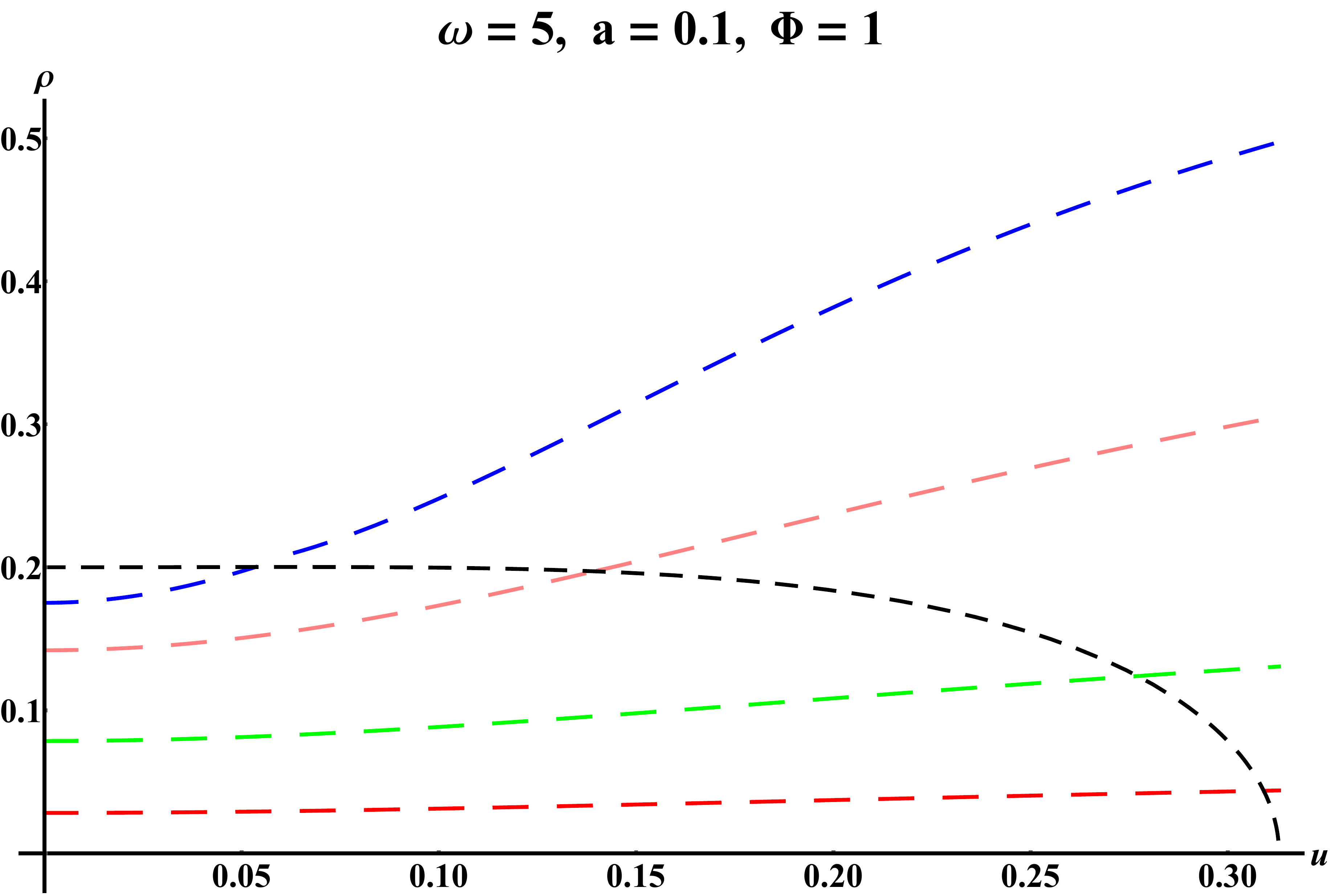}}
	\hspace{.3in}\subfigure[]{\includegraphics[scale=0.08]{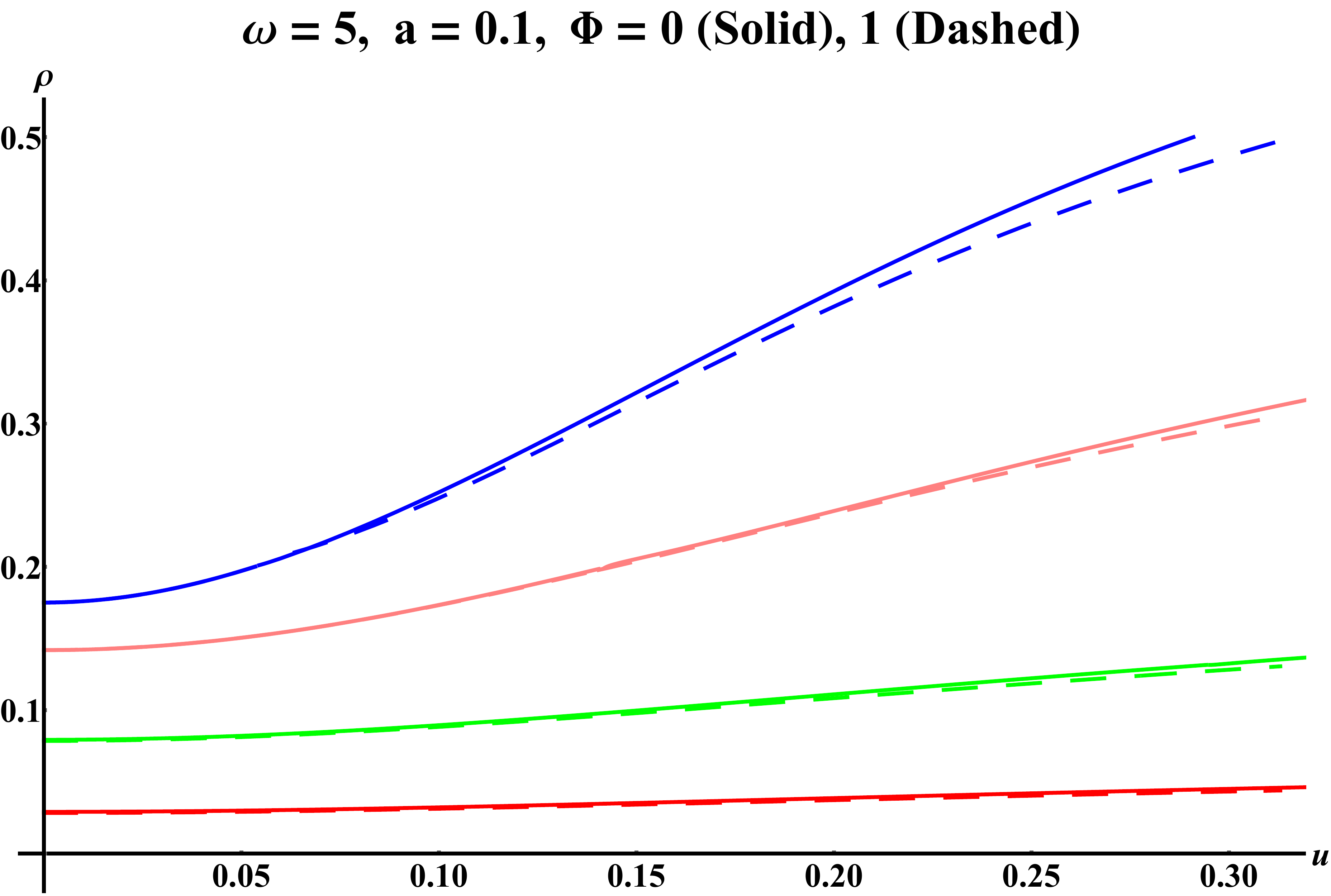}}
	\caption{
	$\rho(u)$ vs $u$ plot for fixed $T = 1$, $\omega = 5$, $a=0.1$, $\Phi=0$ (plot a) and $\Phi=1$ (plot b). Coloured lines (bottom to top) corresponds to $\Pi = .1$, $1$, $10$ and $70$ respectively. Plot (c) comparison of $\rho(u)$ vs $u$ for $\Phi=0$ (solid) and $\Phi=1$ (dashed).}
	\label{Plot_energy_loss_of_rotating_quark_a_point1_q_point1_anda_100_q_1_different_PI_Omega_5}
\end{figure}

\begin{figure}[!h]
	\centering
	\subfigure[]{\includegraphics[scale=0.09]{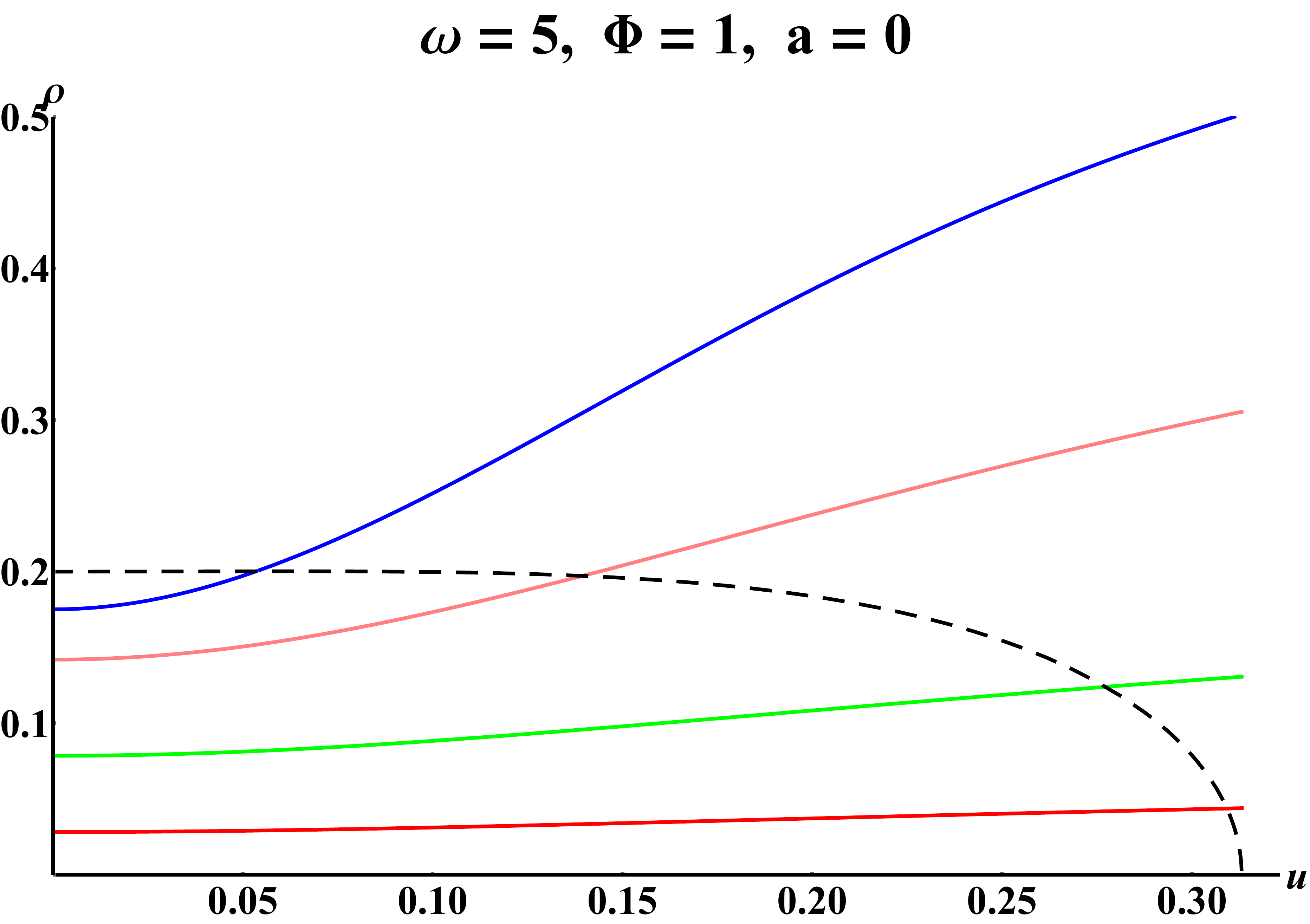}}
	\hspace{.1in}\subfigure[]{\includegraphics[scale=0.09]{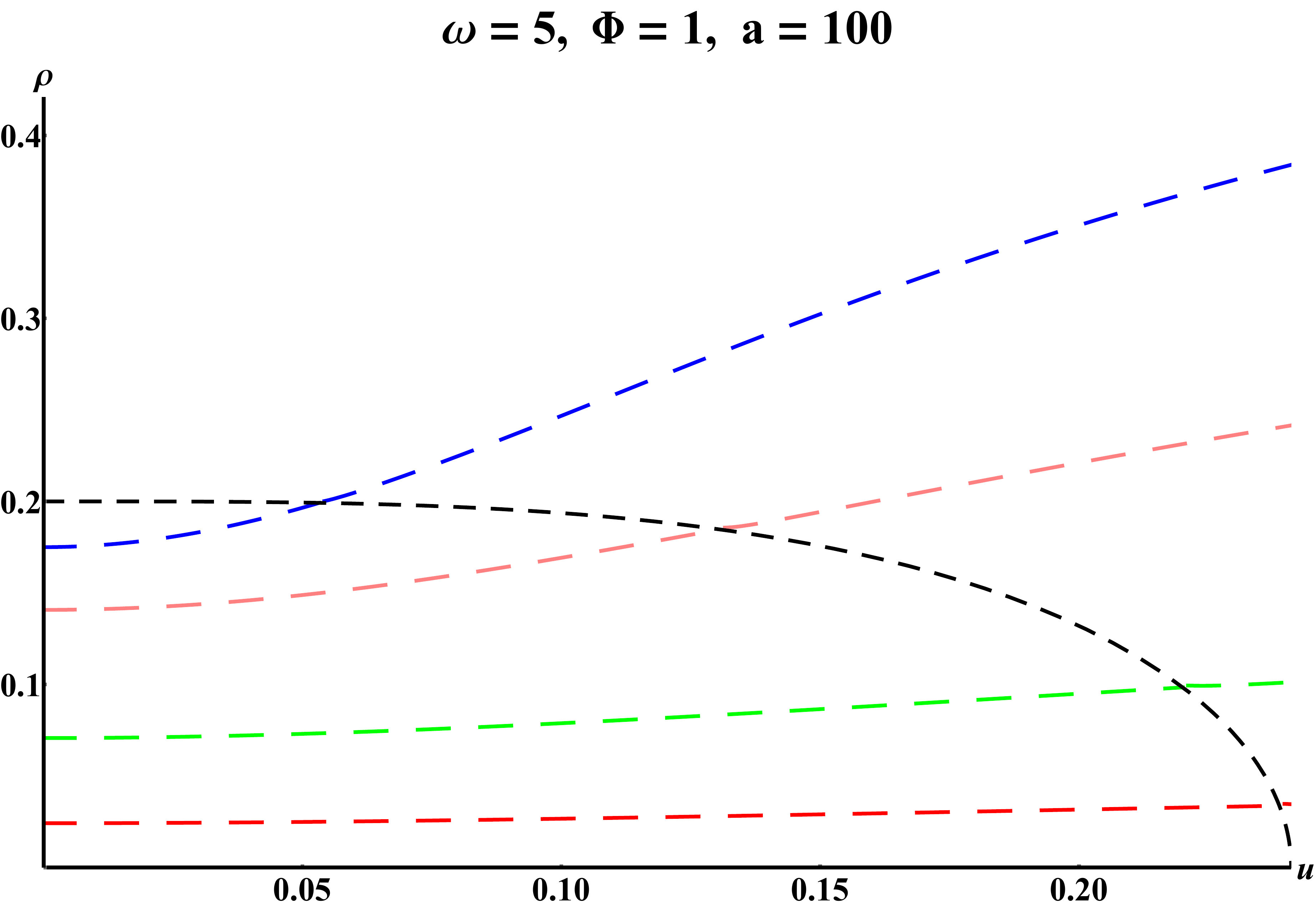}}
	\hspace{.1in}\subfigure[]{\includegraphics[scale=0.09]{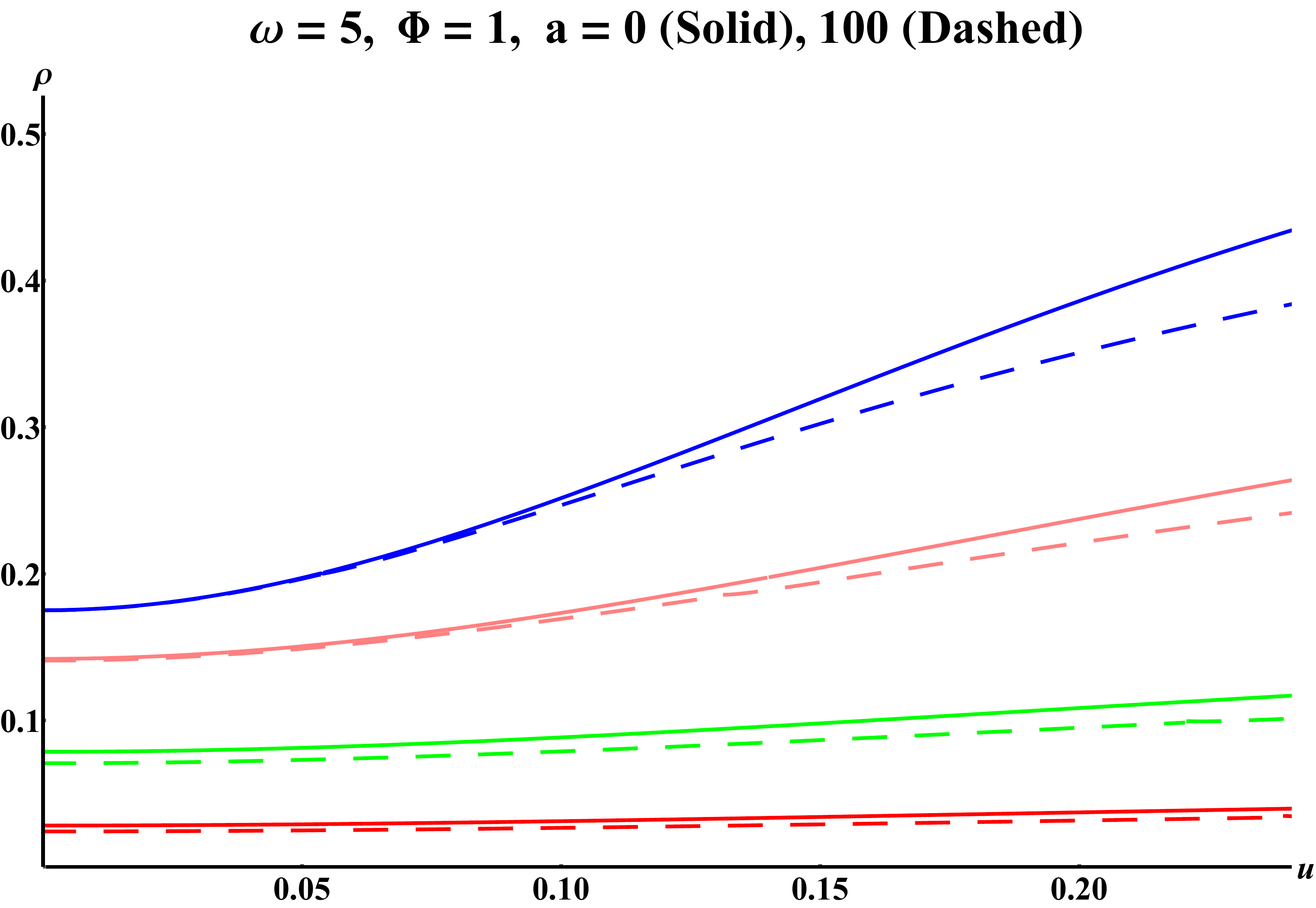}}
	\caption{
	$\rho(u)$ vs $u$ plot for fixed $T = 1$, $\omega = 5$, $\Phi=1$, $a=0$ (plot a) and $a=100$ (plot b). Coloured lines (bottom to top) corresponds to $\Pi = .1$, $1$, $10$ and $70$ respectively. Plot (c) comparison of $\rho(u)$ vs $u$ for $a=0$ (solid) and $a=100$ (dashed).}
	\label{Plot_energy_loss_vs_v_different_omega_different_set_of_a_and_q}
\end{figure}

Here, we discuss the radial profile of a probe string rotating with a constant angular speed in the back reacted charged AdS black hole background. The one end of the string is attached at the boundary of the background and the body is extended up to the horizon of the bulk spacetime. The attached end of the string represents a heavy probe quark rotating with a constant angular speed in the background of static heavy quarks uniformly distributed over $\mathcal{N} =4$ SYM finite temperature plasma with chemical potential. Assuming the probe quark rotates on two dimensional flat space along a circle of radius $\mathcal{R}$ with constant angular frequency $\omega$. The constant speed and acceleration of the rotating quark are given as $ v = \mathcal{R} \omega$ and $a = \omega^2 \mathcal{R}$. Due to the rotational motion, the body of the string experiences a centrifugal force and takes a spiral profile.  The Nambu-Goto action  of the rotating string is useful to study the dynamics of the spiral profile.  Preserving the SO(2) symmetry, we choose the following parameterizations of the string worldsheet,
\begin{equation}
	X^\mu(\tau,\sigma) = (t=\tau, u = \sigma, x = \rho(\sigma)cos(\omega t + \theta(\sigma)), y = \rho(\sigma)sin(\omega t + \theta(\sigma)), z=0),
	\label{equation_energyloss_parameterization}
\end{equation}
where the parameters $\rho(\sigma)$ and $\theta(\sigma)$ represent the radial and angular profiles of the rotating string and follow the boundary conditions,
\begin{equation}
	\rho(0) = \mathcal{R}, \, \theta(0) = 0.
\end{equation}
Following the parameterization  as in (\ref{equation_energyloss_parameterization}), the Lagrangian density for the standard Nambu-Goto action can be written as,
\begin{equation}
	\mathcal{L} = \left[\frac{f^2}{h}(h - \rho^2 \omega^2) + f^2 (h- \rho^2 \omega^2) \rho^{'2} + h f^2 \rho^2 \theta^{'2}\right]^{\frac{1}{2}}.
	\label{equation_energyloss_lagrangian_density}
\end{equation}
Since, the Lagrangian density is not an explicit function of coordinate $\theta$, the corresponding conjugate momentum is obtained as,
\begin{equation}
	\Pi_\theta  = \frac{\partial \mathcal{L}}{\partial \theta'} = \frac{h f^2 \rho^2 \theta'}{\mathcal{L}}.
\label{paitheta}
\end{equation}
By rearranging the above equation, $\theta'$ can be expressed as,
\begin{equation}
	\theta' = \Pi_\theta \sqrt{\frac{(h- \rho^2 \omega^2)(1 + h \rho^{'2})}{h^2 \rho^2 (h f^2 \rho^2 - \Pi_\theta^2)}}.
	\label{equation_energyloss_theta_dashed}
\end{equation}
At the horizon, $h(u=u_h) = 0 $ and the factor in the numerator $h(u) - \rho^2 \omega^2$ becomes negative. However, at the boundary, the factor $h- \rho^2 \omega^2 >0$, since the speed of the quark is less than the speed of the light. Hence this factor must change sign at some critical value of the radial coordinate $u_c < u_h$. Furthermore, the rotating string with radius $\rho = \mathcal{R}$ at the boundary, goes through $(u_c,\rho_c)$ and extends up to the black hole horizon.  Since $\rho^2 \omega^2$ is less than $1$ for $0\le u <u_c$ and becomes greater than $1$ for $u_c < u \le u_h$, the strings in the range $u<u_c$ are causally disconnected from the part of the string in $u>u_c$. The strings embedded in the range $u<u_c$ are physically relevant part of the string which moves with the speed slower than the local speed of light. The curve that represents the radial profile $\rho_l$ of a string moving with a velocity  of light is given by $h(u_c) = \rho(u_c)^2 \omega^2$. All other strings moving with velocity slower than the speed of light should intersect the curve $\rho_l$ at the critical point $u = u_c$. To avoid the imaginary value of $\theta'$, we impose the following conditions at the critical point,
\begin{equation}
	h(u_c) - \rho(u_c)^2 \omega^2 =0,
	\label{equation_energyloss_condition1}
\end{equation}
\begin{equation}
	h(u_c) f(u_c)^2 \rho(u_c)^2 - \Pi_\theta^2 =0.
	\label{equation_energyloss_condition2}
\end{equation}
Solving these two equations we get,
\begin{equation}
	\rho(u_c) = \sqrt{\frac{\Pi_\theta}{f(u_c) \omega}},
\end{equation}
\begin{equation}
	f(u_c) = \frac{\Pi_\theta \omega}{h(u_c)}.
\label{fuc}
\end{equation}
Now, by using (\ref{equation_energyloss_theta_dashed}) we eliminate $\theta'$ from the equation of motion for $\rho$ coordinate which is calculated as 
\begin{equation}
	\frac{\partial}{\partial u}\left(\frac{\partial \mathcal{L}}{\partial \rho'}\right)- \frac{\partial \mathcal{L}}{\partial \rho}=0
\end{equation} 
and obtain the equation of motion for $\rho$ as,
\begin{align}
	& 2(\Pi_\theta^2 - \omega^2 f^2 \rho^4) - \left[2 f h \rho^3 (h -\omega^2 \rho^2) f' - \rho \{\Pi_\theta^2 - f^2 ( 2 h \rho^2 - \omega^2 \rho^4)\}h'\right]\rho'\nonumber \\ & + 2h (\Pi_\theta^2 - \omega^2 f^2 \rho^4 ) \rho^{'2} - \rho^3 \left[ 2 f h^2 (h- \omega^2 \rho^2) f' - (\Pi_\theta^2 \omega^2 - f^2 h^2 )h'\right]\rho^{'3} \nonumber \\ & + 2 \rho(h-\omega^2 \rho^2) (\Pi_\theta^2 - f^2 h \rho^2) \rho^{''} = 0.
	\label{equation_energyloss_final_EOM_for_rho}
\end{align} 
 The radial profile has a direct connection to study the energy loss, hence we are interested to extract the radial profile here. However, the analytic solution of the equation of motion (\ref{equation_energyloss_final_EOM_for_rho}) is difficult to obtain, hence we opt for a numerical solution. We specify the boundary condition by fixing the values of $\rho$ and $\rho'$ at $u=u_c$ for some fixed values of $\omega$, $\Phi$ and $\Pi_\theta$.\\
To determine $\rho'(u_c)$, we expand the radial coordinate $\rho(u)$ around $u_c$ and consider terms up to linear order as,
\begin{equation}
	\rho(u) = \rho(u_c) + \rho'(u_c) (u-u_c) + \cdots.
	\label{equation_energyloss_rho_u_expansion}
\end{equation}
Now, using (\ref{equation_energyloss_rho_u_expansion}) in (\ref{equation_energyloss_final_EOM_for_rho}) and keeping terms up to linear order, we observed that the zeroth order coefficient turns out to be zero, whereas the linear order coefficient sets a quartic equation in $\rho'(u_c)$  and given as,
\begin{align}
	\rho'_c = &\frac{1}{(4 u_c h_c^3 \rho_c)}\left[- 4 u_c^2 h_c^2 - h_c^3 \rho_c^2 + \omega^2 h_c^2 \rho_c^4 + 2 u_c h_c^2 \rho_c^2 h'_c- \Pi^2 u_c^4 h^{'2}_c\right.\nonumber \\ & \left. + \left(16 u_c^2 h_c^5 \rho_c^2 + \{ h_c^3 \rho_c^2 + \Pi^2 u_c^4 h^{'2}_c + h_c^2 ( 4 u_c^2 - \omega^2 \rho_c^4 - 2 u_c \rho_c^2 h'_c)\}^2\right)^{\frac{1}{2}}\right],
\end{align}
where $\rho_c = \rho(u_c)$ and $h_c = h(u_c)$. Furthermore, $u=u_c$ is a singular point for both the Lagrangian (\ref{equation_energyloss_lagrangian_density}) and equation of motion (\ref{equation_energyloss_final_EOM_for_rho}), hence we solve (\ref{equation_energyloss_final_EOM_for_rho}) in the ranges defined from $u_c-\delta$ to the boundary and $u_c+\delta$ to the horizon and finally consider $\delta \rightarrow 0$ limit.\\
In figures (\ref{Plot_energy_loss_of_rotating_quark_a_point1_q_point1_anda_100_q_1_different_PI}), (\ref{Plot_energy_loss_of_rotating_quark_a_point1_q_point1_anda_100_q_1_different_PI_Omega_point5}), (\ref{Plot_energy_loss_of_rotating_quark_a_point1_q_point1_anda_100_q_1_different_PI_Omega_5}), (\ref{Plot_energy_loss_vs_v_different_omega_different_set_of_a_and_q}), the characteristics of the radial profile of the rotating spiral string has been plotted for different choices of the parameters such as temperature $T$, string density $a$, potential $\Phi$, conserved string momenta $\Pi_\theta$ and the angular speed $\omega$. For each profile there is a unique limit $\rho(u\rightarrow 0) = \mathcal{R}$ which signifies the radius of the rotating quark in the boundary theory. The intersection between the black dotted profile $\rho_l$ and each of the radial profiles $\rho(u)$ fixes the value of the critical point $u_c$ and corresponding radius at  the turning point is $\rho(u_c)$. For some fixed $\omega$, string density $a$ and potential $\Phi$, as we increase the value of $\Pi_\theta$, the value of intersecting point $u_c$ reduces and $\rho(u_c)$ increases. Further, it is observed that for $\omega<1$, the radial profile is almost constant, whereas for $\omega>1$ the bending of the string profile is significant and the radius at the horizon is bigger than the radius at boundary $\mathcal{R}<\rho(u=u_c)$. For some fixed $\Pi_\theta$ and angular frequency $\omega$, increasing the back reaction and potential, the radius of rotation decreases accordingly, which is more evident for $\omega<1$.

With the generic discussion of the radial profile of the rotating string, we now move towards the study of the rate of energy loss of a heavy probe quark rotating in the baryon rich back reacted $\mathcal{N}=4$ SYM thermal plasma dual to charged AdS black hole with cloud of strings.


\section{Energy Loss of Rotating Heavy Quark}
\label{energyloss}
\begin{figure}[!h]
	\centering
	\subfigure[]{\includegraphics[scale=0.08]{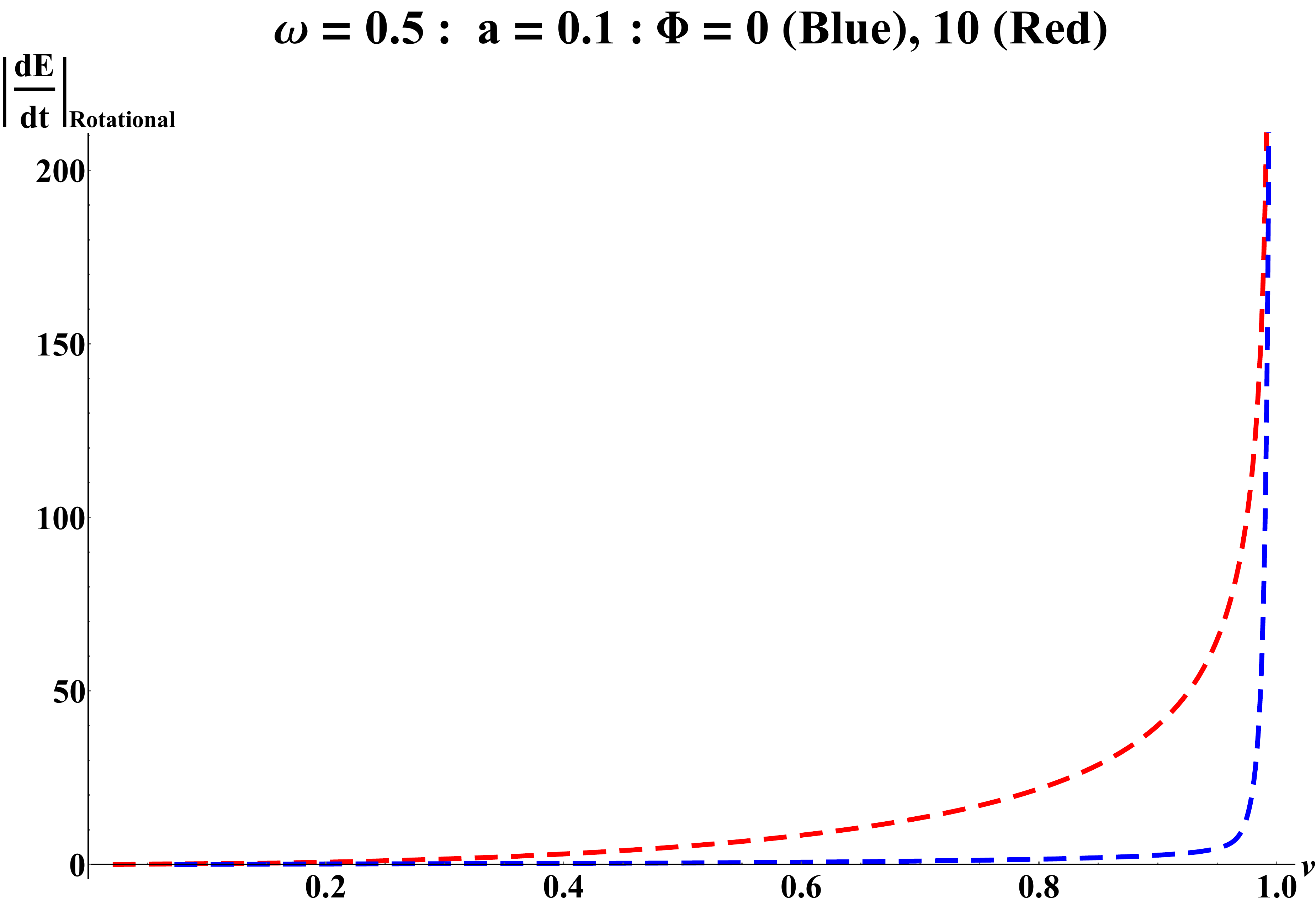}}
	\hspace{.1in}\subfigure[]{\includegraphics[scale=0.08]{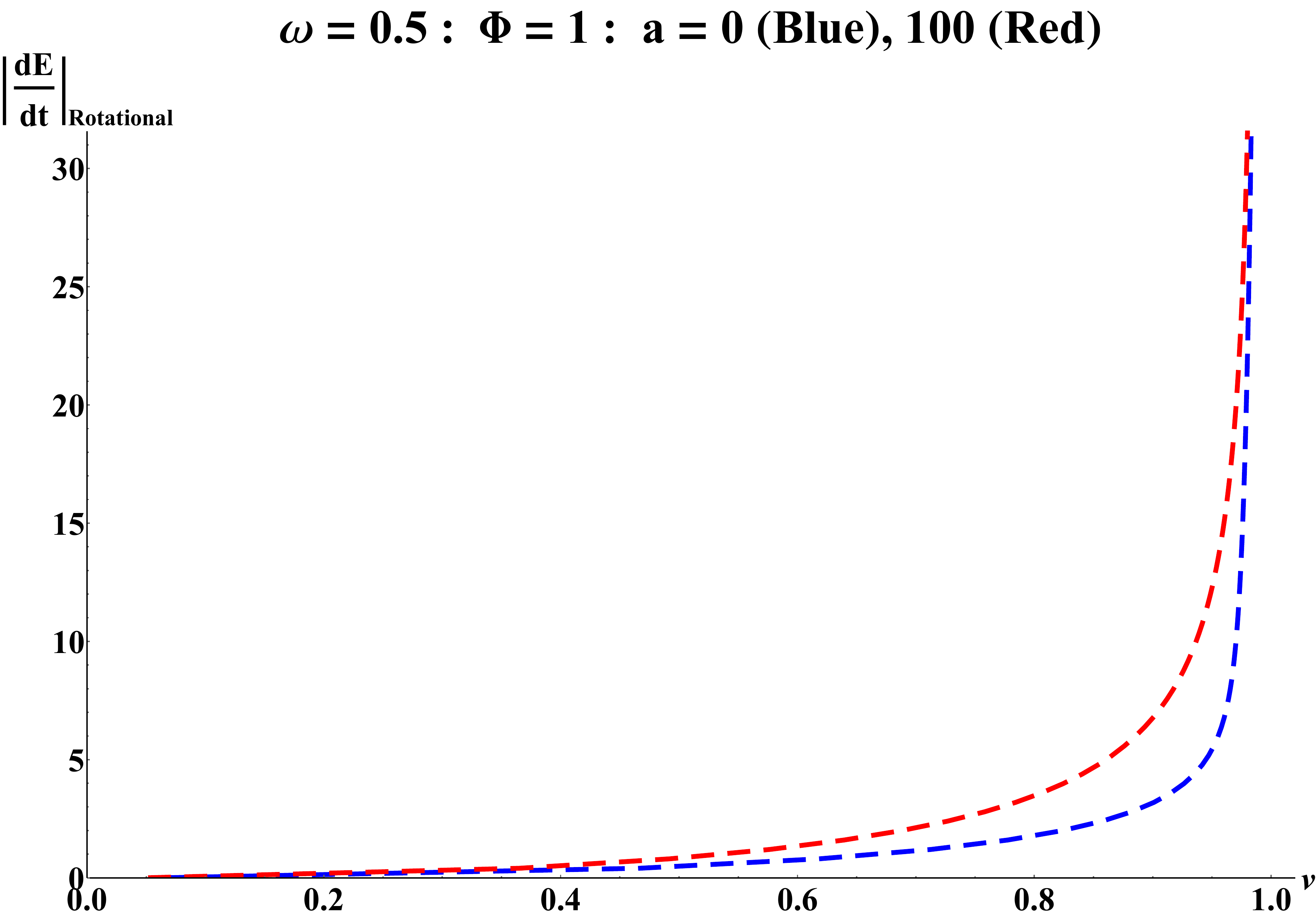}}
	\hspace{.1in}\subfigure[]{\includegraphics[scale=0.08]{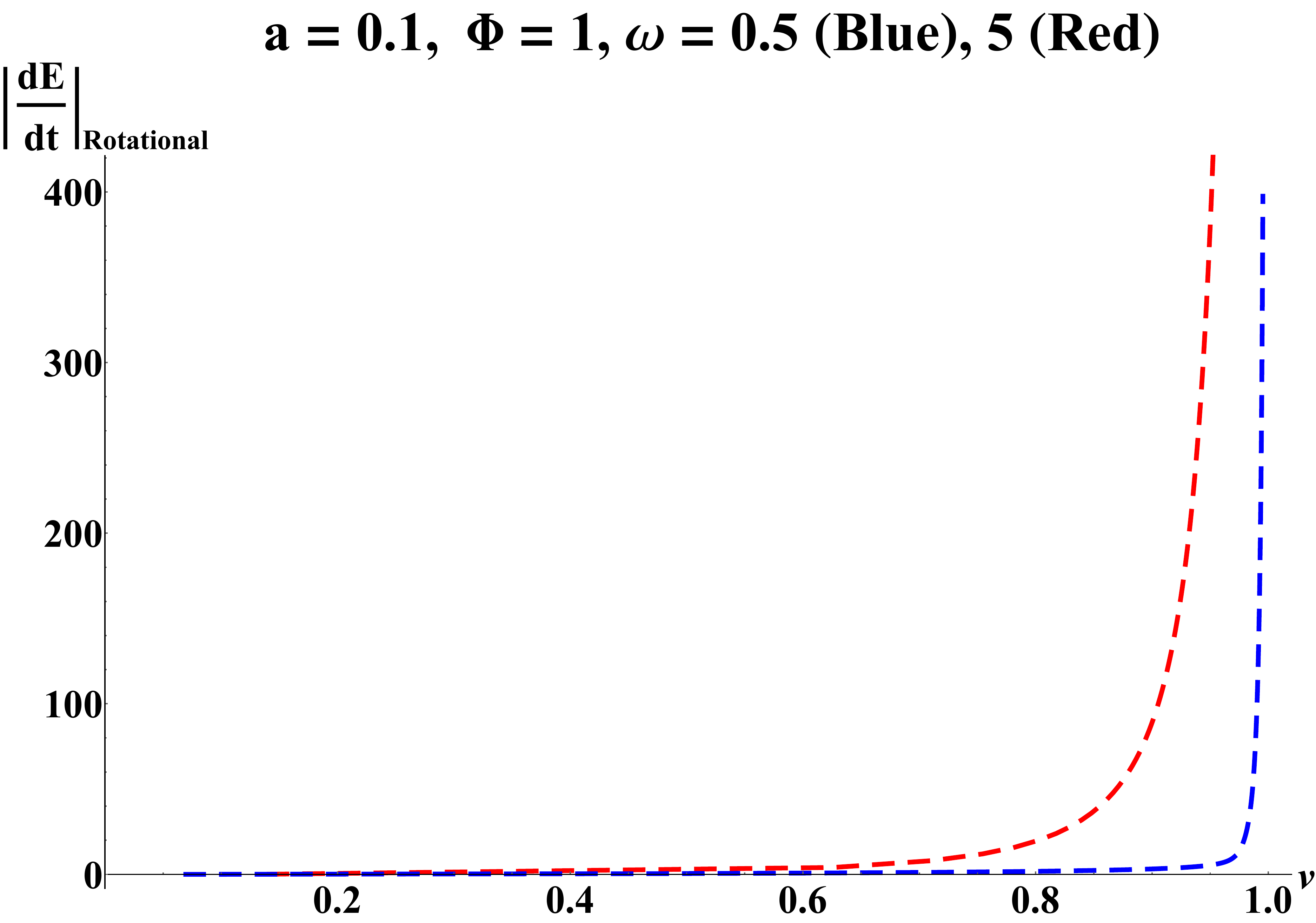}}
	\caption{Energy loss ($\frac{dE}{dt}$) with respect to quark speed ($v$) for Plot (a) $\omega = 0.5$, $a = 0.1$ and $\Phi = 0$ (blue) and $10$ (red). Plot (b) $\omega = 0.5$, $\Phi = 1$ and $a = 0$ (blue) and $100$ (red). Plot (c) $a = 0.1$, $\Phi = 1$ and $\omega = 0.5$ (blue) and $5$ (red).}
	\label{Plot_energy_loss_vs_v_omega_Point_5}
\end{figure}

 The rotating probe quark interacts with the back reacted finite temperature plasma with chemical potential and as a result the energy loss take place either in the form of radiation or in medium dissipation. Directly, the study of energy loss in boundary theory is very difficult. Following \cite{Athanasiou:2010pv,Herzog:2007kh,Chakrabortty2016a}, holographically, the energy loss can be computed by studying the motion of the rotating spiral probe string in the weakly coupled deformed charged AdS black hole spacetime. The one end of this spiral string is representing the heavy probe quark in the boundary theory. Using the dictionary of gauge/gravity duality, the energy loss associated with the rotating string can be calculated and it is expressed as,
\begin{equation}
	\frac{dE}{dt} = -\frac{\delta S}{\delta (\partial_\sigma X^0)} = \Pi_t^\sigma,
\end{equation}
where $S$ is the Nambu-Goto action. Using equations (\ref{metricsol}), (\ref{equation_energyloss_parameterization}), (\ref{paitheta}) and (\ref{fuc}), we can rewrite the above expression in the following way,
\begin{equation}
	\frac{dE}{dt} = \frac{h f^2 \omega \rho^2 \theta'}{2 \pi \alpha \sqrt{-g}} = \frac{\Pi_\theta \omega}{2 \pi \alpha} = \frac{f(u_c) h(u_c)}{2 \pi \alpha} = \frac{h(u_c)}{ 2\pi \alpha u_c^2},
	\label{equation_energyloss_total}
\end{equation}
where, $l=1$ and from the above equation it is clear that the energy loss of a rotating string depends on the critical value $u_c(\Pi_\theta, \omega, a,\Phi)$. We are interested to study the effect of flavour quarks and baryons on the energy loss. Hence, we study the energy loss for different values of back reaction and chemical potential for different set of angular velocities with respect to the boundary quark speed $v$ as given in figure (\ref{Plot_energy_loss_vs_v_omega_Point_5}). To calculate the energy loss, first we choose a set of $\omega$'s, $a$'s, $\Phi$'s and $\Pi_\theta$'s, and for each combination of these parameters we calculate ($\rho_c, \rho'_c$) and solve the differential equation (\ref{equation_energyloss_final_EOM_for_rho}), then compute the velocity $v = \rho(u\rightarrow 0) \omega$ and calculate the energy loss from equation(\ref{equation_energyloss_total}). In figure (\ref{Plot_energy_loss_vs_v_omega_Point_5}), we have shown the energy loss against the quark speed $v$, keeping two parameters out of three parameters ($\omega,\, a,\, \Phi$) fixed and change the third parameter.  It is observed from these plots that the energy loss enhanced with the increase of any parameter in the medium at higher value of speed. The effect of chemical potential on the enhancement of energy loss is more than the other parameters. But when the quark moves with slow speed, the energy loss does not take place even for the finite value of chemical potential, back reaction and angular momentum. 

\begin{figure}[!h]
	\centering
	\subfigure[]{\includegraphics[scale=0.09]{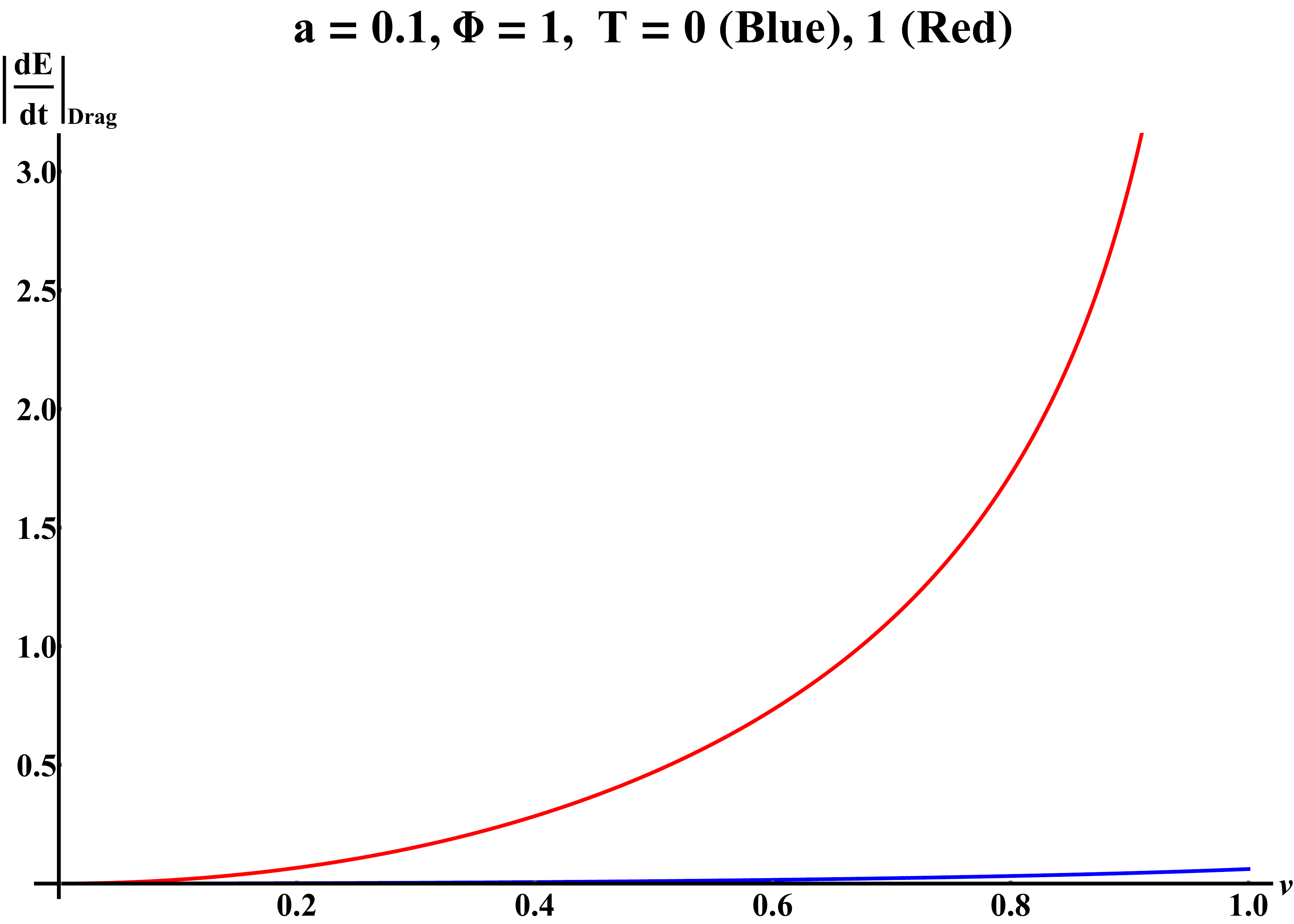}}
	\subfigure[]{\includegraphics[scale=0.09]{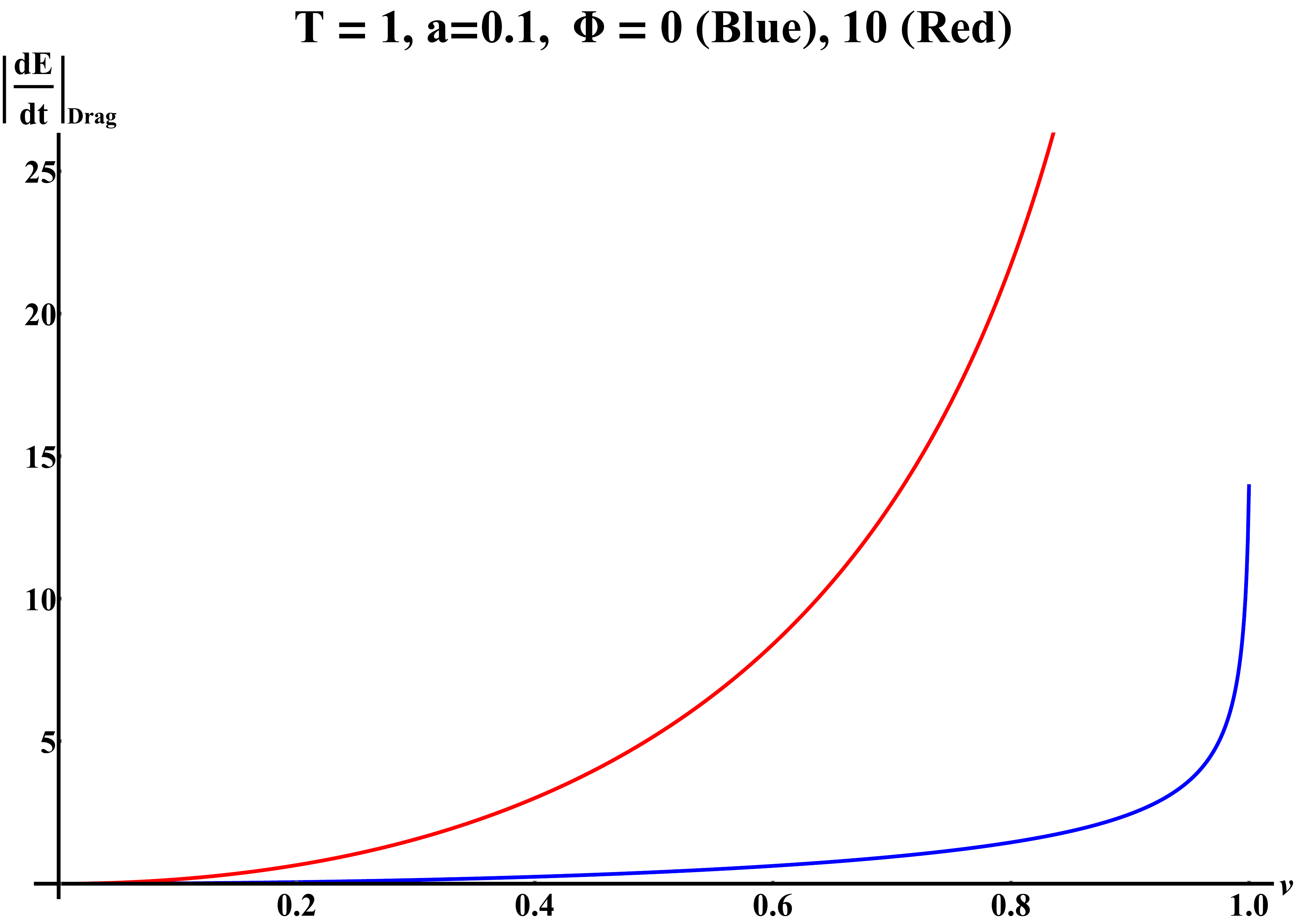}}
    \subfigure[]{\includegraphics[scale=0.09]{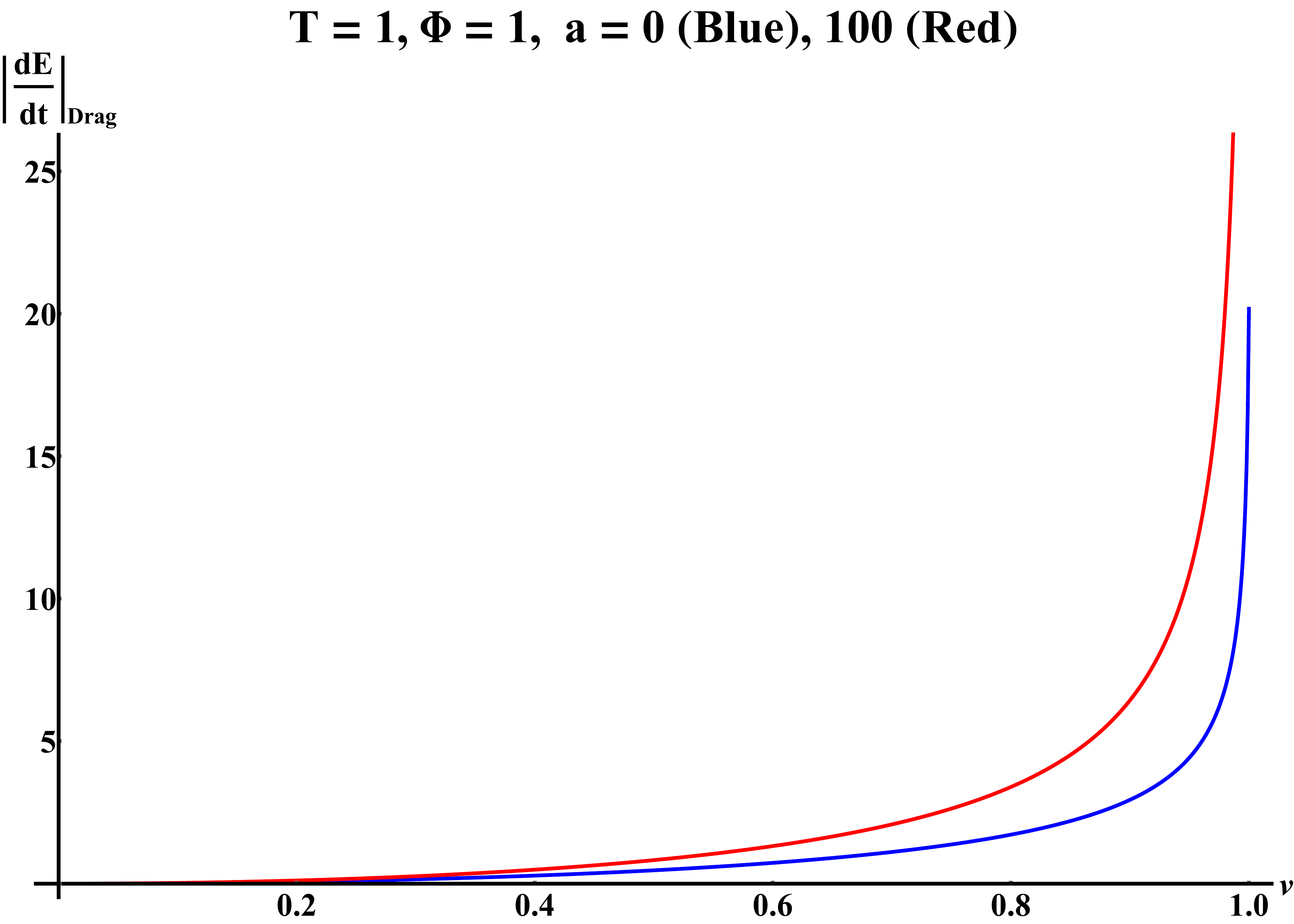}}
	\caption{
	Drag energy loss ($dE/dt$) vs $v$ for fixed (a) $a = 0.1$, $\Phi=1$ and different temperature $T=0$ (blue) and $1$ (red). (b) $T = 1$, $a=0.1$ and different potential $\Phi=0$ (blue) and $10$ (red). (c) $T = 1$, $\Phi=1$ and different string density $a=0$ (blue) and $100$ (red).}
	\label{dEdt_drag_Vs_v_plots}
\end{figure}

Further, we are interested to study the energy loss due to the effect of drag on the string. The energy loss due to the drag is given as,
\begin{equation}
	\frac{dE}{dt} = -\frac{\delta S}{ \delta(\partial_\sigma X^0)} = \left.\frac{h(u_c)}{2 \pi \alpha u_c^2}\right|_{drag}
\end{equation}
Following \cite{Chakrabortty2016a} 
we can say that $u_c$ is the solution of the equation,
\begin{equation}
	\frac{u_c^4 \left(a u_h^3-6 q^2 u_h^6-6 \pi  T u_h+3\right)}{3 u_h^4}-\frac{2 a u_c^3}{3}+q^2 u_c^6+u_c^2+1-v^2 = 0
\end{equation}
In figure (\ref{dEdt_drag_Vs_v_plots}), we have plotted the behaviour of rate of energy loss of a rotating quark due to the drag force acting on it with respect to the velocity $v$. In the plot we keep two variables fixed and other one taking two different values. Since drag force increases for all the parameters, the rate of energy loss due to drag also increases for all the parameters.

\begin{figure}[!h]
	\centering
	\subfigure[]{\includegraphics[scale=0.09]{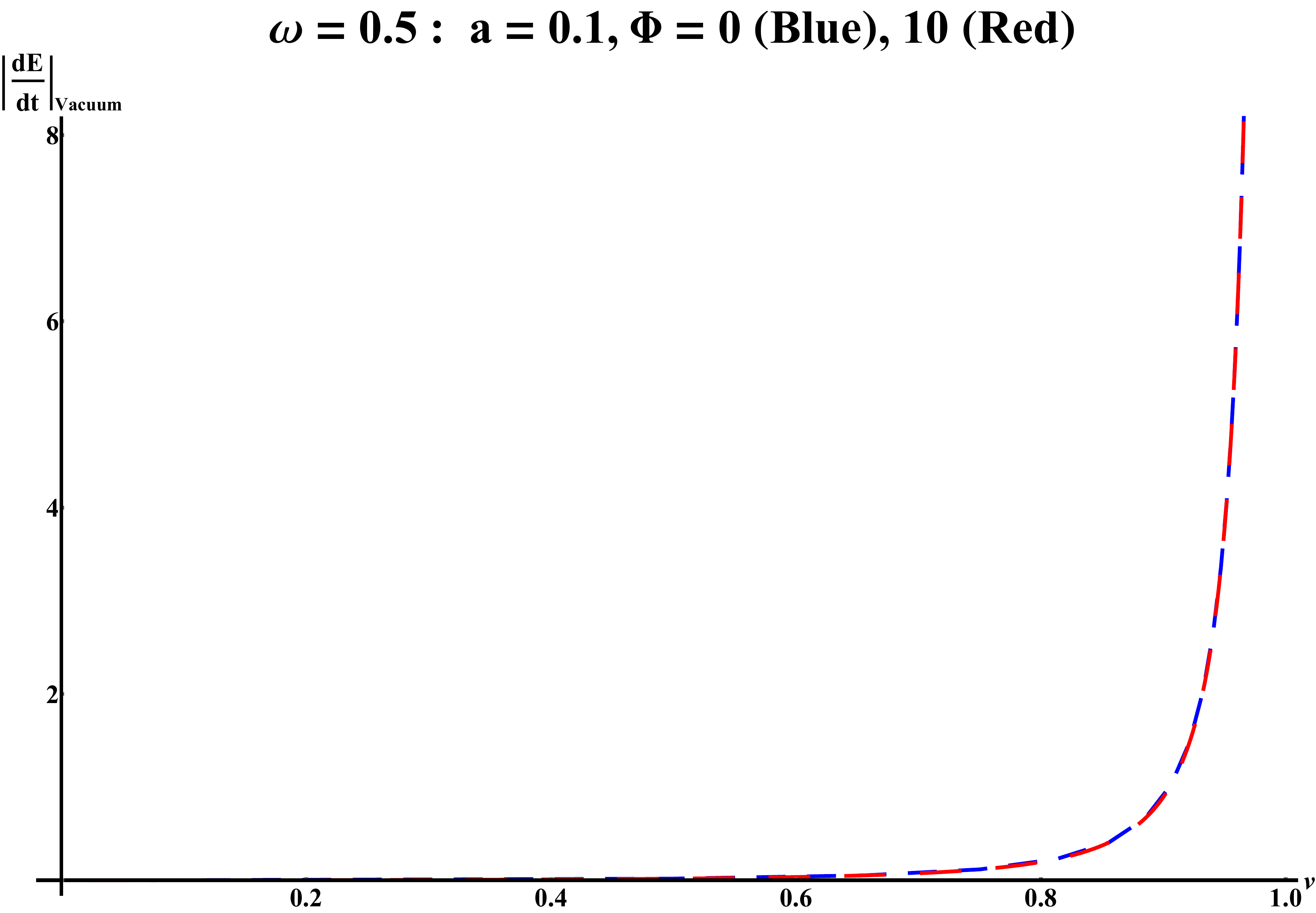}}
	\subfigure[]{\includegraphics[scale=0.09]{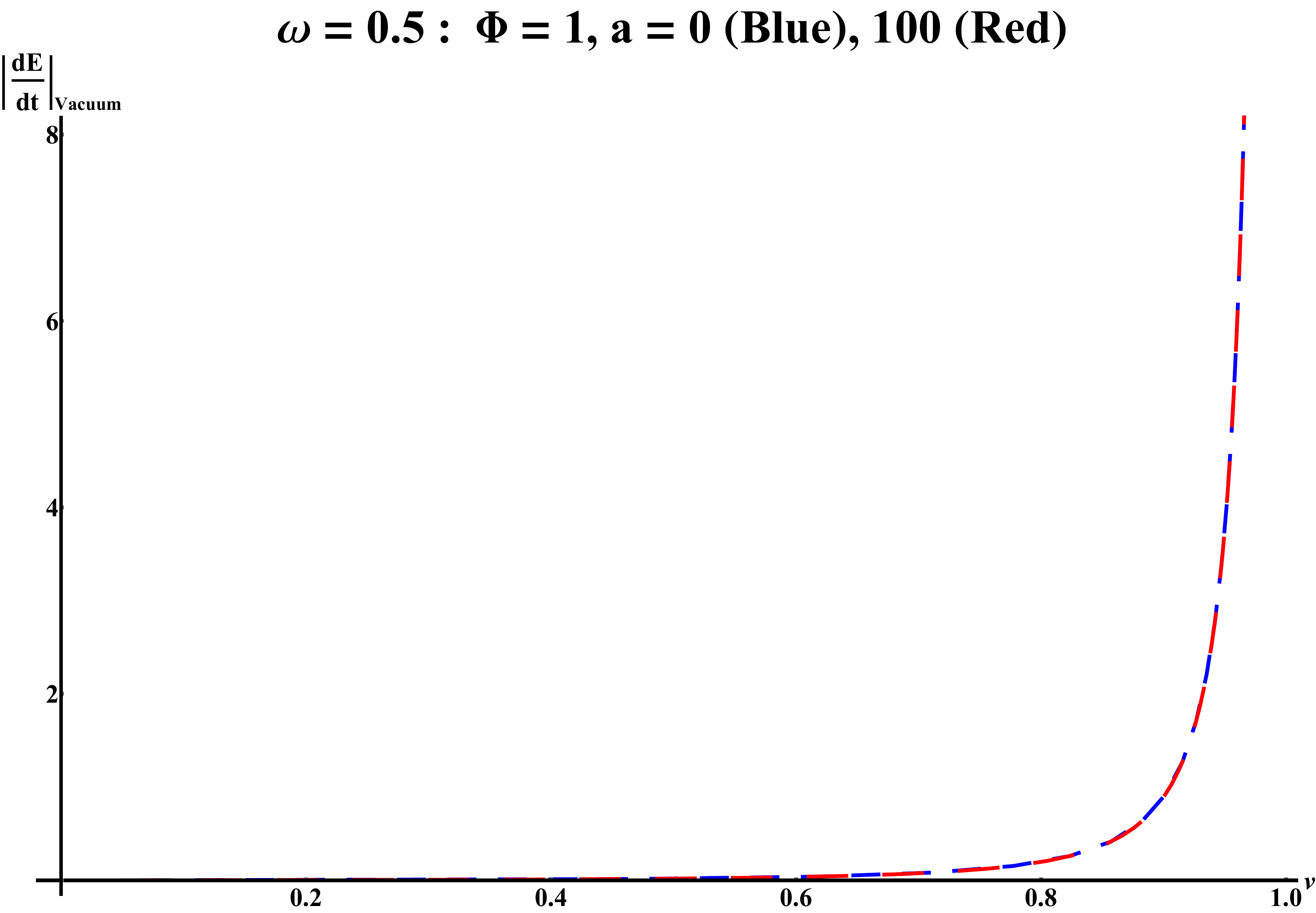}}
	\subfigure[]{\includegraphics[scale=0.09]{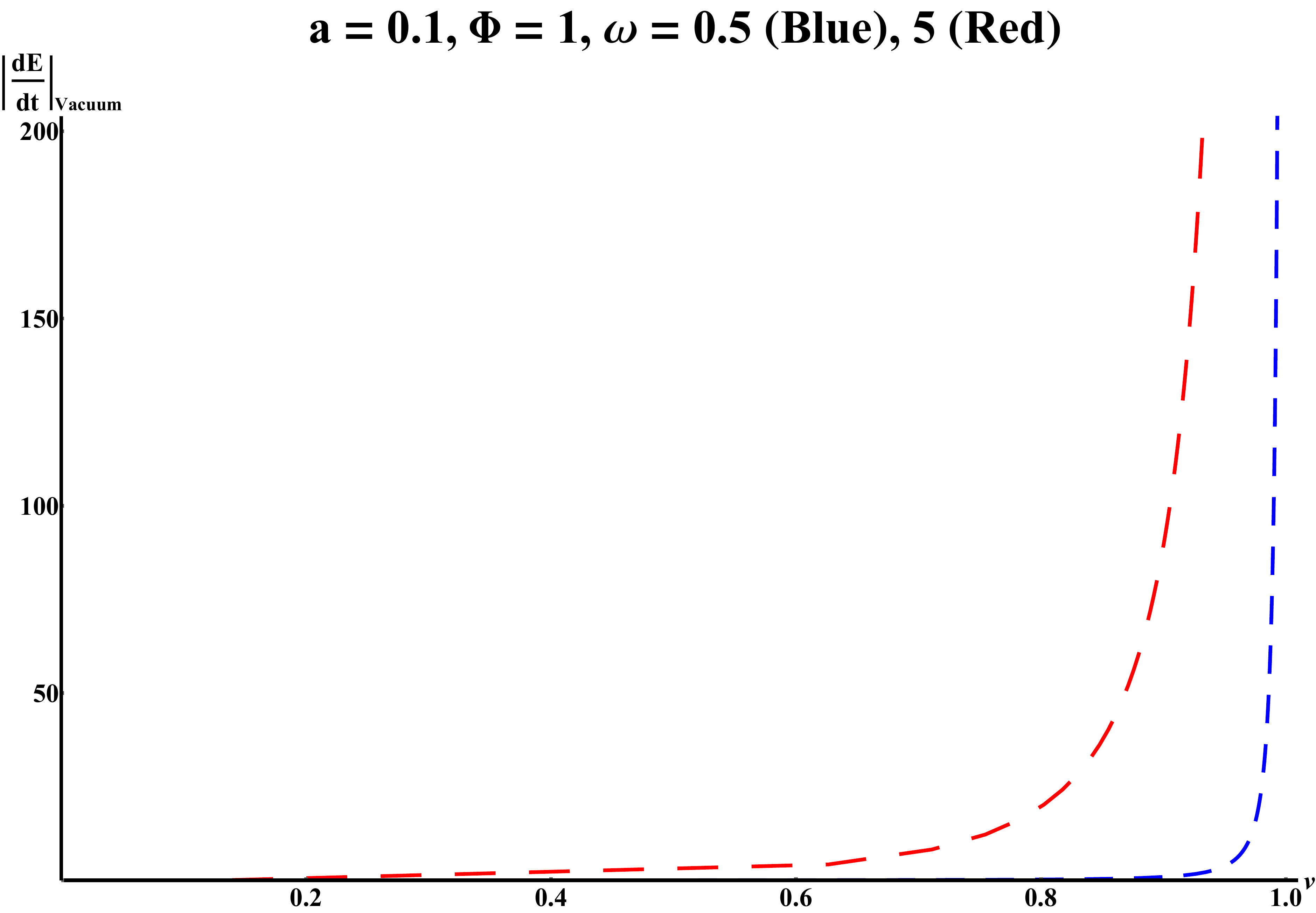}}
	\caption{Plot of vacuum energy loss with respect to quark speed ($v$) for plot (a) fixed $\omega = 0.5,\,a=0.1$ and different $\Phi = 0$ (blue)and $10$ (red). Plot (b). fixed $\omega = 0.5,\,\Phi=1$ and different $a = 0$ (blue) and $100$ (red). Plot (c). fixed $a = 0.1,\,\Phi=1$ and different $\omega = 0.5$ (blue) and $5$ (red).}
	\label{Plot_vacuum_vs_v}
\end{figure}

\begin{figure}[!h]
	\centering
	\subfigure[]{\includegraphics[scale=0.09]{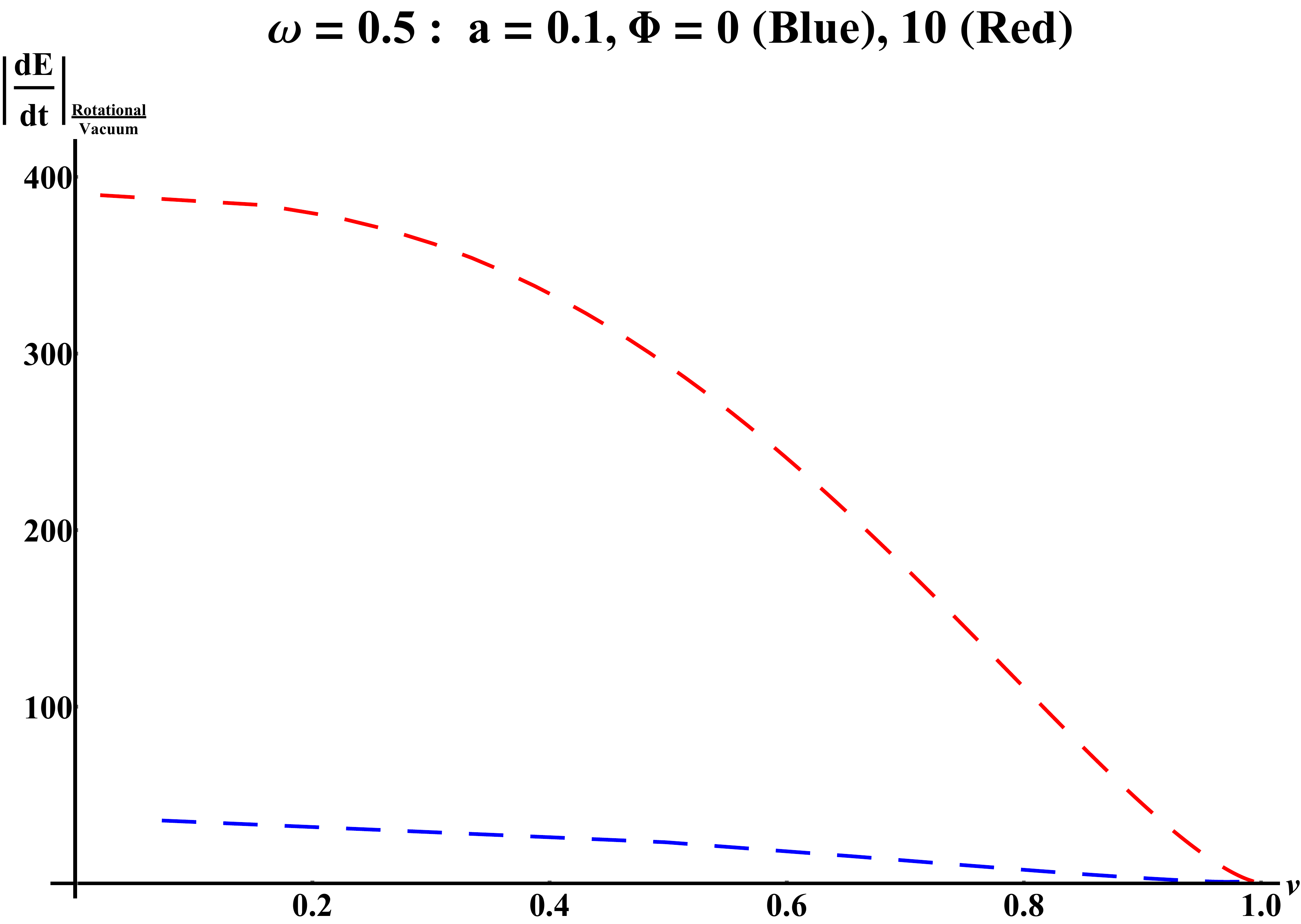}}
	\subfigure[]{\includegraphics[scale=0.09]{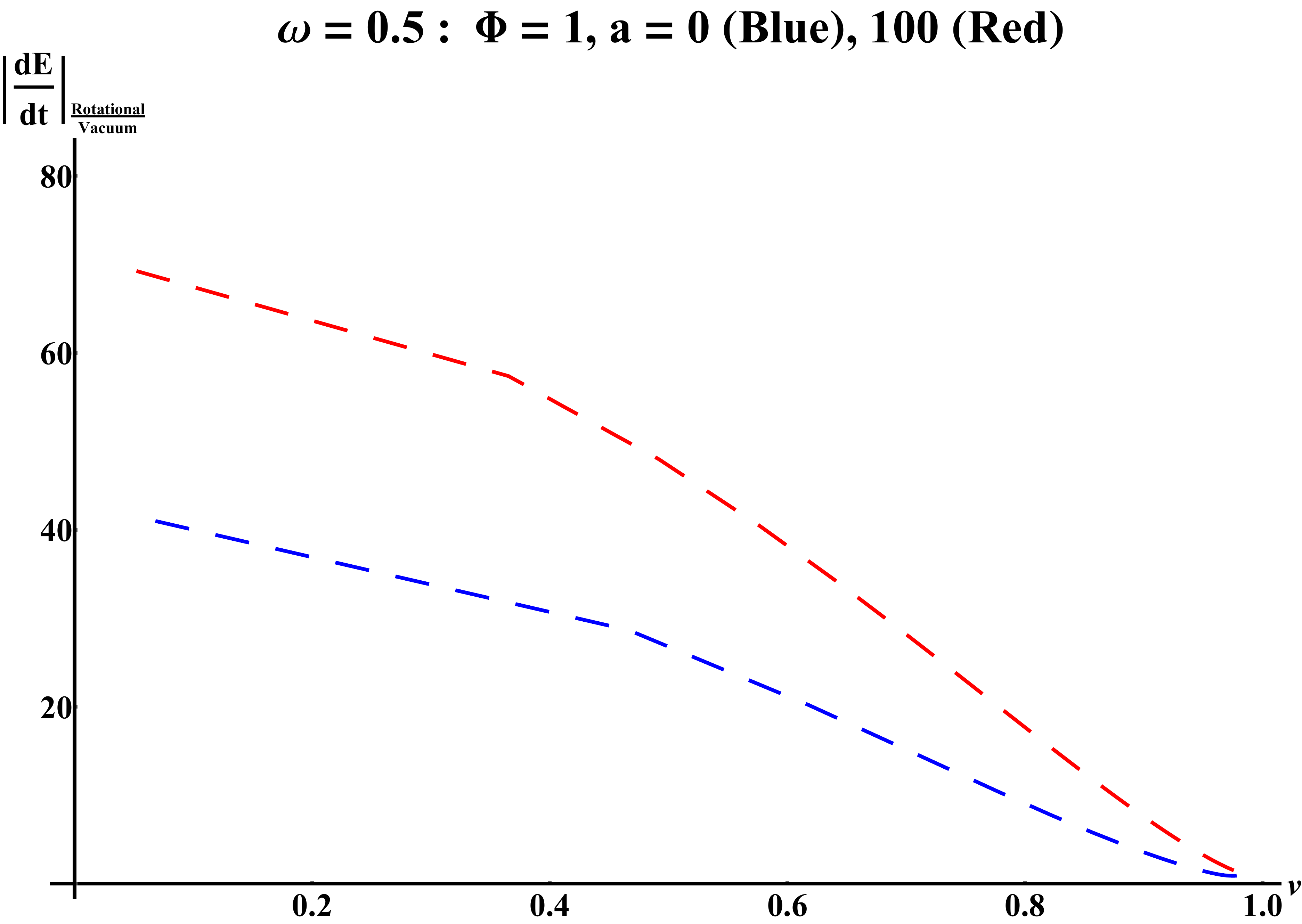}}
	\subfigure[]{\includegraphics[scale=0.09]{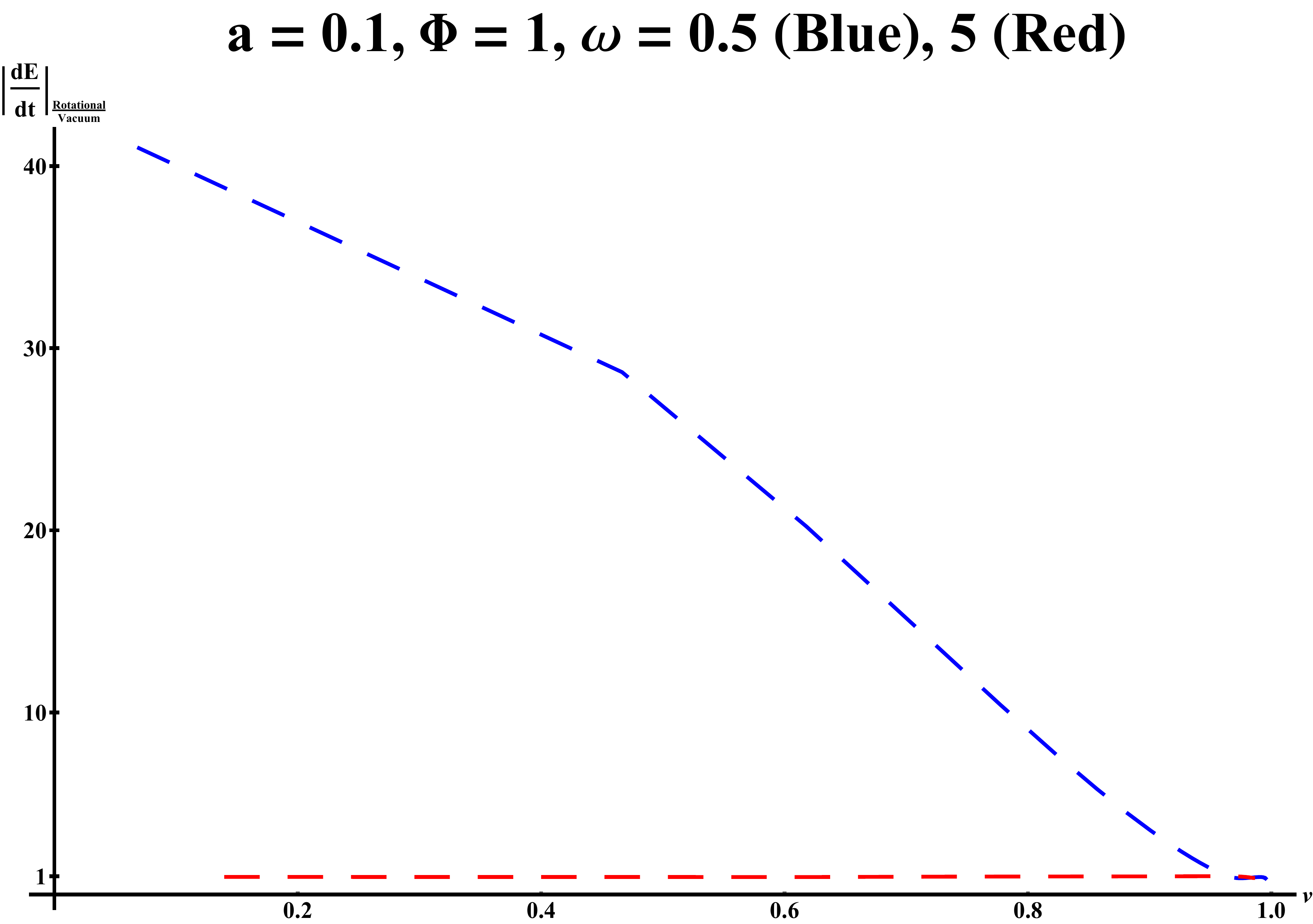}}
	\caption{Plot of the ratio of rotational to vacuum energy loss with respect to quark speed ($v$) for plot (a) fixed $\omega = 0.5$, $a=0.1$ and different $\Phi = 0$ (blue)and $10$ (red). Plot (b) . fixed $\omega = 0.5$, $\Phi=1$ and different $a = 0$ (blue) and $100$ (red).  Plot (c) . fixed $a = 0.1$, $\Phi=1$ and different $\omega = 0.5$ (blue) and $5$ (red).}
	\label{Plot_ratio_of_total_energy_loss_to_vacuum_vs_v}
\end{figure}

\begin{figure}[!h]
	\centering
	\subfigure[]{\includegraphics[scale=0.09]{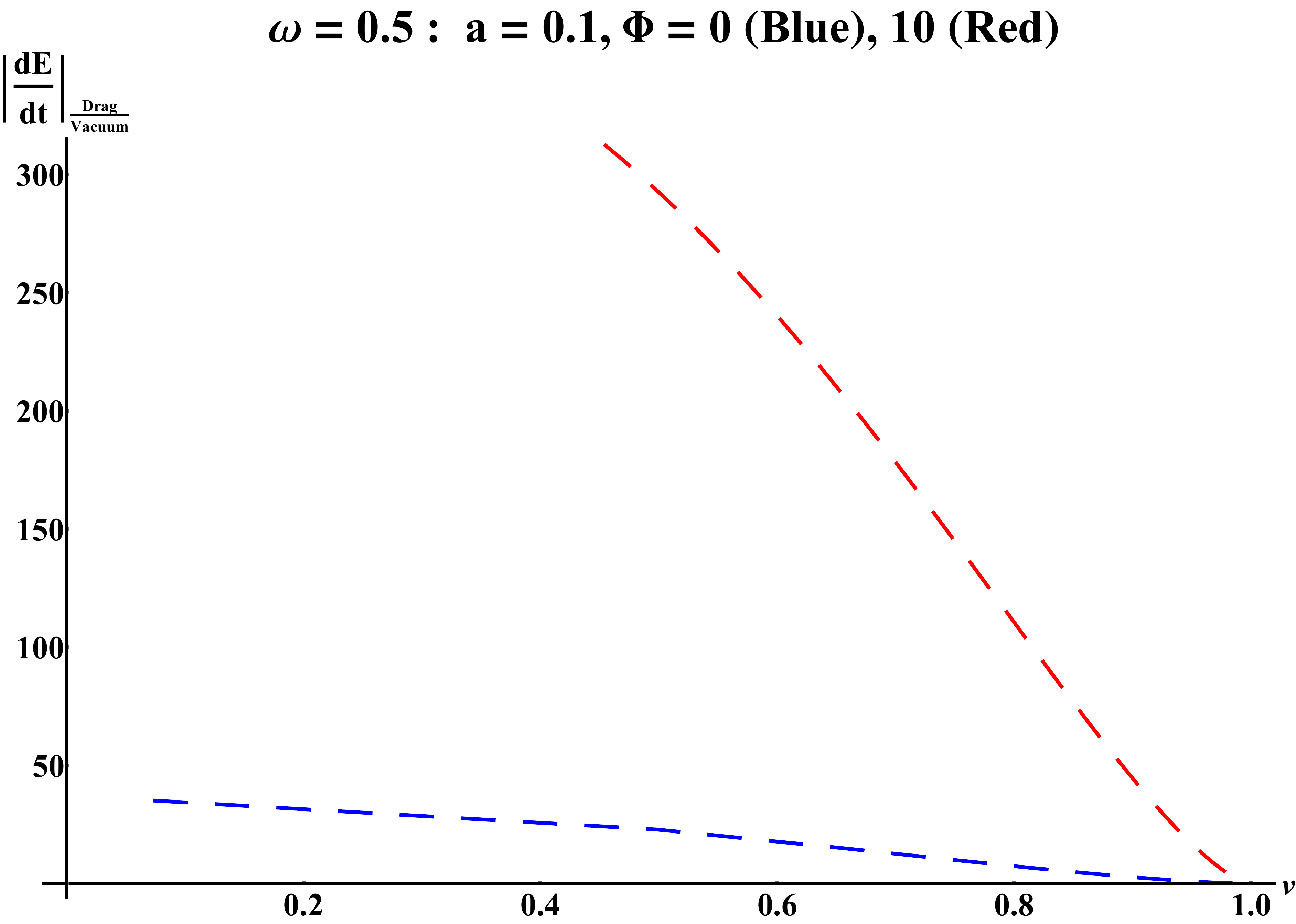}}
	\subfigure[]{\includegraphics[scale=0.09]{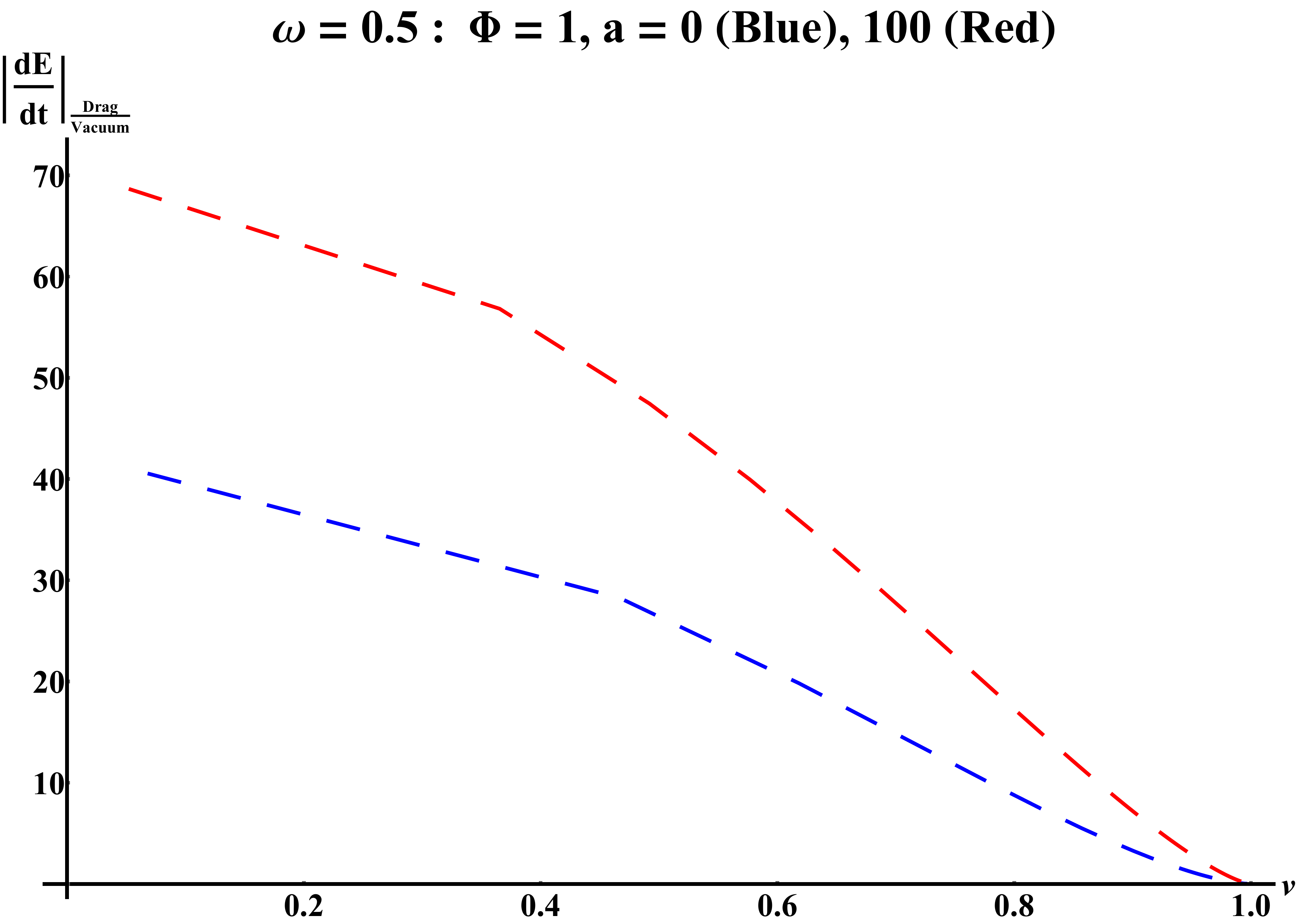}}
	\subfigure[]{\includegraphics[scale=0.09]{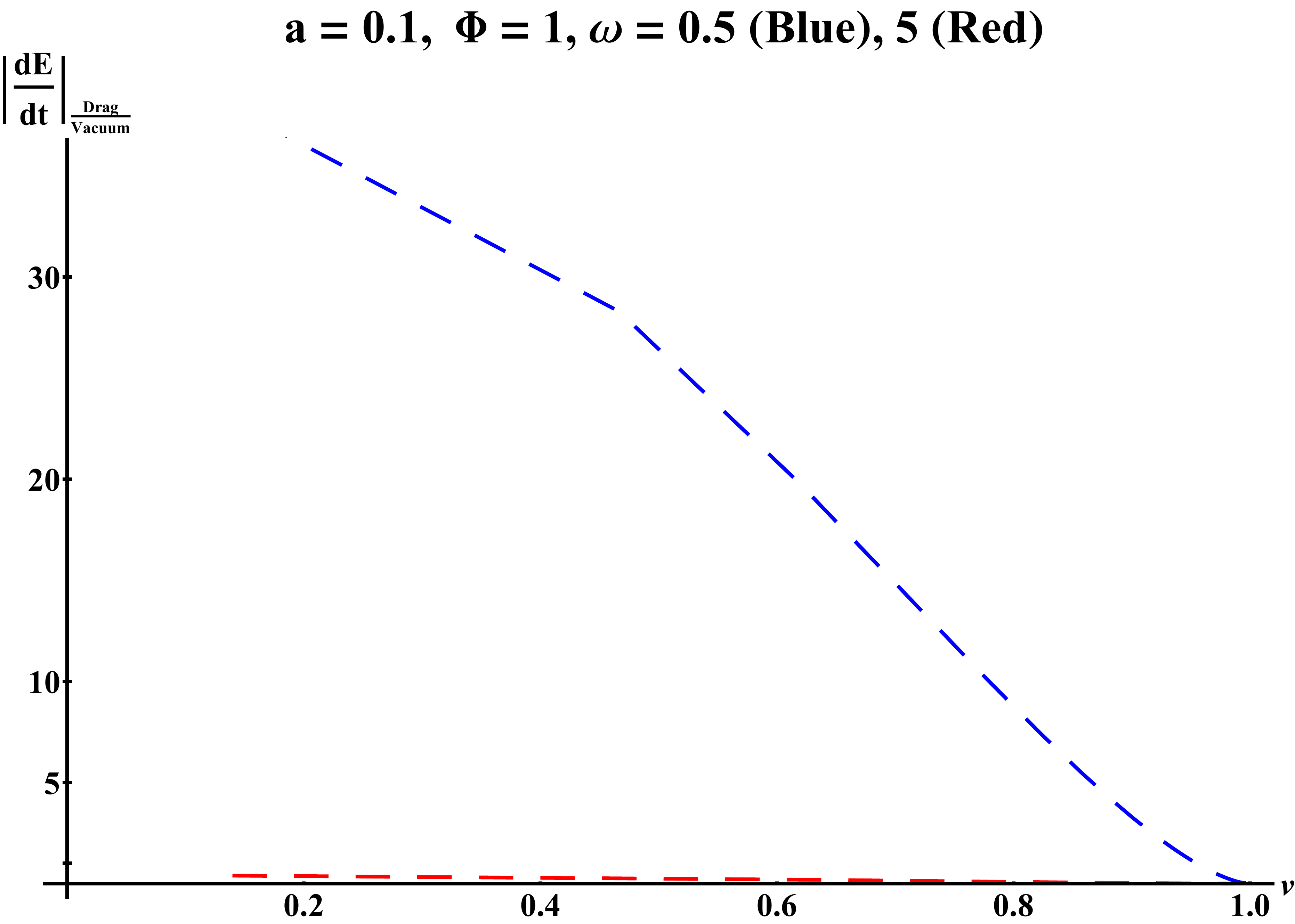}}
\caption{Plot of the ratio of drag energy loss to vacuum energy loss with respect to quark speed ($v$) for, plot (a) fixed $\omega = 0.5,\,a=0.1$ and different $\Phi = 0$ (blue)and $10$ (red). Plot (b). fixed $\omega = 0.5,\,\Phi=1$ and different $a = 0$ (blue) and $100$ (red). Plot (c). fixed $a = 0.1,\,\Phi=1$ and different $\omega = 0.5$ (blue) and $5$ (red).}
	\label{Plot_ratio_of_drag_energy_loss_to_vacuum_vs_v}
\end{figure}

\begin{figure}[!h]
	\centering
	\subfigure[]{\includegraphics[scale=0.08]{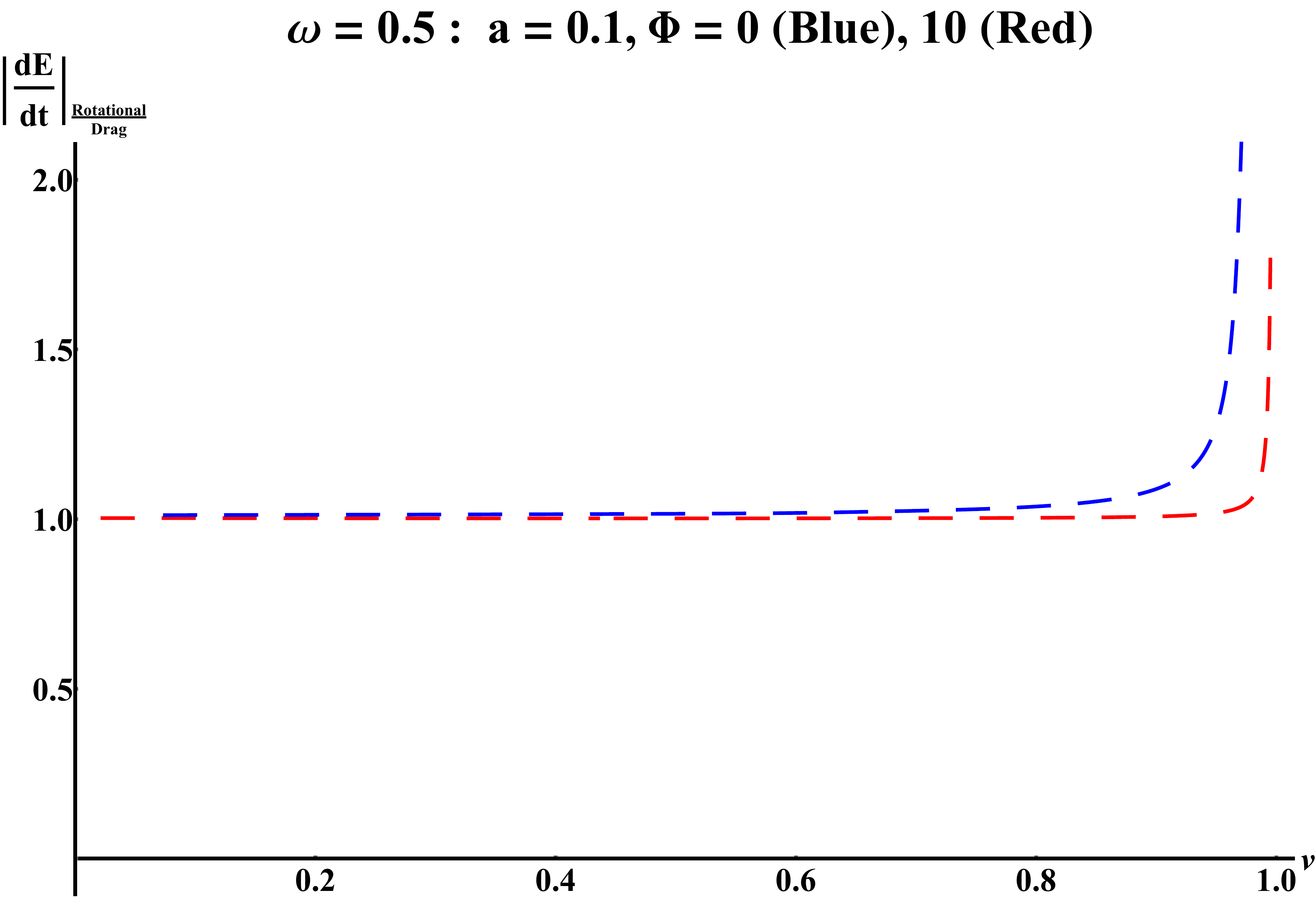}}
	\hspace{.2in}\subfigure[]{\includegraphics[scale=0.08]{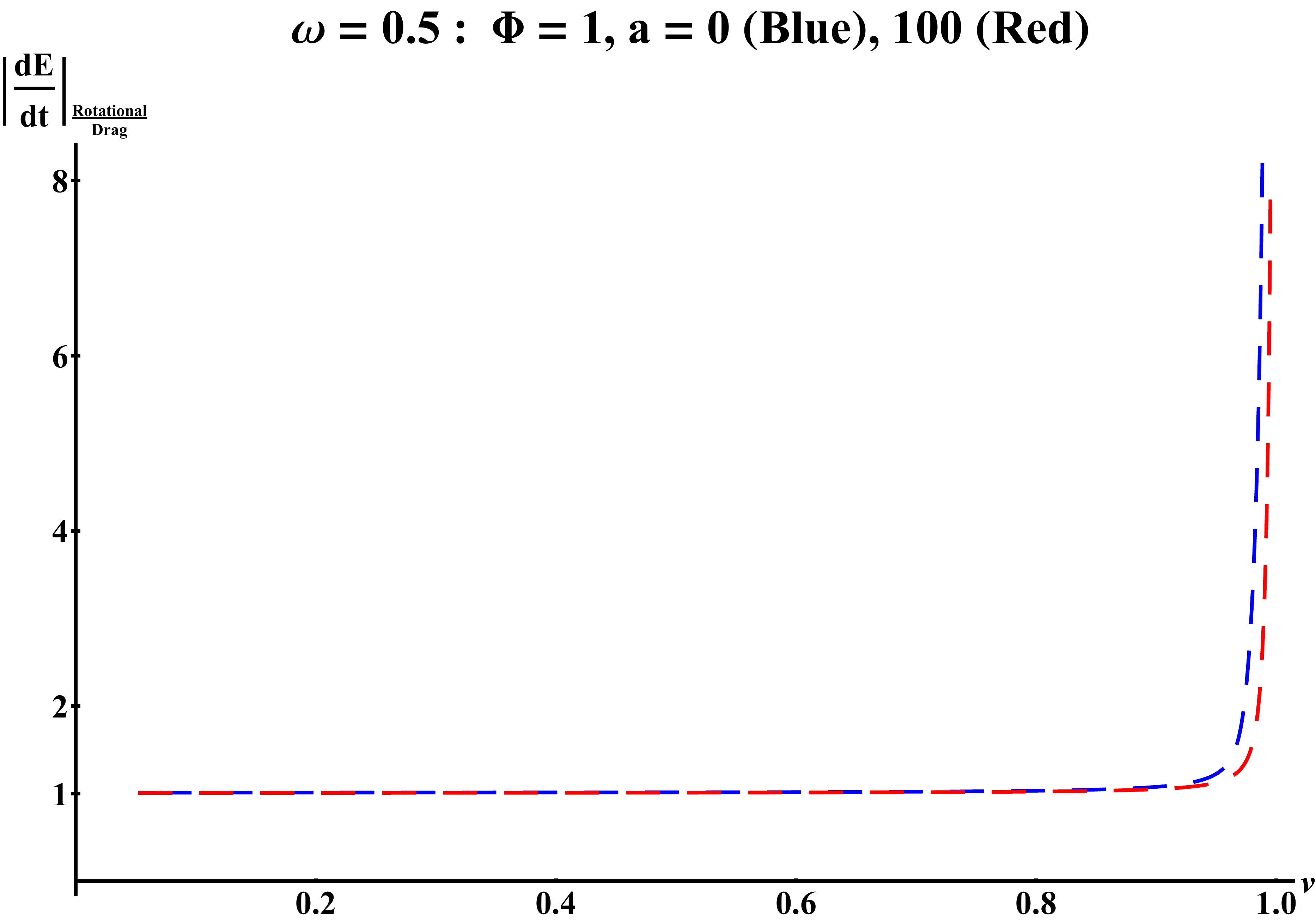}}
	\hspace{.2in}\subfigure[]{\includegraphics[scale=0.08]{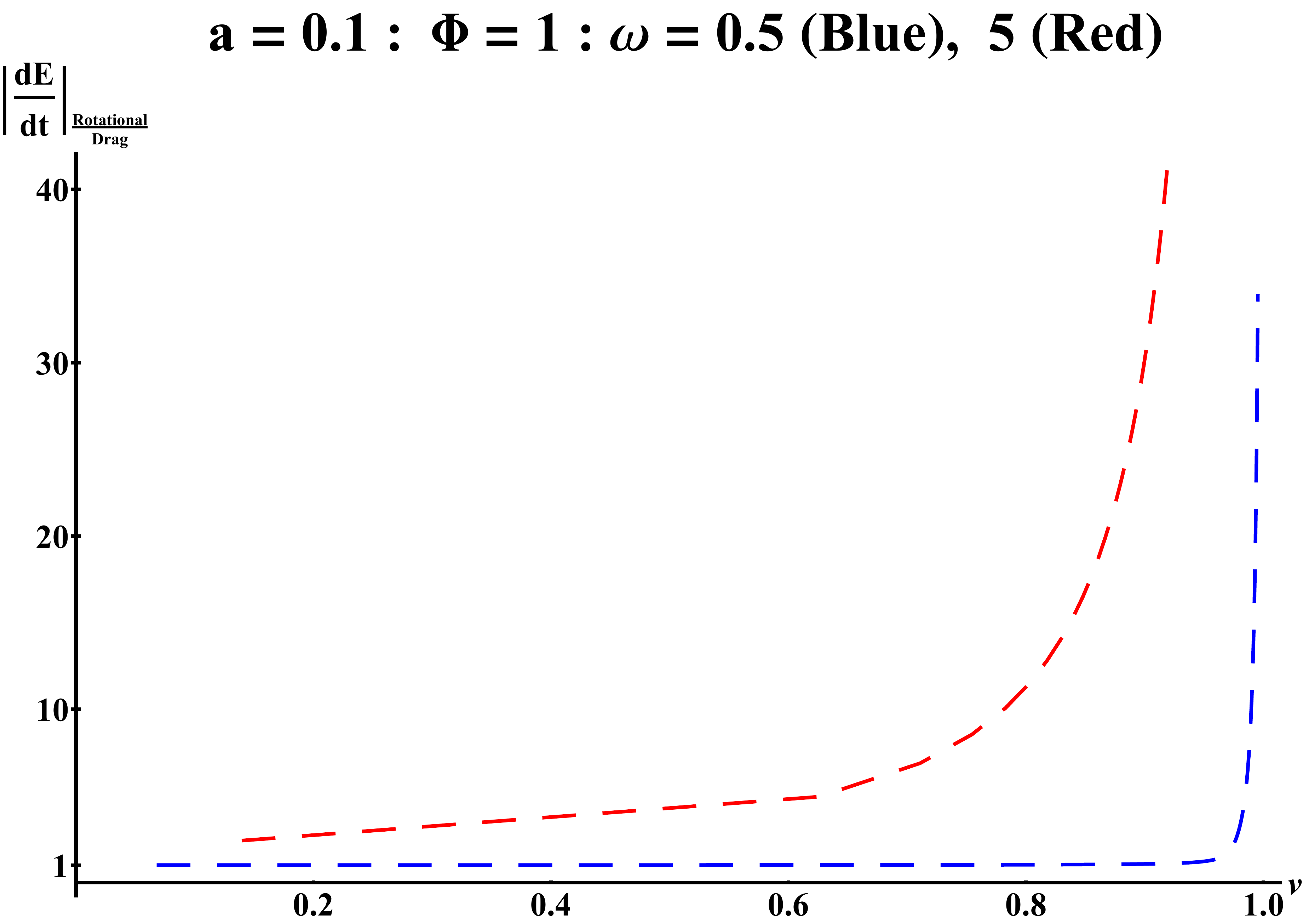}}
	\caption{Plot of the ratio of rotational to drag energy loss with respect to quark speed ($v$) for plot (a) fixed $\omega = 0.5$ $a=0.1$ and different $\Phi = 0$ (blue) and $10$ (red). Plot (b) . fixed $\omega = 0.5$, $\Phi=1$ and different $a = 0$ (blue) and $100$ (red). Plot (c) . fixed $a = 0.1$, $\Phi=1$ and different $\omega = 0.5$ (blue) and $5$ (red).}
	\label{Plot_ratio_of_total_energy_loss_to_drag_vs_v}
\end{figure}

The energy loss of a rotating heavy quark in the vacuum has also been studied \cite{Mikhailov2003}. Here, vacuum means the zero temperature gauge theory without plasma, baryons and flavour quarks. Therefore, the radiation is nothing but the Lienard's electromagnetic radiation due to the acceleration of the charge of the probe quark and it depends only on the velocity of the quark.  The vacuum radiation can be calculated and it obtains the form as,
\begin{equation}
	\left.\frac{dE}{dt}
	\right|_{vacuum} \approx \frac{v^2 \omega^2}{(1-v^2)^2}
\end{equation}
The nature of the rate of energy loss in the vacuum against the linear velocity is shown in figure (\ref{Plot_vacuum_vs_v}). The plot is also depicting that the radiation does not depend on chemical potential and string density but depends only on velocity of the probe quark.
In figure (\ref{Plot_ratio_of_total_energy_loss_to_vacuum_vs_v}), we have plotted the ratio of rotational energy loss to the vacuum energy loss against the linear speed of quark $v$. It is observed that irrespective of the values of angular speed, chemical potential and string density, the rotational energy loss is dominating over the vacuum energy loss if the linear speed is less than the speed of light. The dominance is higher for higher values of chemical potential and string density. But the dominance decreases for the increase of angular speed. Once the linear speed of quark approaches to the speed of light both the radiation radiates equally with opposite phase and interfering destructively.
In figure (\ref{Plot_ratio_of_drag_energy_loss_to_vacuum_vs_v}), we have drawn the ratio of energy loss due to drag to the energy loss due to vacuum against the linear speed of quark $v$. The nature is same as the ratio of rotational energy loss and vacuum energy loss.
Comparing the figure (\ref{Plot_ratio_of_total_energy_loss_to_vacuum_vs_v}) and (\ref{Plot_ratio_of_drag_energy_loss_to_vacuum_vs_v}) it can be concluded that the radiation due to the rotation and drag are in same phase and interferes constructively when the quark moves with speed of light. To validate the conclusion we plot the rate of energy loss for the rotation and drag in figure (\ref{Plot_ratio_of_total_energy_loss_to_drag_vs_v}). 
  Notice that for small angular speed or the linear velocity the rate of energy loss is equal for both the ways since the ratio of rotational energy loss to drag energy loss approaches unity. As the angular speed or linear velocity approaches velocity of light, the ratio of energy loss takes values higher than unity, implying that energy loss due to rotation and the drag energy loss are in same phase and interferes constructively and appears as dominance of the rotational energy loss over the drag energy loss. Further, it is observed that for a higher value of angular speed or linear velocity, the rate of dominance of the rotational energy loss over the drag energy loss is reduced for the increase of chemical potential $\Phi$ and back reaction or string density $a$. The reduction enhanced effectively for the increase of chemical potential than the increase of string density.


\section{Summary and Conclusion}
\label{discussion}
This work studied the hydrodynamical properties of the baryon rich strongly coupled $\mathcal{N} = 4$ SYM finite temperature plasma with flavour quarks using the gauge/gravity duality. The gravity dual is the five dimensional charged AdS black hole with uniformly attached cloud of strings. The attached end points of the strings are representing the flavour quarks and the potential at the boundary due to the charge of the black hole represents the baryon chemical potential density of the gauge theory. For the study of hydrodynamical properties, we have considered an external heavy probe quark moving in the thermal plasma and calculated the dissipative force acting on it and energy loss due to this force and rotational motion.  Quark-antiquark pair is considered as an external probe to study the jet quenching parameter, screening length. In the gravity dual the probe quark is represented by a long string hanging from the four dimensional boundary and elongated up to the horizon of the five dimensional bulk spacetime. The attached end point of the string moves along the boundary with $SU(N)$ fundamental charge. The probe quark-antiquark is represented by a long string whose both ends are attached to the boundary and the body of the string is hanging in the bulk.\\
The dissipative force experienced by the external quark is a function of the velocity, temperature of the black hole, potential due to the charge and the density of the uniformly distributed string cloud. The drag force has been plotted with respect to the velocity of the probe quark, keeping the two parameters out of the other three parameters fixed and the remaining one assigned two different values. The dissipative force experienced by the external probe quark gets enhanced for the increment of the any parameters.\\
The jet quenching parameter connecting to the energy caused by the suppression of the heavy probe quark-antiquark pair with high transverse momentum is computed. We observed from our study that the jet quenching parameter monotonically increases with increase of the temperature, chemical potential and string cloud density, signifying the enhancement of heavy quark suppression.\\
The screening length of the quark-antiquark pair has been analysed with respect to the constant of motion $W$ for different fixed values of the rapidity parameter, string density, temperature and chemical potential. It is noticed from the study that the screening length reduces and binding energy raises with increase of the value of any parameter and triggers the instability of the bound state of the $q\bar{q}$ pair. More precisely due to the presence of baryon chemical potential, the phase transition from confined to deconfined state occurs at lower temperature which is consistent with the QGP phase diagram.\\
Finally, the energy loss associated with the heavy quark rotating with constant angular speed in the thermal plasma has been studied. The energy loss crucially depends on the radial profile of the probe string. For a fixed string density and chemical potential, the radius of the probe string increases as we go away from the boundary of the bulk spacetime if the angular speed is greater than one. Otherwise the radius remains same. For fixed angular velocity, the radius reduces for the increase of chemical potential and string density. The rotational energy loss has been plotted with respect to the quark velocity, for different values of angular velocity, string density and chemical potential. It is observed that the energy loss gets enhanced on raising any of the parameters for the higher velocity. The effect of chemical potential on the enhancement of energy loss is more than the other parameters. On the other hand energy loss reduces to zero when the probe quark moves slowly. Then we studied the energy loss due to the drag force by plotting it against the velocity and it is also enhanced for the increase of any parameters. The enhancement rate is more accelerated due to the  increase of temperature and chemical potential than the string density. We also discussed the energy loss of the probe quark against velocity in vacuum or at zero temperature without the presence of baryons and flavour quarks. In this condition the radiation becomes Lienard’s electromagnetic radiation due to the acceleration of the probe quark and it depends only on the velocity of the quark. Finally, we found the ratio among these energy losses. First we found the ratio between rotational energy loss and vacuum energy loss and we observed that the rotational energy loss dominating more over the vacuum energy loss for small velocity of the probe quark and slowly reducing the dominance and becoming zero at the speed of light resulting destructive interference between the two radiation with opposite phase. The dominance at the lower velocity is more for the higher value of chemical potential and string density. Whereas the dominance is reduced for higher value of angular speed. Similarly, we found the ratio between drag and vacuum energy loss and observed the similar behaviour. Finally, we found the ratio of rotational energy loss and drag energy loss and we observed that both the energy loss are radiating same amount of radiation at the lower velocity of probe quark and interfering constructively when quark speed approaches towards speed of light for the lower angular speed. But for the higher angular speed the rotational energy loss is always dominating over the drag energy loss.\\

From the overall study, we can conclude that except the screening length, the other hydrodynamical properties of the thermal plasma are enhanced due to the  presence of baryon and flavour quarks in the thermal plasma.

\section{Acknowledgement}
We would like to thank K. P. Sherpa, I. K. Pandey and B. Sharma for going through the literature and making valuable comments on it. R.P. acknowledges the TMA-PAI research grant provided by Sikkim Manipal University.

\printbibliography
\end{document}